\documentclass[useAMS,usenatbib,usegraphicx]{mn2e}

\usepackage[british]{babel}
\usepackage{times}
\usepackage{latexsym,amssymb}
\usepackage[fleqn]{amsmath}

\usepackage{graphics,graphicx}
\usepackage{epstopdf}
\usepackage{epsfig}
\usepackage[absolute]{textpos}

\usepackage{natbib}
\usepackage{journalnames}
\usepackage{color}

\usepackage[T1]{fontenc}
\usepackage{aecompl}

\usepackage{caption}
\usepackage{comment}
\usepackage{lipsum}
\usepackage{array}
\usepackage{dcolumn}
\newcounter{subfigure}
\usepackage{amsmath} 

\usepackage[draft]{hyperref}
\usepackage{longtable}

\newcommand{\kms}{km\,s$^{-1}$}

\newcommand{\atlas}{\texttt{ATLAS$^\mathrm{3D}$}} 

\newcommand{\du}{\mathrm{d}}               

\newcommand{\Vrms}{\ensuremath{V_\mathrm{rms}}}

\title[Towards a new classification of galaxies]{
Towards a new classification of galaxies: principal component analysis of CALIFA circular velocity curves}
\author[V. Kalinova et al.]{V. Kalinova$^{1,2}$\thanks{E-mail: kalinova@mpifr.de},
D. Colombo$^{1,2}$,
E. Rosolowsky$^{1}$,
R. Kannan$^{3}$,
L. Galbany$^{4}$,
\newauthor{R. Garc{\'i}a-Benito$^{5}$,
R. Gonz{\'a}lez Delgado$^{5}$,
S. F. S{\'a}nchez$^{6}$,}
T. Ruiz-Lara$^{7,8,9,10}$,
\newauthor{J. M{\'e}ndez-Abreu$^{11}$,
C. Catal{\'a}n-Torrecilla$^{12}$,
L. S{\'a}nchez-Menguiano$^{5,7}$,}
\newauthor{A. de Lorenzo-C{\'a}ceres$^{7,11}$,
L. Costantin$^{13}$,
E. Florido$^{7,8}$,
K. Kodaira$^{2}$,}
\newauthor{R. A. Marino$^{14}$,
R. L{\"a}sker$^{15}$,
J. Bland-Hawthorn$^{16}$}\\
\\
$^{1}$ Department of Physics 4-181 CCIS, University of Alberta, Edmonton AB T6G 2E1, Canada\\
$^{2}$ Max Planck Institute for Radio Astronomy, Auf dem H\"ugel 69, D-53121 Bonn, Germany\\ 
$^{3}$ Department of Physics, Kavli Institute for Astrophysics \& Space Research, Massachusetts Institute of Technology, Cambridge, MA 02139, USA \\ 
$^{4}$ PITT PACC, Department of Physics and Astronomy, University of Pittsburgh, Pittsburgh, PA 15260, USA \\    
$^{5}$ Instituto de Astrof\'isica de Andaluc\'ia, CSIC, Apartado de correos 3004, E-18080 Granada, Spain \\
$^{6}$ Instituto de Astronom\'\i a,Universidad Nacional Auton\'oma de M\'exico, A.P. 70-264, 04510 M\'exico, D.F., M\'exico \\
$^{7}$ Departamento de F\'isica Te\'orica y del Cosmos, Universidad de Granada, Campus de Fuentenueva, E-18071 Granada, Spain \\
$^{8}$ Instituto Carlos I de F\'isica Te\'orica y Computacional, Universidad de Granada, E-18071 Granada, Spain \\
$^{9}$ Instituto de Astrof\'isica de Canarias, Calle V\'ia L\'actea s/n, E-38205 La Laguna, Tenerife, Spain \\
$^{10}$ Departamento de Astrof\'isica, Universidad de La Laguna, E-38200 La Laguna, Tenerife, Spain \\
$^{11}$ School of Physics and Astronomy, University of St. Andrews, SUPA, North Haugh, St. Andrews, KY16 9SS, UK \\
$^{12}$ Departamento de Astrof\'isica y CC. de la Atm\'osfera, Universidad Complutense de Madrid, E-28040 Madrid, Spain\\
$^{13}$ Dipartimento di Fisica e Astronomia 'G. Galilei', Universit\'a di Padova, vicolo dell'Osservatorio 3, I-35122 Padova, Italy\\
$^{14}$ Department of Physics, Institute for Astronomy, ETH Z\"urich, CH-8093 Z\"urich, Switzerland\\
$^{15}$ Finnish Centre for Astronomy with ESO (FINCA), University of Turku, V\"ais\"al\"antie 20, FI-21500 Kaarina, Finland \\
$^{16}$ Sydney Institute for Astronomy, School of Physics A28, University of Sydney, NSW 2006, Australia\\
}

\date{Accepted 2017 April 8. Received 2017 April 4; in original form 2016 December 12}

\pagerange{\pageref{firstpage}--\pageref{lastpage}} \pubyear{2017}

\begin{document}


\label{firstpage}

\maketitle

\begin{abstract}

We present a galaxy classification system for 238 (E1--Sdm) CALIFA (Calar Alto Legacy Integral Field Area) galaxies based on the shapes and amplitudes of their circular velocity curves (CVCs). We infer the CVCs from the de-projected surface brightness of the galaxies, after scaling by a constant mass-to-light ratio based on stellar dynamics - solving axisymmetric Jeans equations via fitting the second velocity moment $V_{\mathrm{rms}}=\sqrt{V^2+\sigma^2}$ of the stellar kinematics.
We use principal component analysis (PCA) applied to the CVC shapes to find characteristic features and use a $k$-means classifier to separate circular curves into classes.  This objective classification method identifies four different classes, which we name slow-rising (SR), flat (FL),  round-peaked (RP) and sharp-peaked (SP) circular curves.  

SR are typical for low-mass, late-type (Sb--Sdm), young, faint, metal-poor and disc-dominated galaxies. SP  are typical for high-mass, early-type (E1--E7), old, bright, metal-rich and bulge-dominated galaxies. FL and RP appear presented by galaxies with intermediate mass, age, luminosity, metallicity, bulge-to-disk ratio and morphologies (E4-S0a, Sa-Sbc). The discrepancy mass factor, $f_d=1-M_{*}/M_{dyn}$,  
have the largest value for SR and SP classes ($\sim$ 74 per cent and $\sim$ 71 per cent, respectively) in contrast to the FL and RP classes (with $\sim$ 59 per cent and $\sim$ 61 per cent, respectively). Circular curve classification presents an alternative to typical morphological classification and appears more tightly linked to galaxy evolution.


\end{abstract}


\clearpage
\section{Introduction}
\label{S:intro4}
Circular velocity curves (CVCs) are considered one of the best tools to trace the mass distribution of galaxies. Since the earliest studies \citep{Rubin1980, Bosma1981, Persic1988}, 
the non-Keplerian shape of the curves implied that the mass derived with this technique was much higher than the total luminous mass (accounting for stars and gas).  
This inconsistency suggested that a significant fraction of galaxies is dark matter (hereafter DM), which becomes progressively more dominant towards the outskirts of the galaxies \citep{Persic1988, Broeils1992}.

Moreover, the shape and the amplitude of CVCs\footnote{When we discuss {\sc Hi} 21-cm CVCs in the paper, we usually refer to their rotation curve, i.e., azimuthal velocity $V_{\phi}$.
We use this approximation here since for cold gas it is valid $V_c\approx V_{\phi}$, where the vertical component of the velocity is negligible (\citealt{Binney1998}).}
are both closely related to the gravitational potential of the galaxy, and therefore, can be used to obtain important information about the different components of the galaxies (\citealt{Noordermeeretal2007}).

In almost all disc galaxies, the stellar disc contribution can be scaled to explain all features of the observed CVCs out to $\sim 2$ disc scale lengths (e.g., \citealt{Kalnajs1983,Palunas2000}). 
This indicates that the total mass density and luminous mass density are closely connected (e.g., \citealt{Sancisi2004,Swaters2009}). 

There are several indications that the {\em shape} of the CVC, in addition to its overall amplitude, is connected to galaxy properties and possibly evolution.  
\cite{Rubin1985} pointed out that, within the same morphological type, spiral galaxies show a progression of central velocity gradients and maximum circular velocities with increasing absolute luminosities. 
The correlation between the light distribution and the inner rise of the CVCs is well known for spiral galaxies (e.g., \citealt{Kent1987, Corradi1990}). 
\cite{Swaters2009} have found that dwarf galaxies with a central concentration of light also have CVCs that rise more steeply in the centre than the CVCs of dwarf galaxies that do not have a central concentration of light. 
They observed a correlation between the light distribution and the inner CVC shape, as seen in both spiral and late-type dwarf galaxies, 
implying that galaxies with stronger central concentrations of light also have higher central mass densities.  
This correlation suggests that the luminous mass dominates the gravitational potential in the central regions, even in low surface brightness (SB) dwarf galaxies.

\cite{Avila-Reese2002} discuss the shapes and decomposition of CVCs of galaxies formed within growing cold dark matter haloes.  
They find that CVC shape correlates mainly with the SB, the luminous mass fraction, and the bulge fraction of the galaxies. 
Further, the galaxy's SB scales with the bulge-to-disc (B/D) ratio, and a steeper slope of decline from a high peak in the CVC.  
Their high-SB models can be maximal discs only when the haloes have a shallow core. The low-SB models possess sub-maximal discs containing DM (\citealt{Sackett1997}). 

The shape and the amplitude of the CVCs reflect the mass distribution of both luminous matter and
DM in galaxies (e.g., \citealt{Zasov1983}; \citealt{Corradi1990}; \citealt{Swaters2009}; \citealt{Martinsson2013}).
The next step is to classify the CVCs by common properties and try to draw global conclusions for the internal structure of the galaxies, 
formation and evolution, through studying large data sets with diversity in morphological types.

There is a suite of dynamics-based classification schemes already present in the literature.  We describe several of these approaches below, focusing on those studies with similar data, 
approaches or outcomes that are relevant to this work.  For example, classification of {\sc Hi} 21-cm CVCs was proposed by \cite{Sofue1999} using a large sample of Sb--Sc galaxies. 
They observe a steep nuclear rise of galaxy  CVC and conclude that is a universal property for massive Sb and Sc galaxies, regardless of the existence of a bar and morphological peculiarities. 
However, less massive galaxies tend to show a rigid-body rise. They classify the observed CVCs into the following three types, according to their behaviour in the central regions. {\it Central Peak} type: 
CVC attains a sharp maximum near the centre at $R\sim$ 100--500 pc, followed by a dip at $\sim$ 1 kpc, then by a broad maximum of the disc component (e.g., Milky Way). 
{\it No-Central Peak} type: The CVC rises steeply at the center, followed immediately by a flat part. {\it Rigid-body} type: the CVC increases mildly from the centre in a rigid-body 
fashion within the central 1 kpc. This type is found in less massive Sc-type galaxies and it has been already reported by \cite{Casertano1991}.

\cite{Wakamatsu1976} proposes a two-dimensional classification of 22 disc galaxies using their optical and/or {\sc Hi} 21-cm CVCs. For the goal, \cite{Wakamatsu1976} 
use two independent parameters: $\emph{k}$ ($=V_{\mathrm{max}}^2/r_{\mathrm{max}}$) and $\mathcal{M}$ (mass of galaxy within Holmberg's radius; \citealt{Holmberg1958}), 
where $\emph{k}$ is the centrifugal force at the radius $r_{\mathrm{max}}$ and maximum value of the 
circular velocity, $V_{\mathrm{max}}$.
The parameter $\emph{k}$ is found to be correlated with the morphological type (decreasing from early-type to late-type galaxies) and the mass ratio of the bulge to the disc, but not correlated with $\mathcal{M}$.
Using dynamical properties of the 22 galaxies, they also find that late-type galaxies have systematically larger angular momenta and smaller rotational energies in comparison to early-type galaxies.

\cite{Shostak1977} shows a relation between a galaxy's Hubble type and the shape of its {\sc Hi} profile. 
They find a type progression from Sbc to Irr, where the CVC (represented by two linear sections) turnover radius increases relative to the full width at half-maximum (FWHM) of {\sc Hi} line.  
They find that Sbc galaxies have CVCs that rise three to five times faster, relatively to their {\sc Hi} radii, than those of irregular galaxies. 
This indicates a greater degree of central mass concentration in the earlier types.

\cite{Biviano1991} find a significant correlation between the velocity gradients of 94 spiral galaxies and their arm classes (as given by \citealt{Elmegreen1982}), 
where galaxies with steeper curves tend to have a flocculent arm structure, and galaxies with flatter curves tend to have a grand design morphology. 
This result is consistent with the predictions of density wave theory for the formation of the spiral structure (\citealt{Lin1964}). 

\cite{Keel1993} classifies a set of spiral galaxies in pairs regarding their velocity structure in a few major classes. {\em Normal} -- galaxies with quickly rising velocity 
from the nucleus to a flat plateau; {\em Solid-body} (or linear) -- galaxies with solid-body rotation, sometimes extended over the entire measured area; 
{\em Slow} -- non-rotating or slow-rotating galaxies with total amplitudes no greater than 50 \kms despite elongated shapes and fairly high luminosities; 
{\em Dist} -- galaxies with disturbed or asymmetric velocity patterns.  Enhanced star formation is found for galaxies with large areas of solid-body rotation 
and for galaxies with more general kinds of disturbed velocity structure, with the highest levels occurring in a set of galaxies distinguished by anomalously small overall velocity amplitude.

\cite{Chattopadhyay2006} carry out an objective classification of four samples of spiral galaxies having extended CVCs beyond the optical radius. 
A multivariate statistical analysis (namely, principal component analysis, PCA) shows that about 96\% of the total variation is due to two components, 
one being the combination of absolute blue magnitude and maximum velocity beyond the optical region and the other being the central density of the halo. 
On the basis of PCA, a Fundamental Plane has been constructed that reduces the scatter in the Tully--Fisher relation up to 16\%. 
A multiple stepwise regression analysis of the variation of the overall shape of the CVCs shows that it is mainly determined by the central 
SB, while the shape purely in the outer part of the galaxy (beyond the optical radius) is mainly determined by the size of the galactic 
disc.

\cite{Marquez1999} also apply PCA to the properties of 22 isolated spiral galaxies and found a tight correlation between the 
B/D ratio of the galaxies and the gradient of the solid-body rotation region of the optical rotation curve, $G$. This parameter is defined as $G=(v^{ob}_G /\sin i)/r_G$, where $r_G$ (in kpc) is the radius of the inner region of solid-body rotation, $v^{ob}_G$  (in km $s^{-1}$) is the observed velocity amplitude at $r_G$ and $i$ is the galaxy disc inclination.
Furthermore, they found that their sample of isolated galaxies can be described by two eigenvectors representing 95 \% of the total variance: a scale parameter set by the size of the galaxy (or the total luminosity or the total mass) and the B/D ratio (or the G parameter). 

\cite{Wiegert2014} present a kinematic classification of 79 non-interacting spiral galaxies using their neutral hydrogen ({\sc Hi}) CVCs. 
Their method employs a simple parametrized form for the CVC to derive the three parameters: the maximum rotational velocity, 
the turnover radius, and a measure of the slope of the CVC beyond the turnover radius. \cite{Wiegert2014} also use the statistical 
hierarchical clustering method to guide their division of the resultant 3D distribution of galaxies into five classes. 
The class that contains galaxies with the largest rotational velocities has a mean morphological type of Sb/Sbc while the other classes tend to later types.  
They confirm correlations between increasing maximum rotational velocity and the following observed properties such as increasing brightness in the $B$-band, 
increasing size of the optical disc ($D_{25}$) and increasing star formation rate. The analysis also suggests that lower velocities are associated with a higher ratio of the {\sc Hi} mass over the dynamical mass. 

Other classification approaches have been forwarded in various studies  (\citealt{Spitzer1951,vandenBergh1976,Poggianti1999}) 
and modern observations are providing reason to raise the question of galaxy classification again (e.g., \citealt{Bertin2000}, section 18.2).  
For example, the recent study of 260 early-type galaxies (E/S0) by \cite{Cappellari2011} as part of \atlas\ project gives an overview of the limitations of the classic \cite{Hubble1936} tuning-fork diagram. 
They show, instead, the usefulness of a scheme similar to the one proposed by \cite{vandenBergh1976} to properly understand the morphology of early-type galaxies. 
The authors consider two classes of objects: (i) slow rotators, which are consistent with being genuinely elliptical-like objects with intrinsic ellipticity $\epsilon \gtrsim 0.4$; 
and (ii) the fast rotators, which are generally flatter than $\epsilon \lesssim 0.4$ and are morphologically similar to spiral galaxies, 
or in some cases to flat ellipticals with discy isophotes, and span the same full range of bulge sizes of spirals. 
They argued for a revised comb-shaped scheme to represent the morphology of nearby galaxies, which overcomes the limitations of the tuning-fork diagram.

Despite the rich previous literature, most past work has relied on CVCs to infer other galaxy properties, using by-eye classification schemes, or simple approximations to CVC shapes.  
In this work, we use PCA to fully describe the shape and amplitude of the CVCs across its full extent. 
This work uses CVCs derived from the stellar disc, in contrast with most of previous work that usually relies on 
interstellar medium (ISM) tracers, typically {\sc Hi} 21-cm data. 
This allows us to study the internal structure of the galaxies from all Hubble types (E1--Sdm), since not all galaxies contain significant amount of neutral gas ({\citealt{Longair2008}). 
In addition, stellar dynamics is more sensitive to dynamically hot stellar components like bulges and may provide a tighter connection to the evolution of the galaxies. 

We introduce the data and sample, needed to probe the shape and amplitude of the galaxy circular curves in Section \ref{S:data}. 
Section \ref{S:analysis} describes the analysis approach used in our study - dynamical modelling, 
PCA applied to circular curves and $k$-means statistics. We propose a CVC classification of our sample of galaxies in Section \ref{S:dynclass}, and examine a possible genesis of the proposed CVC classes in Section \ref{S:discussion}. The main uncertainties of our results are discussed in Section \ref{S:caveats}. An example application of our classification is presented in Section \ref{S:apply}, and we make our conclusions in Section \ref{S:summary}.

\section{Sample and Data}
\label{S:data}
\subsection{Sample selection}
\label{S:sample}
\begin{figure}
\begin{center}
{\includegraphics[width=0.45\textwidth]{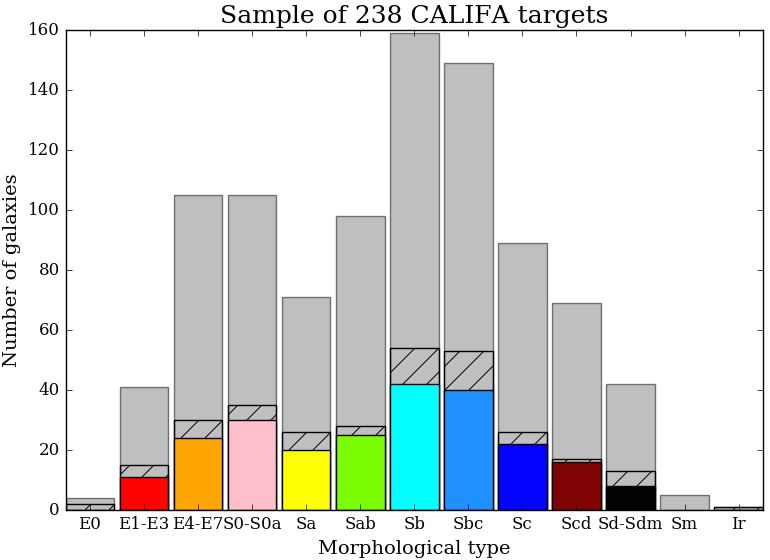}}
\caption{Distribution of our 238 galaxy sub-sample through Hubble type (colour)  
compared with CALIFA mother sample of 939 targets
(grey, \citealt{Walcher2014}) and with stellar kinemtics galaxy sample of 300 galaxies (hatched lines, \citealt{Falcon-Barroso2017}).
The sub-sample of 238 targets well represents through morphology both CALIFA mother sample and stellar kinematics sample. 
}
\label{fig:histo}
\end{center}
\end{figure}
We select our galaxies from the sample of \cite{Falcon-Barroso2017}, who provide reliable stellar kinematics of 300 galaxies from the 
Calar Alto Legacy Intergral Field Area (CALIFA) survey\footnote{\url{http://califa.caha.es/}} (\citealt{Sanchez2012}) observed until 2014 June. From the original sample of 300 galaxies, we discard galaxy mergers and those cases where the dynamical models of the galaxies were not of sufficient quality for our study (see Section \ref{SS:dyn}). The final sub-sample consists of 238 CALIFA targets with various masses and morphologies from elliptical (E1) to late-type (Sdm) galaxies. 
The distribution of the Hubble type (in colour) for the 238 CALIFA galaxies sample is shown in Fig. \ref{fig:histo}, which is  
compared with CALIFA mother sample of 939 targets
(grey; \citealt{Walcher2014}) and with stellar kinematics galaxy sample of 300 galaxies (hatched lines; \citealt{Falcon-Barroso2017}). The sub-sample of 238 targets well represents through Hubble sequence both the CALIFA mother sample and stellar kinematics sample. 
The morphology of all CALIFA galaxies has been defined after visual examination by several members of the CALIFA team. 
We refer the reader to \cite{Walcher2014} for further details related to the selection procedure and statistical 
properties of the CALIFA mother sample\footnote{The sample characterization tables 
(e.g., morphological type, effective radius, etc.) from \citet{Walcher2014} can be found 
here: \url{http://www.caha.es/CALIFA/public_html/?q=content/sample-characterization-tables}}. 

\subsection{Stellar kinematics and imaging}
\label{SS:sauronifs}
The observations of the $238$ CALIFA galaxies have been made by using the integral-field spectroscopic instrument $\texttt{PMAS}$ (\citealt{Roth2005}) in $\texttt{PPaK}$ mode (\citealt{Verheijen2004}), mounted on the 3.5 m telescope at the Calar Alto Observatory. 

To obtain well resolved stellar kinematics for these galaxies, \cite{Falcon-Barroso2017} used the mid-resolution prism data (V1200 with R $\sim$ 1650) having the coverage of the nominal wavelength range $3850$--$4600$~\AA~at an FWHM spectral resolution of $\sim\,2.3$ \AA, i.e., $\sigma \sim 85$~km\,s$^{-1}$ and a $74^{\prime\prime} \times 64^{\prime\prime}$ hexagonal field of view (FoV). The exposure time per pointing has been fixed to $1800$~s, split into two or three individual exposures (\citealt{Sanchez2012}, \citealt{Husemann2013}, \citealt{Garcia-Benito2015}). 
The stellar kinematics used in this study is based on data from CALIFA v1.4 reduction pipeline (\citealt{Sanchez2012, Husemann2013, Garcia-Benito2015}). 
\cite{Falcon-Barroso2017} extracted the stellar kinematic maps from the CALIFA data cubes using the $\texttt{PPXF}$ fitting procedure (\citealt{Cappellari2004}) with a  
subset of $\sim$ 300 stars from the $\texttt{INDO-US}$ (\citealt{Valdes2004}) spectral templates library. In the case the full library is applied, the stellar kinematic results will be reproducible (see section 4 of their paper).
Furthermore, \cite{Falcon-Barroso2017} spatially binned the data cube using the Voronoi 2D binning algorithm of \cite{Cappellari2003} to obtain signal-to-noise ratio of $\sim\,20$ and, hence, reliable stellar kinematics.
For most of the galaxies, the FoV covers a radial extent $R_{\mathrm{max}}$ of one to two times of the galaxy's half-light radius, $R_e$ (see Table 1 of \citealt{Falcon-Barroso2017}).
We also calculate the systemic velocity of the 238 CALIFA galaxies ($V_{\mathrm{sys}}$) as the median value of the velocity 
field (see Tables \ref{tab:DS1}--\ref{tab:DS6}).
 
To obtain the SB of the 238 CALIFA galaxies, we use the $r$-band photometric images from the Sloan Digital Sky Survey\footnote{{http://www.sdss.org/}} (SDSS; \citealt{Yorketal2000}) using Data Release $12$\footnote{{http://data.sdss3.org/}} (DR12; \citealt{Alam2015}), which have already been flux calibrated and sky subtracted.

\section{Analysis}
\label{S:analysis}
\begin{figure*}
\centering
\includegraphics[width=0.98\textwidth]{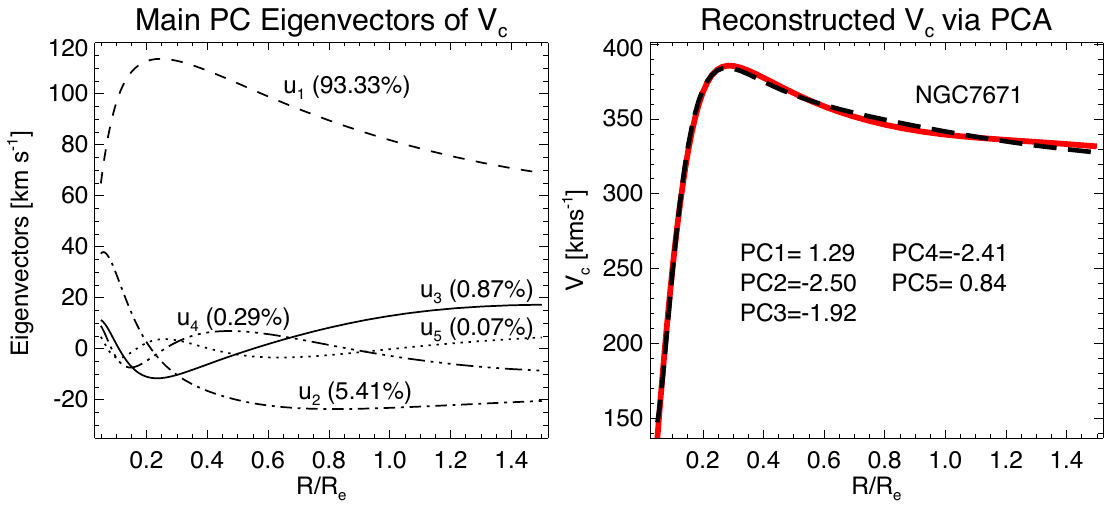}
\caption{PCA. \emph{Left:} the main five principal component eigenvectors $\textbf{\textit{u}}$, needed to reconstruct the CVCs of the 
238 CALIFA galaxies shown with the corresponding fraction of power in that component. \emph{Right:} example of the reconstructed circular velocity (red curve) of 
the galaxy NGC7671 after applying PCA, where $V_{\mathrm{c}}$ (black dashed curve) is the linear combination between the five main PC eigenvectors $\textbf{\textit{u}}$ and projections PC$_i$.}
\label{fig:pcaradial}
\end{figure*}
%

In this paper, we aim to explore a new classification method based on the dynamics of the 238 CALIFA galaxies given 
their $V$ and $\sigma$ stellar kinematic maps and photometric images. We use the maps and the images of the galaxies to construct a reliable dynamical model and derive the CVCs.  
We then use the suite of 238 CVCs to define a basis set of generalized circular curve 
components using PCA.  
These basis curves are used to project the CVCs into a basis space, where each curve can be represented by a linear 
combination of the basis set.  We use two-dimensional $k$-means 
clustering of points in the space of the two main components of the PCA 
to define four classes of CVCs, providing a new galaxy classification framework.

\subsection{JAM-MCMC CVC derivation}
\label{SS:dyn}
To reliably derive the total CVCs of our sample of galaxies from the CALIFA stellar kinematic fields, we use the axisymmetric Jeans anisotropic multi-Gaussian expansion (JAM; \citealt{Cappellari2008}) dynamical model with the basic assumptions of a constant velocity anisotropy in the meridional plane $\beta_z=1-\sigma_z^2/\sigma_R^2$ and a constant  mass-to-light ratio 
$(M/L)_{\mathrm{dyn}}$ of the galaxies.
We obtain the values of these two parameters by fitting the observed second-order velocity moment 
$V^2_{\mathrm{rms}}=V^2+\sigma^2$ using the $\texttt{PYTHON}$ version of the JAM\footnote{http://purl.org/cappellari/software} method. 
Further, we apply a Markov chain Monte Carlo (MCMC) technique to constrain the parameter space in 
JAM model (\citealt{Cappellari2008}) between the two fitting parameters -- $\beta_z$ and $(M/L)_{\mathrm{dyn}}$ (see \citealt{Kalinova2017}).  
We also symmetrized our observed $V^2_{\mathrm{rms}}$ fields before applying JAM-MCMC code in order to reduce outliers in the data (see \citealt{Cappellari2015}). In this case, our $\chi^2$-statistics for fitting the $V^2_{\mathrm{rms}}$ field will be less influenced by a few deviant values. This is especially important when one applies MCMC analysis. 

Due to the large computational expense of evaluating JAM-MCMC model, we run the code with only 30 walkers, though the model still converged with only 30 steps for burn-in phase and 80 steps for sampling. The prior distributions of the parameters were taken to be uniform with, e.g., $\beta_z$ $\in[-1,1]$ and $(M/L)_{\mathrm{dyn}}$ $\in[0,20]$ $M_{\odot}/L_{\odot}$. 
For the calculation of the dynamical models, we fixed the inclination $i$ of the galaxies to the photometric value, estimated from the ellipticity ($\epsilon$) of the galaxies using equation 1 in \citet[where $\epsilon$ is the avarage ellipticity, determined from $\texttt{findgalaxy}$ \texttt{python}  procedure of \citealt{Cappellari2002}]{Kalinova2017}. For those galaxies where the calculated photometric inclination was below the minimum allowed multi-Gaussian expansion (MGE) inclination ($i^{MGE}$) of JAM approach (see section 2.2.2 in \citealt{Cappellari2002}), we adopt $i=i^{MGE}$ 
(see Tables \ref{tab:DS1}$-$\ref{tab:DS6}). 
 
In Appendix \ref{A:maps}, Figs \ref{fig:maps1}--\ref{fig:maps8}, 
we present the second moment maps $\Vrms = \sqrt{V^2+\sigma^2}$ of the data ($V^{OBS}_{\mathrm{rms}}$; first columns) and
the best-fitting models of the JAM-MCMC method ($V^{MOD}_{\mathrm{rms}}$; second columns) with overplotted fluxes and MGE contours, respectively. The third columns correspond to the residual field between the observed and modelled $\Vrms$ maps, where 
$RES=|1-(V_{\mathrm{rms,OBS}}/V_{\mathrm{rms,MOD}})|$. 
On average, the median value of the residual maps $\overline{RES} \sim 0.10$ for most of the 238 CALIFA galaxies, 
which corresponds to around 10 per cent error.
We initially discarded galaxies with not well-defined minima of the prior chain distributions for the parameters $\beta_z$ or 
$(M/L)_{\mathrm{dyn}}$  (see section 4.3.2 and appendix B of \citealt{Kalinova2017} for examples).

Next, we derived the CVCs of the 238 CALIFA galaxies by applying Poisson's equation to the best fitting gravitational potential and calculating $V^2_{c}=R \frac{\partial\Phi}{\partial R} |_{z=0}$. We calculate the $\Phi(R,z)$ through 
the MGE method \citep[MGE;][]{Monnet1992,Emsellem1994}. It describes the observed SB of the galaxies as a sum of $N$ Gaussian components, which also allows us to reproduce the photometry of the galaxies in detail,
\begin{equation}
  \label{eq:mgeSB}
  I(x',y') = \sum_{j=0}^{N} I_{0,j} \exp\left\{ -\frac{1}{2{
{\bf \xi}
'_j}^2} \left[ x'^2 + \frac{y'^2}{{q'_j}^2} \right] \right\},
\end{equation}
where $I_{0,j}$ is the central SB, ${\bf \xi}'_j$ is the dispersion along the major $x'$-axis
and $q'_j$ is the flattening.  
The intrinsic dispersion and flattening, $\xi_j$ and $q_j$, are related to their observed (i.e., plane-of-sky) quantities, as
\begin{equation}
  \label{eq:qintr}
  	\xi_j = \xi'_j
  	\quad \mathrm{and} \quad
  	{q'_j}^2 = \cos^2i + q_j^2 \sin^2i, 
\end{equation}
where the inclination $i$ ranges from $i=0^\circ$ for face-on viewing to $i=90^\circ$ for edge-on viewing.

We apply the MGE fitting method to the $r$-band SDSS images using the software implementation of \cite{Cappellari2002}.
We also take into account the point spread function (PSF) convolution, where the median PSF FWHM is $1.3$ arcsec for the $r$-band SDSS images. 
The resulting analytically deconvolved MGE model for each galaxy was corrected for extinction as given by NASA/IPAC Extragalactic Database\footnote{{The NASA/IPAC Extragalactic Database (NED) is operated by the Jet Propulsion Laboratory, California Institute of Technology, under contract with the
National Aeronautics and Space Administration, (\url{https://ned.ipac.caltech.edu/})}} (NED). 
It was converted to an SB in solar units assuming the $r$-band absolute magnitude for the Sun of $M_{r,\odot}$ = 4.68 mag (table 1.3 of \citealt{Sparke2007}). Additionally, we adopt the `bulge-disc decomposition' keyword in the software implementation of \cite{Cappellari2002} when performing the MGE fits. In this case, the Gaussians were separated into two groups, where each of them described more efficiently the bulge and disc of the spiral/lenticular galaxies, or the nuclear disk of the ellipticals. The MGE fit parameters are listed in Appendix \ref{A:MGEs}.

The MGE parametrization can be deprojected analytically into an intrinsic luminosity density $\nu(R,z)$ when the viewing direction is given. Furthermore, the calculation of $\overline{v_\mathrm{los}^2}$ reduces from the (numerical) evaluation of a triple integral to a straightforward single integral \citep[][equation 27]{Cappellari2008}.
Similarly, the gravitational potential $\Phi(R,z)$ can be calculated by means of one-dimensional integral \cite[][equation 39]{Emsellem1994}.

Given the latter, the circular velocity from the JAM model in the equatorial plane then follows upon (numerical) evaluation of 

\begin{multline}
  \label{eq:mgevcirc}
  v_{c,\mathrm{JAM}}^2(R) = \sum_{j=0}^{N} 
  \frac{2 G L_j (M/L)_j}{\sqrt{2\pi}\xi_j} \frac{R^2}{\xi_j^2} 
   \times \\
 \int_0^1 \exp\left\{ -\frac{u^2 R^2}{2\xi_j^2} \right\}
\frac{u^2\,\du u}{\sqrt{1-(1-q_j^2)u^2}},
\end{multline}
where $L_j \equiv 2 \, \pi \xi_j^2 q'_j I_{0,j}$ and $(M/L)_j$ are the total luminosity and the mass-to-light ratio of the $j$th Gaussian. 
Since we assume that mass follows light in our dynamical model,  $(M/L)_j$ is the same for all Gaussians, 
i.e., $(M/L)_{\mathrm{dyn}}=(M/L)_{j}$ (see section 3.2 of \citealt{Kalinova2017}).

\subsection{Principal Component Analysis}
\label{SS:pca}
%
\begin{figure*}
\centering
{\includegraphics[width=1.0\textwidth]{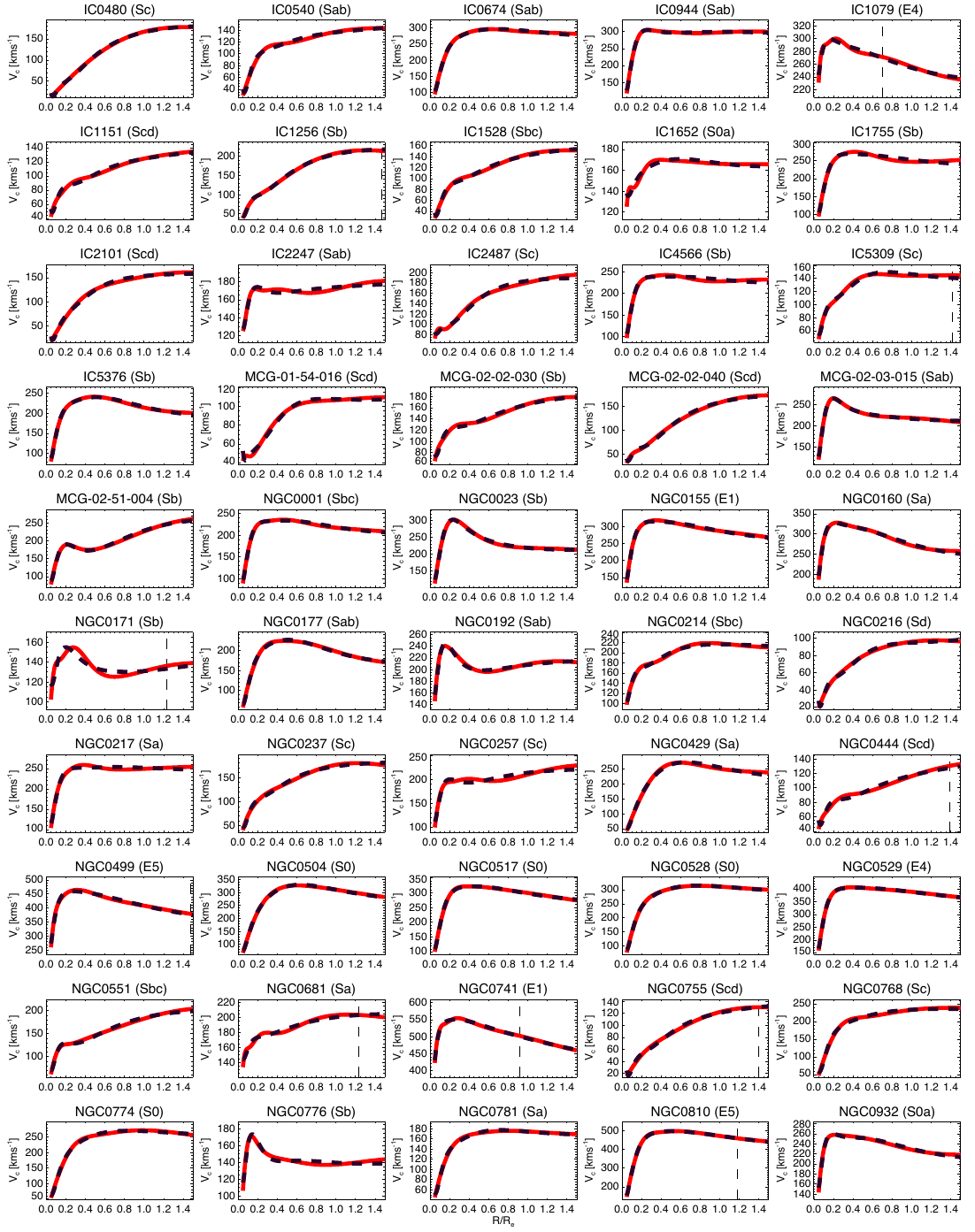}}
\caption{CVCs $V_{\mathrm{c}}$ of the 238 CALIFA galaxies (red curve), normalized to the effective radius ($R_e$) and their reconstruction (thick black dashed curve) from applying PCA, which is the linear combination between the five main PC eigenvectors $\textbf{\textit{u}}$ and five main PC projections PC$_i$. The vertical thin dashed line indicates the maximum radial extent of our observations in the cases where the data do not cover all $R< 1.5 R_e$.
}
\label{fig:califa}
\end{figure*}
%
\addtocounter{figure}{-1}
\addtocounter{subfigure}{1}
%
\begin{figure*}
\centering
{\includegraphics[width=1.0\textwidth]{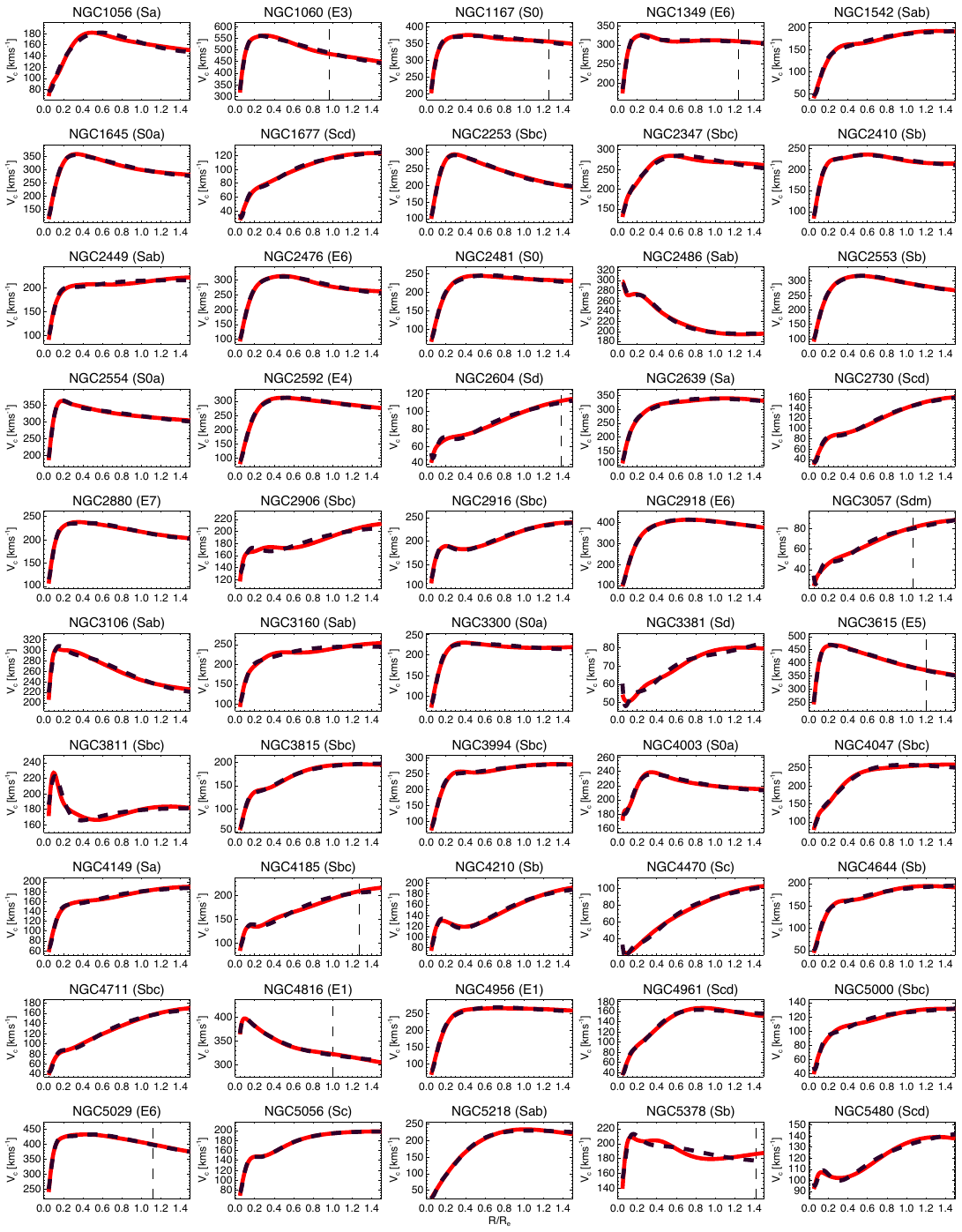}}
\caption{{\it -- continuation}}
\label{fig:califa}
\end{figure*}

\addtocounter{figure}{-1}
\addtocounter{subfigure}{1}
%
\begin{figure*}
\centering
{\includegraphics[width=1.0\textwidth]{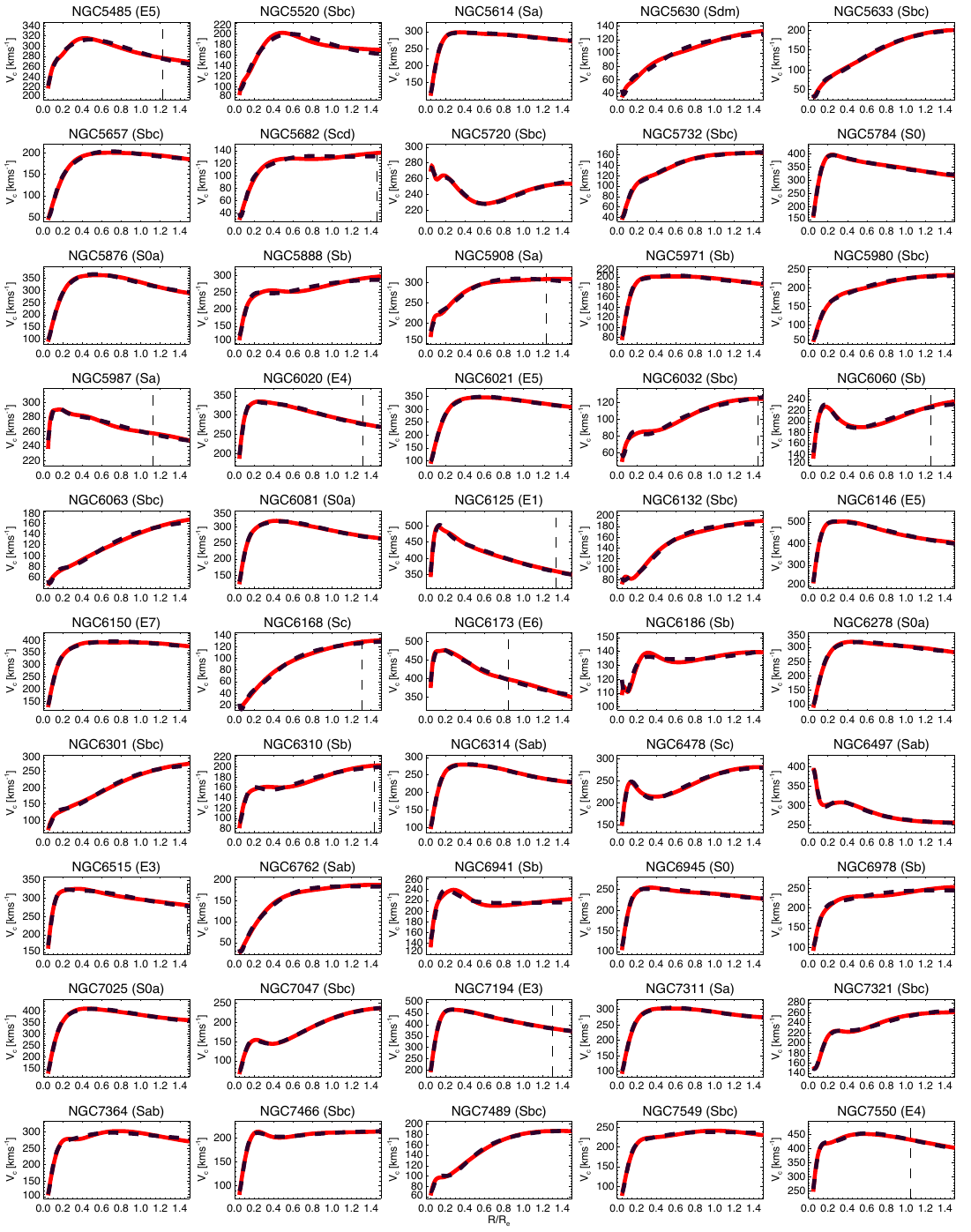}}
\caption{{\it -- continuation}}
\label{fig:califa}
\end{figure*}
\addtocounter{figure}{-1}
\addtocounter{subfigure}{1}
%
\begin{figure*}
\centering
{\includegraphics[width=1.0\textwidth]{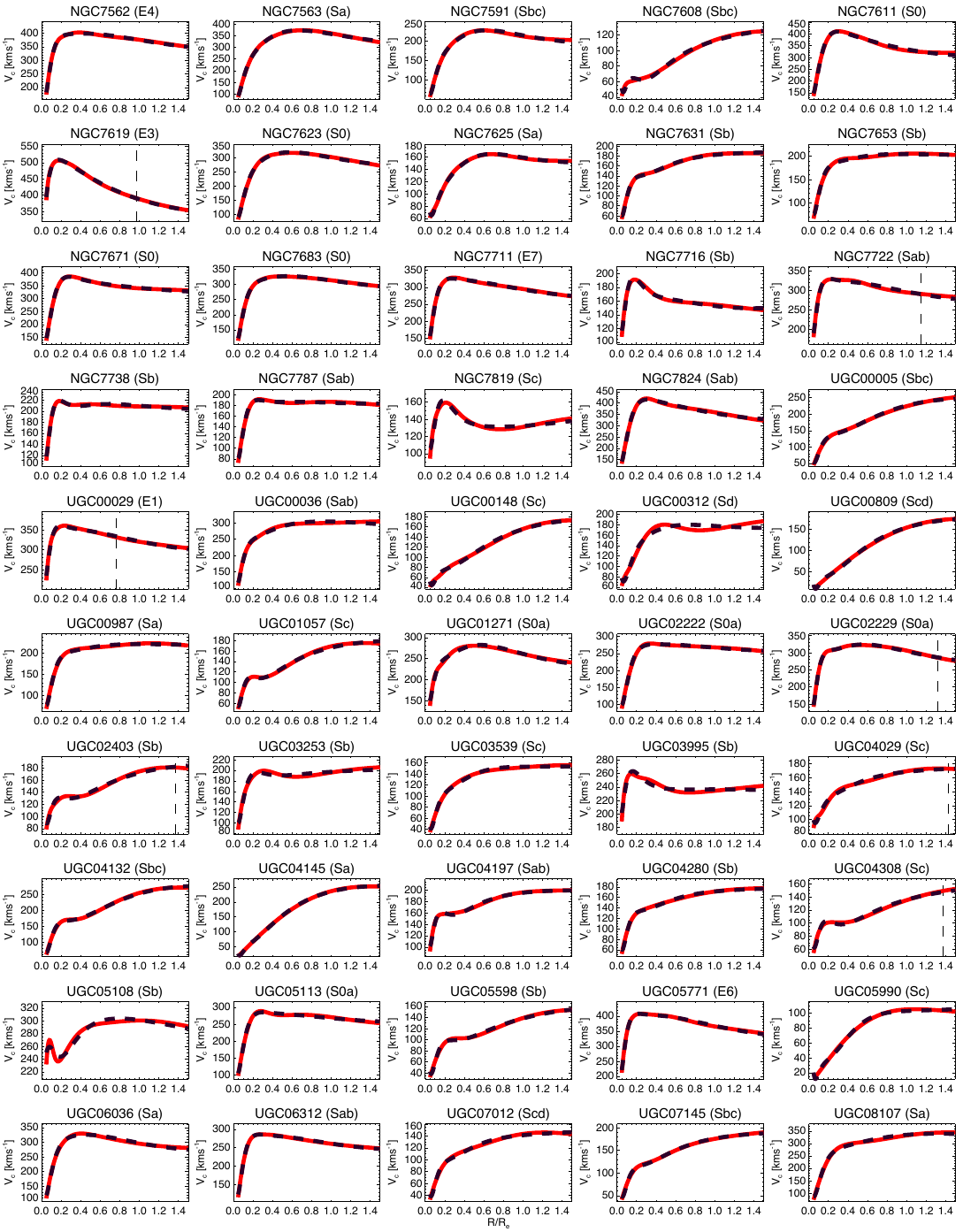}}
\caption{{\it -- continuation}}
\label{fig:califa}
\end{figure*}
\addtocounter{figure}{-1}
\addtocounter{subfigure}{1}
%
\begin{figure*}
\centering
{\includegraphics[width=1.0\textwidth]{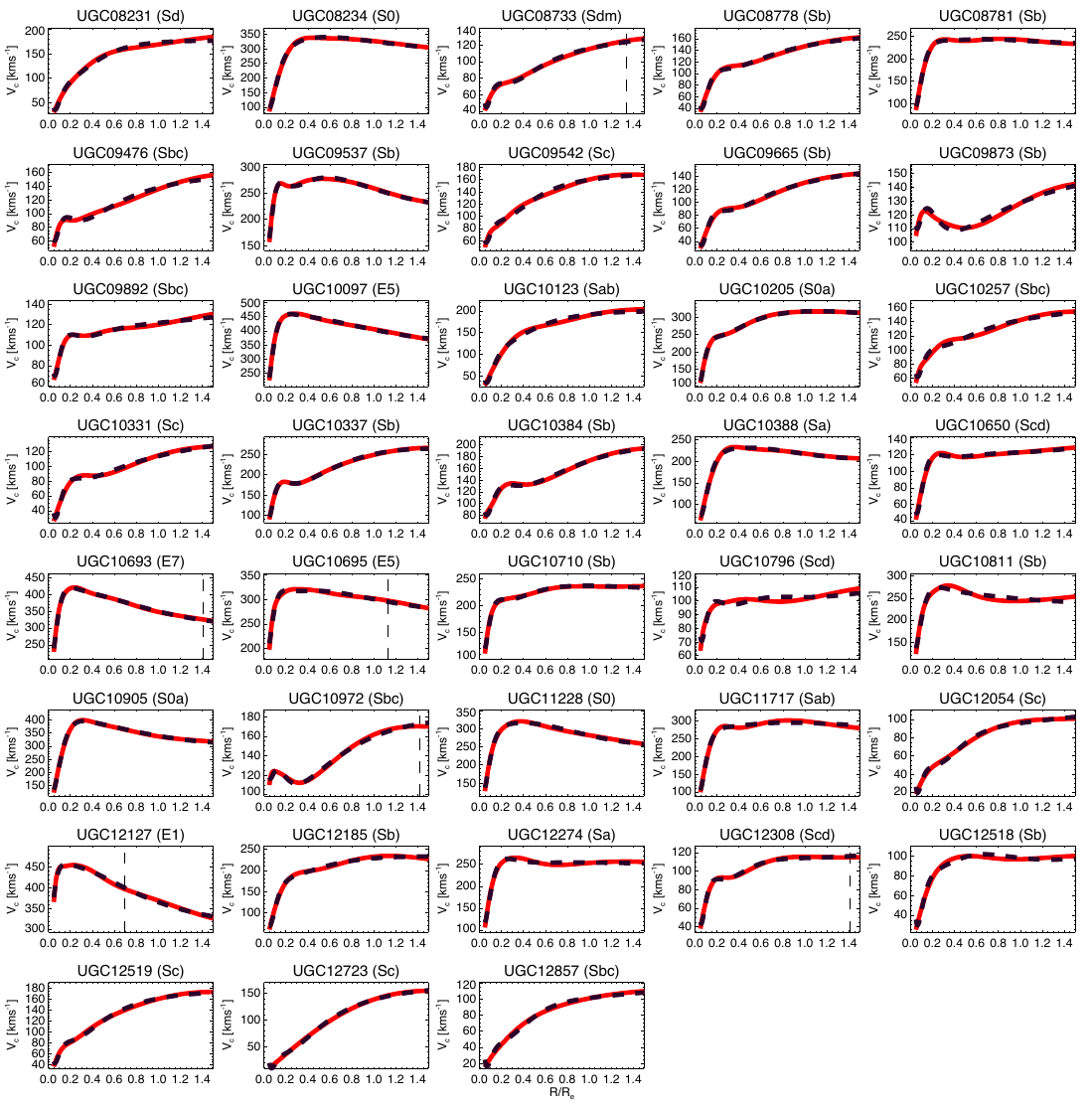}}
\caption{{\it -- continuation}}
\label{fig:califa}
\end{figure*}
A common criticism of the visual classifications in general is that the criteria for assigning galaxies to classes are subjective, leading to different observers assigning galaxies to different classes.  To develop an objective classification scheme, 
we use PCA (\citealt{Pearson1901}) applied to the CVCs to classify the 238 E1--Sdm galaxies by their dynamics.  PCA is a powerful approach to identify patterns in broad range of data, and expressing the data in such a way as to highlight their similarities and differences 
(e.g., \citealt{Shlens2014}). 

PCA operates by constructing a covariance matrix between data sets, diagonalizing that matrix and using the resulting eigenvectors as a basis set for the new space.  
We normalize the radial coordinate to the effective radius $R_e$ and interpolate each circular curve $V_c$ to a common radial sampling, using 50 points spanning $R/R_e \in [0, 1.5]$.

We construct a covariance between the $j$th and $k$th sample on the velocity curve as 
\begin{equation}
\sigma_{jk} = \sum_i (v_{c,i} (R_j)-\overline{V_{\mathrm{c}}}) (v_{c,i} (R_k)-\overline{V_{\mathrm{c}}}),
\end{equation}
where the sum is carried out over the sample of galaxies $N=238$ and $\overline{V_{\mathrm{c}}}$ is the mean circular speed for all galaxies across the whole sample.  Note that, in our application of PCA, we do not subtract the mean of each circular curve but rather the mean of this sample (for our sample $\overline{V_{\mathrm{c}}}$=196.336 \kms). Such a formulation is needed in the cases where we are trying to develop a representative basis (\citealt{ Koch2013}).  We diagonalize the covariance matrix, obtaining a set 
of $N$ eigenvectors ($\textbf{\textit{u}}_n$) and the eigenvalues represent the relative power of those eigenvectors in describing the population.  

Fig. \ref{fig:pcaradial}, left-hand panel, shows the first five principal component (PC) eigenvectors $\textbf{\textit{u}}$, needed to reconstruct the CVCs of the 238 galaxies shown with the corresponding fraction of the power in that eigenvector. The first dominant feature of the 238 CVCs is shown by the first eigenvector $\textbf{\textit{u}}_1$ with 93.33 per cent (long-dashed line).  
Eigenvectors $\textbf{\textit{u}}_2$ to $\textbf{\textit{u}}_5$ contain 5.41 per cent, 0.87 per cent, 0.29 per cent and 0.07 per cent of the power respectively.  The reconstructed circular velocities, using the five eigenvectors and five principal components, contain 99.97 per cent of the total information of the 238 CVCs. 

In Fig. \ref{fig:pcaradial}, right-hand panel,  we show an example of reconstructed circular velocity ($V_{\mathrm{c,rec}}$) of the galaxy NGC\,7671 (red curve) after applying PCA. $V_{\mathrm{c,rec}}$ (black dashed curve) is the linear combination between the five main PC eigenvectors $\textbf{\textit{u}}$ and five projections PC$_i$, adding the mean value $\overline{V_{c}}$ of the circular velocities calculated among all 238 galaxies:
\begin{equation}
V_{c,\mathrm{rec}}=(\mathrm{PC}_1 \textbf{\textit{u}}_1+\mathrm{PC}_2 \textbf{\textit{u}}_2+
\mathrm{PC}_3 \textbf{\textit{u}}_3+\mathrm{PC}_4 \textbf{\textit{u}}_4+\mathrm{PC}_5 \textbf{\textit{u}}_5)+\overline{V_{c}}.
\label{eq:eigen}
\end{equation}

\noindent
For this galaxy, the projections are PC$_1=+1.29$, PC$_2=-2.50$, PC$_3=-1.92$, PC$_4=-2.41$, PC$_5=+0.84$.
 
In Fig. \ref{fig:califa}, we show the reconstructed circular velocities of the 238 CALIFA galaxies. The fits are nearly perfect for all galaxies. The five main PC eigenvectors and the five main projections PC$_i$ are listed in Appendices \ref{A:PC_vect} and  \ref{A:PCs}, respectively.  
  
\subsection{$k$-means clustering}
\label{SS:kmeans}
%
\begin{figure*}
\centering
\includegraphics[width=1.0\textwidth]{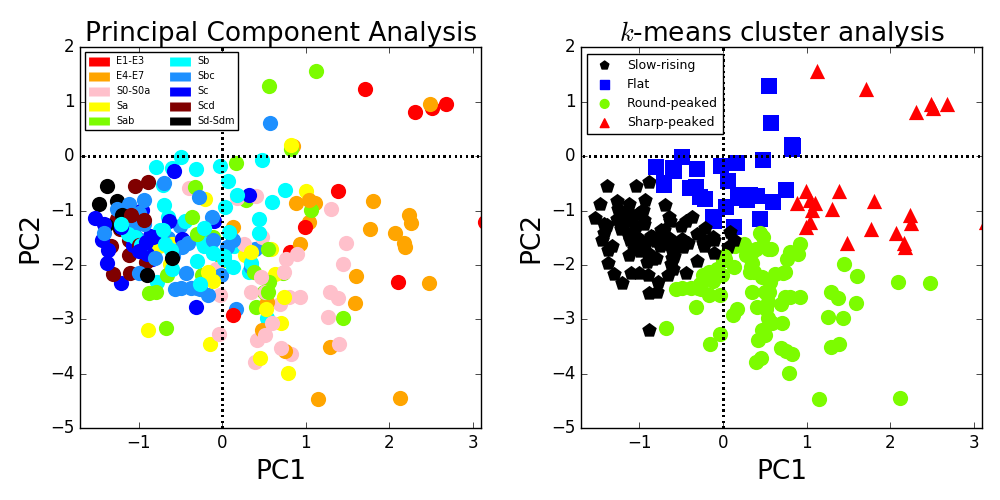}
\caption{PCA of the 238 galaxies.
 \emph{Left:} PC projections for each circular curve coded by Hubble type.
\emph{Right:} applied $k$-means clustering method on the two PC projections carrying $\sim$ 99 per cent of the power, where the colour coding corresponds to the proposed four groups of galaxies.
}
\label{fig:kmeans}
\end{figure*}

\begin{figure*}
\centering
\includegraphics[width=1.0\textwidth]{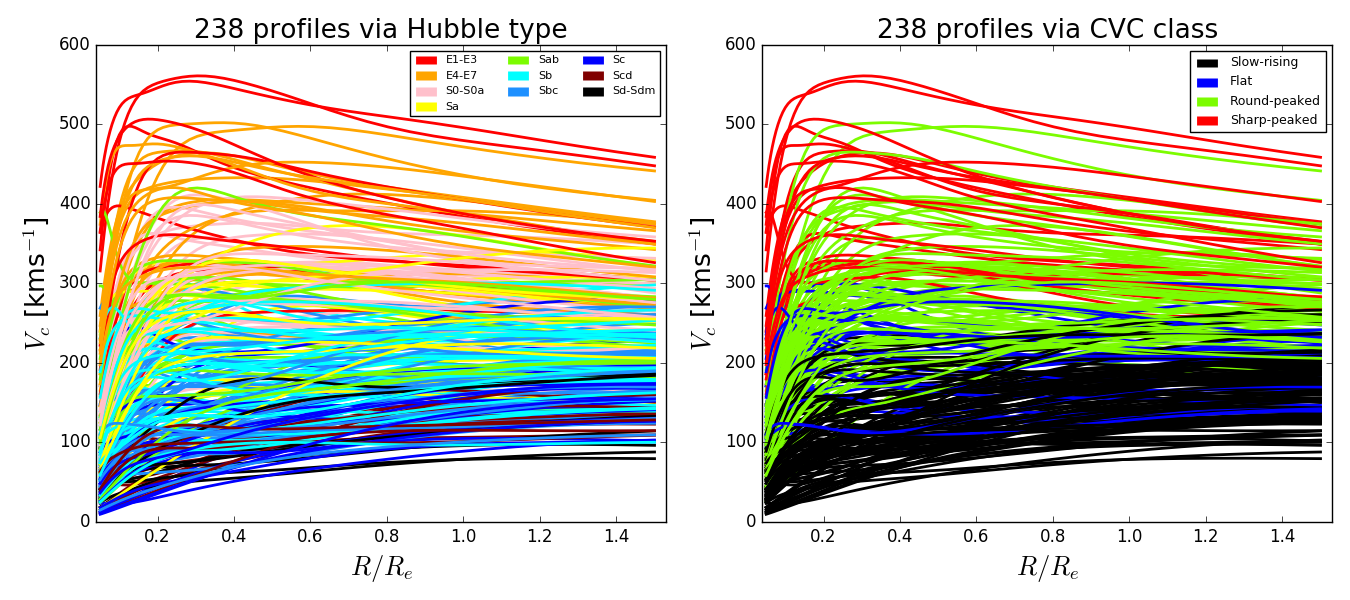}
\caption{CVCs of the 238 (E1--Sdm) CALIFA galaxies colour coded by morphological type (left) and  CVC class (right). CVCs form a continuum of different shapes, amplitudes and slopes. Thus, there will be always borderline cases for our CVC classification like other schemes for galaxy classification (Section \ref{SS:kmeans}).}
\label{fig:morphvsdyn}
\end{figure*}

%
\begin{figure*}
\centering
\includegraphics[width=1.0\textwidth]{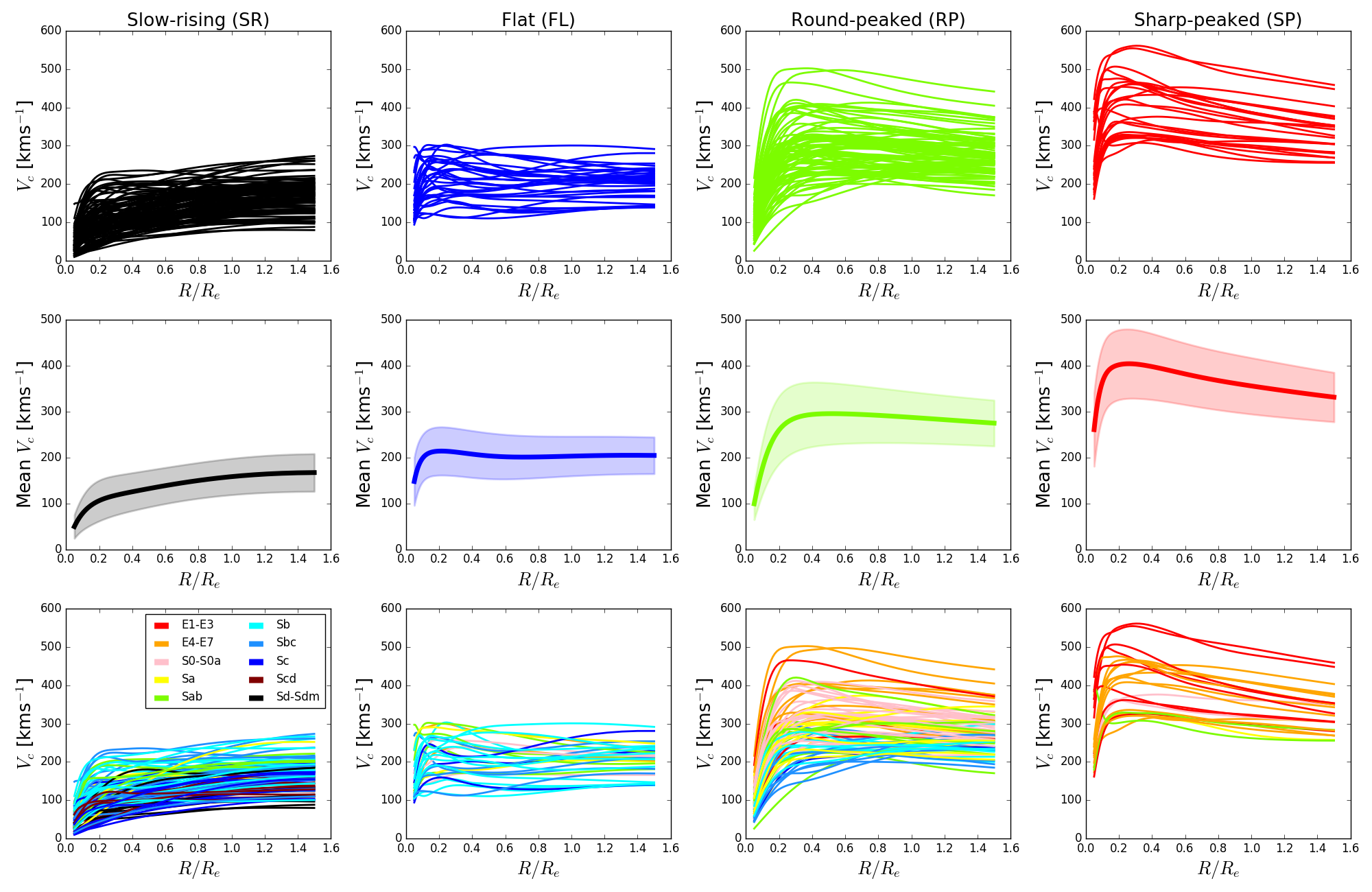}
\caption{PCA of the 238 galaxies. 
\emph{First row:} CVCs of the four groups classified using the $k$-means clustering method (Section \ref{SS:kmeans}). \emph{Second row:} 
prototype circular curve, which corresponds to the mean CVC of each $k$-means cluster group with the corresponding uncertainty band. 
\emph{Third row:} $V_{c}$ from each $k$-means cluster group, where the colour coding corresponds to the morphology of the galaxies. 
}
\label{fig:dynclass}
\end{figure*}
%

With the eigenvector decomposition, each circular curve can now be represented in a reduced-dimensionality space 
as the PC projections required for a reconstruction. We use $k$-means clustering to identify classes within that space.
The $k$-means clustering approach is a vector classification method (\citealt{macqueen1967}) based on Lloyd's algorithm (\citealt{lloyd1982}).
It is commonly used in data mining due to its simplicity (for a review of clustering methods see \citealt{Everitt1993,jain1999}).
The $k$-means clustering method partitions $n$ observations into $k$ clusters in which each observation belongs to the cluster with the nearest mean,  considered as the ``centroid" (prototype) of the cluster.
Given $k$, the method optimizes, by several iterations, the mean positions that minimize intracluster variation and maximize intercluster variation. The final result is a partitioning of the data space to regions based on the closest mean.
We use the \texttt{IDL} implementation of $k$-means algorithm:
the \texttt{CLUSTER\_WTS}\footnote{\url{https://www.harrisgeospatial.com/docs/clust\_wts.html}} task calculates the weights and
the centres of the clusters, and 
the \texttt{CLUSTER}\footnote{\url{https://www.harrisgeospatial.com/docs/cluster.html\#C\_854643309\_862474}} task
defines the final clusters based on those weights.

In Fig. \ref{fig:kmeans}, we present the $k$-means clustering applied to the first and second PCs of our circular curves.  In $k$-means clustering, the choice of clusters $k$ 
is established by the user, though some algorithmic approaches exist (e.g., \citealt{Everitt1993}).  
In our case of the 238 galaxies, we choose $k=4$, 
because we visually see four distinct  sets of CVC shapes that appear to represent a useful classification approach. 
Additionally, our choice is guided by the general $k$-means heuristic method for determining the maximum number of the clusters within a given data set: the square root of the number of data points ($N$) divided by two (\citealt{Trupti2013}), i.e., $k\approx\sqrt{N/2}$, where $N=238$ for our sample, 
and the calculated maximum number of clusters is $k\approx11$. Thus, our choice for $k=4$ is also consistent with this rule.
For our CVCs classification, like other schemes for galaxy classification, the circular curves form a continuum of different shapes and there will be always borderline cases (see Fig. \ref{fig:morphvsdyn}).  
We have experimented with variation of both the numbers of PCs included and the number of clusters. 
We settle on two components and four clusters and arrive at the groupings shown in Fig. \ref{fig:centroid}, which form a basis for classification.  
However, PCA also provides dimensionality reduction, which enables the continuum of galaxy properties to be quantified.

\section{Towards a new classification of galaxies}
\label{S:dynclass}

\subsection{Shape of the CVCs across Hubble sequence}
\label{SS:rcshape}
In the left-hand panel of Fig. \ref{fig:kmeans}, we show the two projections of the PCA to the stellar circular curve of the 238 E1--Sdm galaxies (Section \ref{SS:pca}) 
with colour coding corresponding to the morphological types of the galaxies. 
The right-hand panel shows the cluster labels that are derived after applying the two-dimensional $k$-means clustering technique, 
considering the projection of the two main PC Eigenvectors, plotted in the panels. The colour coding and symbols correspond to the four CVC groups identified by this method. 
Given the appearance of these four classes, we describe them as: \emph{slow-rising} (SR, black filled pentagons), \emph{flat} (FL, blue filled squares), {\it round-peaked} (RP, green filled circles), 
and \emph{sharp-peaked} (SP, red filled triangles). 

To investigate the properties of the four classes,  in Fig. \ref{fig:dynclass}, first row, we plot together the CVCs from each class (from left- to the right-hand panel): SR-CVCs (black curves), FL-CVCs (blue curves), RP-CVCs (green curves) and SP-CVCs (red curves).  
In the second row of Fig. \ref{fig:dynclass}, we calculate the mean prototype circular curve, which corresponds to the mean value of the CVCs at each $k$-means cluster group, in order for each class to obtain a representative shape and amplitude of the CVCs.
The shaded band represents the standard variation of the CVCs within the cluster group. 
To contrast the CVC and photometric (Hubble) classification, we use the last row of Fig. \ref{fig:dynclass} to show the CVCs from each class with corresponding colour 
representing the morphology of the galaxies. The SR-CVCs have a clear connection to late-type spirals (Sb--Sdm), and the SP-CVCs to early-type galaxies (E1--E7).  
However, the other two CVC classes  are represented by galaxies with different morphologies. Interestingly, the lenticular galaxies fall mainly in one category -- RP class. 

In Fig. \ref{fig:centroid} (left), we combine the four prototype curves together to compare their typical shapes and amplitudes. 
SP-CVCs have the highest and the SR-CVCs have the lowest amplitude. Furthermore, the SR-CVCs and FL-CVCs have tighter mass range than RP-CVCs and SP-CVCs.

We modelled the CVCs of our galaxies with a maximum extent to 1.5$R_e$ since most of the galaxies appear to reach their flat (asymptotic) circular speed ($V_{\mathrm{F}}$) at this radius. 
This allowed us to represent the normalized prototype curves $V_{c}/V_{\mathrm{F}}$ in the right-hand panel of Fig. \ref{fig:centroid}, 
where we can more clearly see the shape of the peak of the CVCs (absent, anaemic/flat, round or sharp). 

Additionally, we note that the different prototype curves differ in the steepness of the rising central parts ($\sim$ 0.1$R_e$) and the declining of the outer parts 
($\sim$ 1.5$R_e$). The dashed line correspond to unity of the ratio $V_{c}/V_{\mathrm{F}}$ for guiding the eye.

\begin{figure*}
\centering
\includegraphics[width=1.0\textwidth]{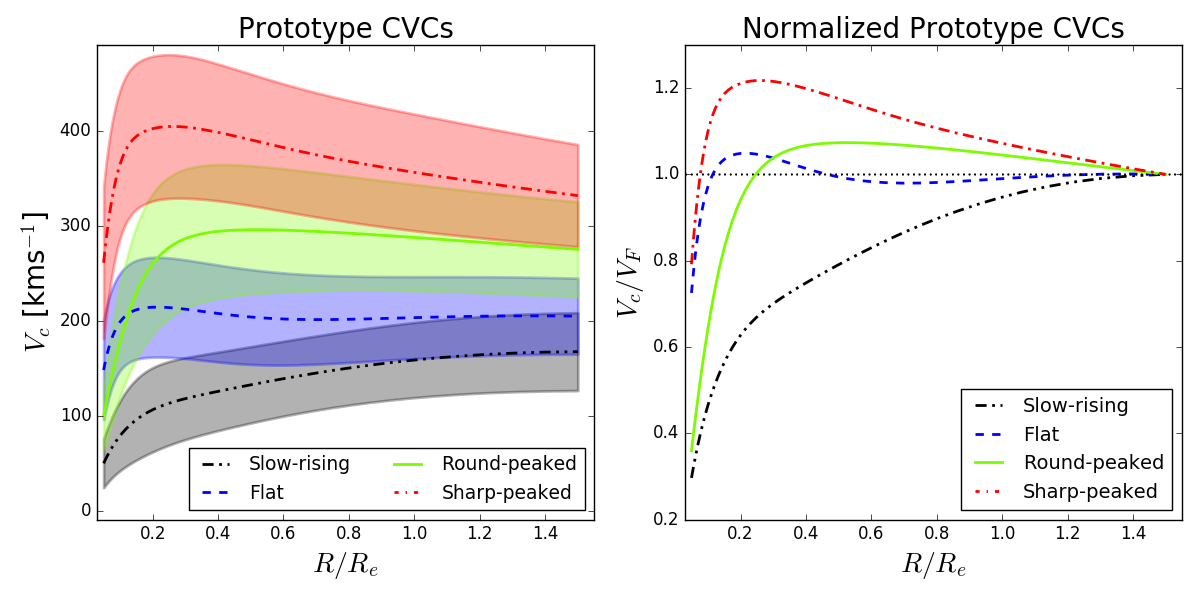}
\caption{CVC classification of the 238 galaxies.
 \emph{Left:} prototype (mean) circular curve, which corresponds to the CVC at center of each $k$-means cluster group.
\emph{Right:} normalized (on the flat/asymptotic circular speed) prototype curve for each $k$-means cluster group.
}
\label{fig:centroid}
\end{figure*}
%

\subsection{CVC classes -- main features}
\label{SS:dynprop}
After analysing the shape and amplitude of the CVCs of the 238 (E1--Sdm) galaxies, 
we list the main properties of the four new CVC classes  (see Figs \ref{fig:centroid} and \ref{fig:histos}).

{\it Slow-rising (SR)}: the circular velocity increases monotonically without  a peak in the centre. 
This class consists mostly of late-type spiral galaxies (Sb$-$Sdm). The typical amplitude of CVCs reaches $\sim$ 150 \kms.

\emph{Flat (FL)}: the circular velocity is approximately constant throughout the whole galaxy. This class is represented mostly by early-type spiral galaxies (Sab$-$Sbc). 
The average amplitude of CVCs reaches $\sim$ 210 \kms.

\emph{Round-peaked (RP)}: the circular velocity rises steeply and has a round peak, gradually changing to the flat part of the CVC at radius $\sim 0.5R_e$. 
This class is represented by early-type galaxies and early-type spiral galaxies (E4$-$S0a, Sa$-$Sbc). The average amplitude of CVCs reaches $\sim$ 300 \kms.

{\it Sharp-peaked (SP)}: the circular velocity rises steeply and has a sharp peak, making a sudden change to the flat part of the CVC at radius $\sim 0.2R_e$. 
This class is  dominated by early-type galaxies (E1$-$E7). 
The average amplitude of CVCs reaches $\sim$ 400 \kms.

In general, the SR class is mostly represented by late-type galaxies, while the SP class is represented by early-type galaxies. 
FL- and RP-CVCs include galaxies with a wide-range of morphologies.

\subsection{Global properties of the galaxies through CVC Class }
\label{SS:box}
To understand how the CVCs of the galaxies are shaped at different stage of their evolution, we explore further 
the relation between the CVC class properties and fundamental galaxy parameters. 

In Fig. \ref{fig:box}, we plot vertical box-and-whisker diagrams of the four CVC classes  with different colour 
(SR$-$black, FL$-$blue, RP$-$green, SP$-$red). Box plots are a convenient way to graphically represent groups of numerical data through their quartiles of the distribution (\citealt{Tukey77}). 
The lines extending vertically from the boxes, called whiskers 
(shown with dashed line in Fig. \ref{fig:box}), indicate variability outside the upper and lower quartiles (i.e., box-and-whisker diagram). 
In this descriptive statistics, the outliers are plotted as individual points (crosses in our case), where the spacings between the different parts of 
the box indicate the degree of dispersion (spread) and skewness in the data. The median values of the distributions are indicated by the white lines in the box plots. 
The box contains 50 per cent of the data. In the case of a normal distribution, this corresponds to two times the median absolute deviation of the distribution, 
where the extent of the whisker corresponds to approximately 3$\sigma$ ($\sigma$ being the standard deviation).
The numbers of the galaxies, included in the statistics of the box-and-whisker diagrams, are printed with the label "stat" in each panel. 
\begin{figure}
\centering
{\includegraphics[width=0.4\textwidth]{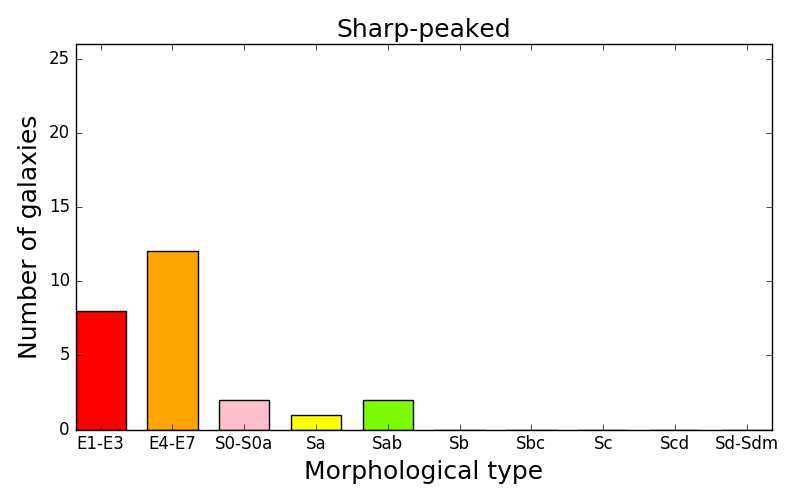}
\includegraphics[width=0.4\textwidth]{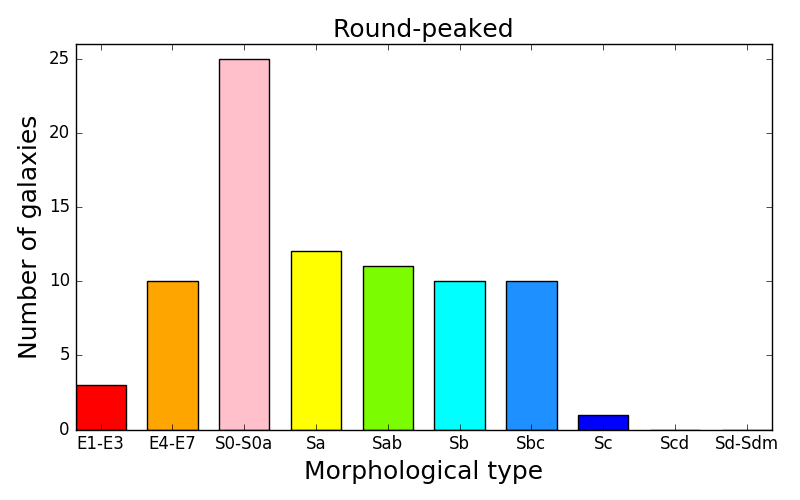}
\includegraphics[width=0.4\textwidth]{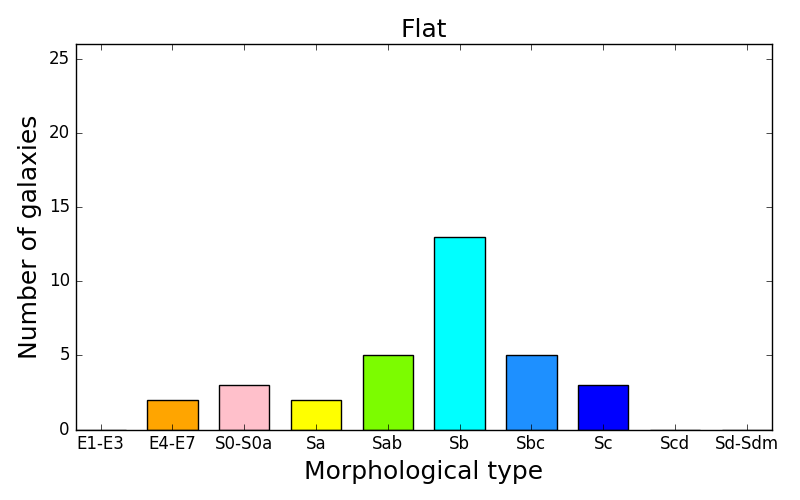}
\includegraphics[width=0.4\textwidth]{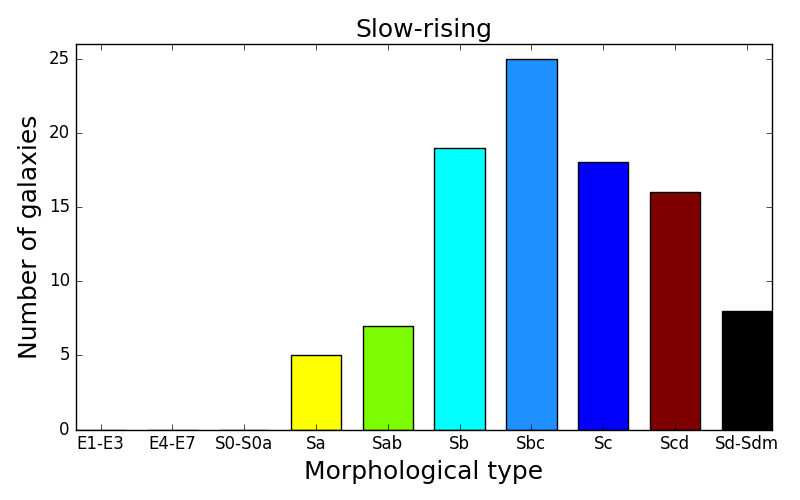}}
\caption{Distribution of the CVC classes by morphology. 
SP-CVCs are mostly presented by early-type galaxies, while SR-CVCs are mainly presented by late-type spiral galaxies. RP-CVCs and FL-CVCs are presented by various morphologies. }
\label{fig:histos}
\end{figure}

In panels `A$-$D' of Fig.\ref{fig:box}, we explore the photometric galaxy properties:
the effective radius ($R_e$) corresponds to the radius, which contains half of the total light of the galaxies and it is measured via growth curve analysis as described in \citet{Walcher2014}; 
the total luminosity ($L_{\mathrm{tot}}$) is obtained from the $r$-band SDSS (DR12; \citealt{Alam2015}) images via MGE method 
(see Section \ref{S:analysis}); 
the bulge-to-total flux  (B/T) ratio  and disc-to-total luminosity (D/T) ratio  are calculated by \cite{Mendez-Abreu2017} from $r$-band SDSS (DR7; \citealt{Abazajian2009}) images.

In panels `E$-$H' of Fig.\ref{fig:box}, we explore the stellar population properties, taken from \citet{Gonzalez-Delgado2015}:
the stellar surface-mass-density, $\mu_*$ (within half-light radius) is extinction corrected; the total stellar mass ($M_*$) is calculated using Chabrier initial mass function (IMF); 
the age (Age) and the metallicity ($Z$) of the galaxies are light-weighted and mass-weighted within half-light radius, respectively. 
\citet{Gonzalez-Delgado2015} apply the fossil record method based on spectral synthesis techniques to derive these physical properties for each spatial resolution element of the integral field unit (IFU).

In panels `I$-$L' of Fig.\ref{fig:box}, we explore the kinematic properties of the galaxies. 
One of them is the mean value of the de-projected rotational velocity of the galaxies $v_{\phi}$, 
calculated within half-light radius ($R=0.5R_e$).  
It is expressed as $v_{\phi}=v_{los}/\sin(i)$, where $i$ is the inclination of the galaxies and $v_{los}$ 
is the line-of-sight velocity derived via \texttt{kinemetry} procedure of \citet{Krajnovic2006}. Further,
the mean value of the line-of-sight velocity dispersion ($\sigma_{los}$) is taken within an aperture of half-light radius ($R=0.5R_e$).   
The ratio $v_{\phi}/\sigma_{los}$ aims to reflect the ordered-over-random motion of the galaxies in their central parts.

However, a more robust representation of the latter parameter is the specific angular momentum $\lambda_{R_e}$, which is 
integrated within one effective radius ($R=1.0R_e$, reached by most of our galaxies) as follows:
\begin{equation}
\lambda_{\mathrm{R}} = \frac{\sum_{i=0}^{N_p}F_iR_i|V_i|}{\sum_{i=0}^{N_p}F_iR_i\sqrt{V^2_i+\sigma^2_i}} 
\label{eq:lambdaR}
\end{equation}
where $F_i$,$R_i$,$V_i$ and $\sigma_i$ are the flux, circular radius, velocity and
velocity dispersion of the $i$-th spatial bin, the sum running on the $N_p$ bins (\citealt{Emsellem2007}).

In panels `M$-$P' of Fig.\ref{fig:box}, we explore the dynamical properties: 
the maximum circular velocity ($V_{c,max}$) is the circular velocity of the galaxy with the largest amplitude up to 1.5$R_e$; 
the total dynamical mass $M^{tot}_{dyn}$ is calculated as the multiplication between the total luminosity of the galaxies ($L_{tot}$), 
derived via MGE method, and the dynamical mass-to-light ratio $(M/L)_{\mathrm{dyn}}$, obtained from JAM-MCMC approach (see Section \ref{SS:dyn});
the discrepancy factor $f_d=1-(M^{tot}_{*,pop}/M^{tot}_{dyn})$, showing the relation between the total stellar and dynamical masses as a fraction value.

The maximum circular velocity, the total $r$-band luminosity, total stellar mass, total dynamical mass, age, metallicity and B/T ratio increase gradually in order: 
SR, FL, RP, SP. This means that galaxies with SR-CVCs are low-mass, faint, young, metal-poor with small bulges in comparison with 
the other CVC classes,  and in particularly with SP class, which is the most massive, bright, old, metal-rich with almost 
100 per cent B/T ratio. 
Instead, the D/T ratio decreases with increasing CVC class. 

The stellar surface mass density of the galaxies ($\mu_*$) also increases with CVC class,  but it saturates to log $\mu_* \sim$ 2.7 $M_{\odot}~\mathrm{pc}^{-2}$ for RP and SP classes. 
They have the highest and approximately identical median values of the distribution.
Interestingly, the median of the distribution for the galaxy effective radius (half-light radius) show that SP- and FL-CVCs have the largest effective radii in contrast to SR- and RP-CVCs.

The random motion of the galaxy $\sigma_{los}$ increases with CVC class. 
Instead, the ordered motion $v_{\phi}$ increases with CVC class up to the RP-CVCs, and then decreases for SP class.
The value of $v_{\phi} / \sigma$ better represents which of the motions of the stars in the galaxies (ordered or random) mostly contribute in the total potential of the galaxy and shape the CVC. For example, FL-CVCs are mostly shaped by the ordered motions, instead SP-CVCs are shaped by the random motions. The angular momentum of the galaxies $\lambda_R$ shows similar trend. 
SL, FL and RP class are fast rotators, while SP are slow rotators. 

The dynamical mass-to-light ratio, $(M/L)_{\mathrm{dyn}}$, is lowest for FL-CVCs, and larger for SR, RP and SP classes. The discrepancy mass factor ($f_d=1-M_{*}/M_{dyn}$) 
follows similar trend: SR and SP class have the largest value
(with 74 per cent and 71 per cent, respectively) in contrast to the FL and RP class (with 59 per cent and 61 per cent, respectively). The large value of $f_d$ might be due to the uncounted gas fraction (most likely for late-types), variation of the IMF or a large DM content in the galaxies.

\begin{figure*}
\centering
\includegraphics[width=1.0\textwidth]{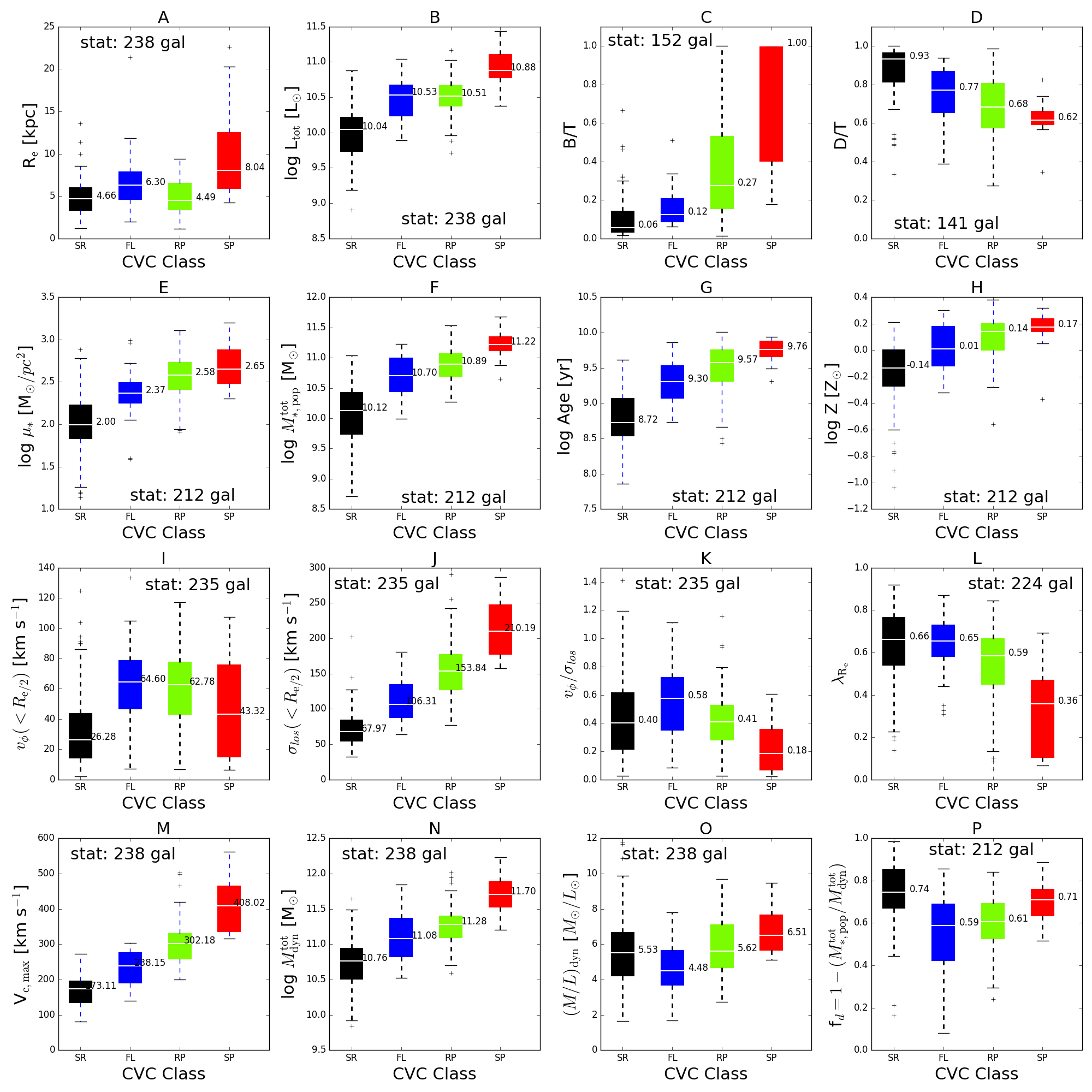}
\caption{Box-and-whisker diagram of fundamental galaxy properties through CVC class (see Section \ref{SS:box}). 
The white lines and the numbers represent the median value of the distributions for each class (SR-red, FL-blue, RP-green, and SP-red). 
The box contains 50 per cent of the data. In the case of a normal distribution this corresponds to two times the median absolute deviation of the distribution, 
where the extent of the whisker (dashed lines) corresponds to approximately 3$\sigma$ ($\sigma$ being the standard deviation). 
The numbers of the galaxies, participating in the statistics of the box-and-whisker diagrams, are printed with the label `stat' in each panel (note here that for panels C and D, biases might occur due to the lower number of galaxies).
We find several global correlations:
SR-CVCs are typical for low-mass, young, faint, metal-poor, disc-dominated galaxies with large value of the discrepancy mass factor ($f_d\sim$ 74 per cent); 
SP-CVCs are typical for high-mass, old, bright, metal-rich, bulge-dominated galaxies with also significant value of the discrepancy mass factor ($f_d\sim$ 71 per cent);
FL-CVCs and RP-CVCs appear presented by galaxies with intermediate mass, age, luminosity, metallicity, 
bulge-to-disc ratio, and value of the discrepancy mass factor ($f_d\sim$ 59 per cent and $f_d\sim$ 61 per cent, respectively). 
These relations suggest that CVCs can be used as a reliable classifier for studying the formation and evolution of the galaxies (see Section \ref{S:discussion}). 
}
\label{fig:box}
\end{figure*}

\section{Discussion}
\label{S:discussion}
Every classification scheme focuses on certain properties of the objects, and consider some features as primary and others as secondary. In our study, the primary focus of the CVC classification scheme is the shape and amplitude of the CVC, tracing the total potential of the galaxies. We aim to understand if the CVC encodes some important information about galaxy's internal structure and merging history. We use stellar dynamics to explore the potential of the galaxies from early- to late-types with one approach, which give us the great opportunity to study this problem uniformly. 

\subsection{The new classification scheme}
\label{SS:scheme}

The Hubble sequence (\citealt{Hubble1936}) is the original classification of galaxies, based on their photometric morphology.  Morphological classification has good utility and clearly groups galaxies with similar properties.  However, the Hubble sequence is only tenuously connected to the long-term evolutionary processes that govern galaxies.  A galaxy's morphology is set by its recent star formation history and gas content. 
For example, elliptical and dwarf spheroidal look morphologically similar but have different formation histories (e.g., \citealt{Kormendy2012}). In contrast, the stellar kinematics are intimately connected to the galactic potential and are thought to probe a longer evolutionary history. Thus, identifying a CVC classification scheme for galaxies will likely highlight different physical features than morphological classifications. Our classification scheme is an effort in this direction.  

We have used PCA analysis for dimensionality reduction and $k$-means clustering to group the circular curves into four distinct groups. 
The choice of four clusters, while only weakly preferred by clustering metrics, is guided by the visual distinction of four different families of circular curves regarding their shape (see Fig.\ref{fig:morphvsdyn}). The four families serve as a useful representation, but we stress that \emph{there is a continuum of CVC types}.
This continuum is quantitatively described by the projections, PC$_i$.

Two features distinguish this approach from past efforts.  First, we use detailed stellar dynamical modelling using the JAM model, 
which capture the dynamical signatures of both hot and cold stellar components.  Secondly, we use a non-parametric approach to generate a 
classification based on the circular curves.  Since a small basis set can reconstruct $\sim$99.97 per cent of the informative content of a wide variety of circular curves,
this indicates the power of the PCA analysis when derived from a large observational set. However, the main features of our sample of galaxies ($\sim$ 99 per cent) are contained 
in the first two PCs; thus we apply $k$-means method only in the two-dimensional parameter space of PC1 and PC2.

The first five PC eigenvectors $\textbf{\textit{u}}_i$ represent an excellent vector base for reconstruction of the CVCs 
(see Figs \ref{fig:pcaradial} and \ref{fig:califa}). 
The main eigenvectors appear similar to the classical 
CVC decomposition on individual radial velocity contributions of 
the components: bulge, disc, gas, DM halo, etc. (e.g., \citealt{vanAlbada1986}).
In upcoming papers, we will explore if such physical connection exists. 

Further, our proposed classification shows very promising relations with the fundamental properties of the galaxies (see Fig. \ref{fig:box}).
SR-CVCs are represented by young, late-type spirals (Sc--Sdm), low-mass, faint, low-metallicity galaxies with small bulges, effective radius, stellar-mass-surface density and maximum circular velocity.  
The SR class is one of the extremes in our classification, where the other extreme is the SP class,  represented by elliptical (E1-E7), old, high-mass, bright, high-metallicity galaxies with 
large bulge/spheroid, large effective radius, stellar mass surface density and maximum circular velocity. SR- and SP-CVCs are the most separate groups in our two-dimensional PC 
clustering space (see Fig. \ref{fig:kmeans}) and opposite in most of the  explored galaxy properties. On the other hand, SR and SP classes are characterized by 
similar small rotation and large dark matter contents in comparison with the other two classes,  FL- and RP-CVCs.

In Fig. \ref{fig:box} we see that FL-CVCs and RP-CVCs appear to be intermediate CVC classes.
By properties distributions, FL class appears close to SR class, while RP appears close to SP. This is also the case in the PC clustering space, where SR and FL have mostly negative values of the PC projections, while RP and SP have mostly positive values (see Fig. \ref{fig:kmeans}).

\emph{In short, by looking at the shape and the amplitude of the CVCs, it appears possible to guess the properties of the galaxies.}
%
\subsection{Link to galaxy evolution: CVC classes on the mass-size plane}
\label{SS:evol}
\begin{figure*}
\centering
\includegraphics[width=1.0\textwidth]{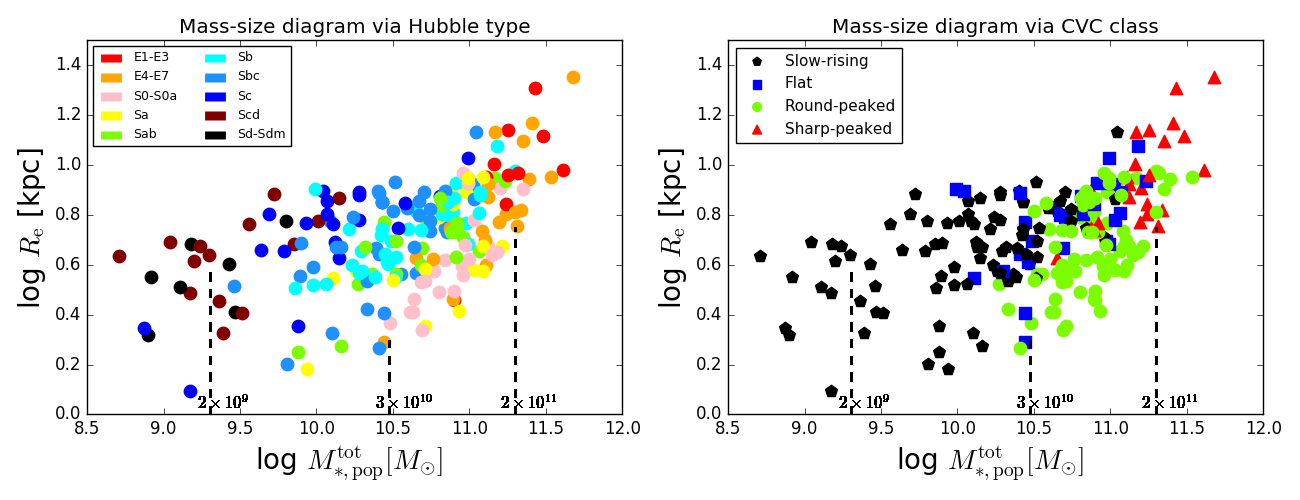}
\caption{Mass-size (MS) plane of the 238 (E1-Sdm) CALIFA galaxies via Hubble type (left) and CVC class (right).
The dashed lines indicate the three characteristic stellar masses in the MS diagram of \citet{Cappellari2013}: 
below around $2\times10^{9}$ the MS is populated by no regular early-type galaxies; around $3\times10^{10}$ the early-type galaxies reach their minimum size and maximum stellar density before a sudden change in the slope at larger masses; 
below $2\times10^{11}$  the MS is mostly populated by fast rotators and rarely by slow rotators; above $2\times10^{11}$ there are no spiral galaxies and MS is dominated by round and triaxial slow rotators. The distribution of our 238 CALIFA galaxies on the MS diagram via Hubble type (left) fit well to these three domains of characteristic masses, described in \citet{Cappellari2013}. 
Galaxies with SR-CVCs occupy the MS diagram below around $M^{\mathrm{tot}}_{\mathrm{*,pop}}\approx3\times10^{10}$, FL-CVCs and RP-CVCs -- mainly the domain between  $M^{\mathrm{tot}}_{\mathrm{*,pop}}\approx3\times10^{10}$ and $M^{\mathrm{tot}}_{\mathrm{*,pop}}\approx2\times10^{11}$, whereas SP-CVCs -- mainly above $M^{\mathrm{tot}}_{\mathrm{*,pop}}\approx2\times10^{11}$. This indicates that the shape and the amplitude of the galaxies are tightly linked to galaxy evolutional processes (see Section \ref{SS:evol}).
}
\label{fig:MassSize}
\end{figure*}

CVCs trace the total potential of the galaxies, and potentially incorporate all information about their single constituents: stars, gas and DM.
In Fig. \ref{fig:box}, we found that the properties of the galaxies are connected to the shape and the amplitude of their CVCs. Therefore, the galaxy evolution should, somehow, be related to the new classification scheme.

The relation between the total stellar mass $M^{\mathrm{tot}}_{\mathrm{*}}$ and the effective radius $R_e$ of the galaxies is called mass-size (MS) diagram, and it is often used to probe galaxy evolution (e.g., \citealt{Cappellari2013}; \citealt{vanDokkum2015}; \citealt{Cappellari2016}). 
Through this diagram, fig.29 of \cite{Cappellari2016} summarizes the possible evolutionary scenario for present-day early-type galaxies.
\cite{Cappellari2016} discuss two channels of galaxy evolution. In the first one, galaxies grow through gas accretion or gas rich minor mergers, where the star-forming discs are the progenitors of the fast rotators early-type galaxies, and with the increase in mass the bulge grows. The second channel is dominated by mostly dry mergers, which move galaxies on the MS diagram by increasing their size proportionally to the stellar mass. This scenario is also postulated by \cite{vanDokkum2015}, who explore the MS diagram of galaxies from $z \sim$ 3 to $z\sim$ 0. They conclude that star-forming galaxies grow mostly
in mass with the gradual increase in their stellar density. Once galaxies reach a threshold of their velocity
dispersion or stellar density, they tend to quench due to AGN feedback or other processes.
The possible growth of the quenching galaxies goes through dry merging with consequent steepness of the MS diagram.

To study how our findings may be connected with these galaxy evolution theories, in Fig. \ref{fig:MassSize}, we investigate the local 
MS diagram, ($M^{\mathrm{tot}}_{\mathrm{*,pop}}$, $R_e$), for our sample through Hubble type (left) and CVC class (right). The total stellar mass ($M^{\mathrm{tot}}_{\mathrm{*,pop}}$) of the galaxies is taken from \cite{Gonzalez-Delgado2015}, while the effective radius ($R_e$), measured in $r$-band SDSS photometry, comes from \cite{Walcher2014}. We overplot three dashed lines, corresponding to the characteristic stellar masses in the MS diagram of \citet{Cappellari2013}. In their fig.9, \citet{Cappellari2013} show that the MS diagram is populated by no regular early-type galaxies below $\approx2\times10^{9}$ M$_{\odot}$, while around $3\times10^{10}$ M$_{\odot}$ the early-type galaxies reach their minimum size and maximum stellar density. After this characteristic mass, MS relation becomes suddenly steeper at larger stellar masses. The MS diagram is mostly populated by fast rotators and rarely by slow rotators below $2\times10^{11}$ M$_{\odot}$; however above $2\times10^{11}$ M$_{\odot}$ is mainly presented by round and triaxial slow rotators with the absence of spiral galaxies.

The distribution of our 238 CALIFA galaxies on the MS diagram via Hubble type, in Fig. \ref{fig:MassSize} (left), fits well to these three domains of characteristic masses, described in \citet{Cappellari2013}. 
Galaxies with SR-CVCs occupy the MS diagram below  $M^{\mathrm{tot}}_{\mathrm{*,pop}}\approx3\times10^{10}$ M$_{\odot}$, FL-CVCs and RP-CVCs mainly cover the mass range between  $M^{\mathrm{tot}}_{\mathrm{*,pop}}\approx3\times10^{10}$ M$_{\odot}$ and $M^{\mathrm{tot}}_{\mathrm{*,pop}}\approx2\times10^{11}$ M$_{\odot}$, and SP-CVCs mostly populate above $M^{\mathrm{tot}}_{\mathrm{*,pop}}\approx2\times10^{11}$ M$_{\odot}$.

Our results appear complementary to the MS diagram presented in figs 9 and 14 of \cite{Cappellari2013}. 
They show that the growth of bulge makes the galaxy old and red with increasing dynamical mass-to-light ratio. This is in agreement with our results in  panels C, G, H and O of Fig. \ref{fig:box}. Furthermore, they discussed that the spiral galaxies do not populate the region of the MS diagram above the characteristic mass $\gtrsim 2 \times 10^{11}$ M$_{\odot}$, and the ($M_*,R_e$) relation become steeper from late-type galaxies to early-type galaxies. We see the same trends through Hubble type for our sample in Fig. \ref{fig:MassSize} (left-hand panel). In addition to the classical Hubble type study of the MS diagram, we illustrate how the galaxy potential, shape and amplitude of the CVC transform on this diagram in the right-hand panel of Fig. \ref{fig:MassSize} with symbols corresponding to the different CVC classes.
Interestingly, the CVC classes  transit in the \emph{sequence} SR-FL-RP-SP from left to right on the MS plane, which represents the direction of growing of the amplitude and steepness of the CVCs. Those two characteristics of the CVCs may represent the mass increase and bulge growth of the galaxies, respectively (e.g., \citealt{Marquez1999}). Therefore, the CVC sequence matches closely with the direction of the galaxy evolution through Hubble type according to \cite{Cappellari2013} and \cite{Cappellari2016}.

A similar galaxy evolution scenario is proposed by \cite{Combes2009}, where the bar-bulge cycle drives the local processes (e.g. secular evolution) in the spiral galaxies, which are considered the progenitors of the early-type galaxies via sequences of minor and major mergers. Additionally, \cite{Khochfar2006} shows that bulges might grow from the local or global instabilities of the disc through secular evolution or minor mergers, respectively, which can drive the gas to the centre and form new stars. This scenario is further supported by \cite{Gonzalez-Delgado2015}, where CALIFA galaxies are described to grow from inside out.

The properties of our four CVC classes can also be linked to recent results from hydrodynamical simulations. In particular, \cite{Tonini2016} present four channels of galaxy evolution. The first channel of their galaxy evolution scheme is presented by a smooth accretion of gas in the DM halo potential, where the stellar disc is growing inside out regulated by local star formation and stellar feedback without formation of bulge. This channel can be related to the SR-CVCs class of the faint, young, metal-poor, bulgeless with low stellar surface mass density CALIFA galaxies (see panels B, C, D, E, G, H of Fig. \ref{fig:box}). 

The second channel is presented by disk perturbations from chaotic gas accretion with high infall rates or gas-rich minor mergers provide large supplies of gas, which sinks towards the galaxy centre and a violent burst of star formation occurs. This channel can be associated with FL or RP class,  which show intermediate properties' values in Fig. \ref{fig:box}. These galaxies already have well-formed bulge and disc, and are characterized by intermediate age and metallicity, significant rotation and stellar surface mass density (see panels C, D, E, G, H of Fig. \ref{fig:box}). For those galaxies, the discrepancy mass factor $f_d$ is the lowest, which might indicate high star-forming galaxies (see panel P of Fig. \ref{fig:box}). 

The third channel is presented by minor mergers of satellites that overtake the disruptive force of the central galaxy's DM halo. In the case of early-type galaxies, there is a formation of a merger-driven bulge with the formation of the new stellar material in shells around it. However, in the case of a disc-dominated galaxy, there is a formation of an instability-driven bulge with the formation of new 
starburst in its regions. This channel can be associated with RP or SP classes.  These galaxies have big spread in the B/T and D/T ratios (panels C and D), which might be the result of previous minor mergers. Additionally, those are the CVC classes dominating the MS relation on the steepness massive end with the largest dynamical masses, dynamical mass-to-light ratios, maximum velocities, which galaxies are also oldest, richest of metals, with the highest stellar surface mass density (see panels M, N, O, E, G, H of Fig. \ref{fig:box}).

The fourth channel is presented by major mergers, where the two colliding galaxies have similar masses. These galaxies are completely pressure supported with the formation of merger-driven bulge. The final equilibrium of the new system depends on the overall gravitational potential. This channel can be presented by the SP class,  which is presented by bright, old, metal-rich, pressure-supported, slow-rotating galaxies with the highest stellar mass surface density (see panels B, E, G, H, I, J, K, L of Fig. \ref{fig:box}), which might indicate those previous major mergers.

Furthermore, \cite{Wang2015}, using hydrodynamical simulations for studying the inefficiency of galaxy formation across cosmic time, found various shapes of the rotation curves of the galaxies with changing halo mass, which represent a gradual transition, similar to our finding in Fig \ref{fig:morphvsdyn}. They also show similar group of shapes and amplitudes of the circular curves as the shapes and amplitudes of our four CVC classes  (fig.8 of their paper), which correspond to four different DM halo masses, i.e., the lowest halo masses $M_{halo} < 2\times 10^{10} M_{\odot}$ have CVCs that rise relatively steeply. When the halo masses increase, the CVCs rise more slowly, and vice-versa, galaxies with halo masses $\approx 6 \times 10^{11} M_{\odot}$ show steeply rising CVCs with growth of high central peak. Therefore, our CVC classes  are expected to be also linked to the DM halo mass of the galaxies, which we expect to increase with CVC class in the \emph{sequence} SR-FL-RP-SP.

Because the above considerations argue that evolutionary processes will leave a signature both on the spatial mass distributions of the disc and on resulting stellar motions, a CVC classification scheme can be well motivated.  
Fortunately, this idea is presented at the advent of a large number of IFU studies of galaxies of all types and can be further extended. The integral-field surveys such as CALIFA (\citealt{Sanchez2012}), \texttt{VENGA} (VIRUS-P Exploration of Nearby Galaxies, \citealt{Blanc2010}), \texttt{SAMI} (Sydney-AAO Multi-object Integral Field Spectrograph, \citealt{Croom2012}) and \texttt{MANGA} (Mapping Nearby Galaxies, \citealt{Bundy2015}) provide or will provide large statistical samples, a variety of FoV and spatial resolutions. This will give the opportunity to further investigate the mass distribution of galaxies with different morphologies and kinematics.

The interpretation of the MS relation (Fig. \ref{fig:MassSize}) and the box-and-whisker diagrams
(Fig. \ref{fig:box}), in the context of the current galaxy evolutionary theories, 
encourage us to explore the possibility if the CVC classification scheme could be a \emph{dynamical evolution sequence of the galaxies} in the direction SR-FL-RP-SP.
In upcoming papers, we intend to explore the properties of the CVC classes  in relation to the ISM -- star formation and gas content of the galaxies (e.g.,  \citealt{Kalinova2016}; \citealt{Bolatto2017}), as well as the radial acceleration traced by the CVCs 
(e.g., \citealt{McGaugh2016}), which will shed more light on a possible dynamical evolution of the galaxies.

\section{Caveats}
\label{S:caveats}
\emph{(i) Sample.} In our study, the selected galaxies cover a variety of morphological types and balance the number of early- versus late-type galaxies, as well as covering a wide range of masses. However, we still might expect a bias in our results due to either our selection criteria, or the CALIFA mother sample selection criteria (see \citealt{Walcher2014}). Additionally, given the data-based nature of the classifier, increasing the sample size would bring this CVC-based classification scheme into clarity and possibly provide a clearer distinction among the proposed CVC classes.  

\emph{(ii) Dynamical modelling assumptions.}
A significant limitation of our work stems from the adopted assumptions in JAM dynamical model.
We rely on two fundamental assumptions: constant mass-to-light ratio $(M/L)_{\mathrm{dyn}}$ and constant velocity anisotropy $\beta_z=1-\sigma_z^2/\sigma_R^2$.
While more general techniques exist, such as Schwarzschild's orbit superposition method (\citealt{Schwarzschild1979}), JAM represents the best available compromise between physical accuracy in deriving stellar dynamics and computational efficiency for a large sample.  
The remain long-standing uncertainties in the dynamical modelling since we do not have good knowledge about the gradient of the mass-to-light ratio and the velocity anisotropy in the different types of galaxies, which might affect the shape and the amplitude of our derived 238 CVCs.  
Once more extended studies address these issues, as well as a set of gas dynamical properties are derived for a large sample of galaxies (e.g. the EDGE sample; \citealt{Bolatto2017}), the classification should be repeated using improved approaches.  In addition, the combination of different kinematical tracers (e.g., stars, CO, HI) may provide better constraints on galaxies' potential without heavy computational cost (e.g., \citealt{Cappellari2011, Cappellari2015}), but large, matched samples of gas and stellar dynamics need to be explored (Kalinova et al., in preparation). 

\emph{(iii) Radial extent of the observations.}
For several galaxies of our sample (e.g., IC\,1079, NGC\,0171, NGC\,6173), the stellar kinematics does not reach $R/R_e=1.5$. 
In this case, the extrapolations of the CVCs based on the $(M/L)_{\mathrm{dyn}}$ can be uncertain.  
Further, we have experimented with classifying the CVCs using smaller sections of $R/R_e$.  
In general, the classification becomes noisier; however, since most of 
the classification power comes from the structure of the inner circular curve, the same basic results remain visible.  The utility of a new approach remains high, 
and this method can be updated as a larger sample becomes available. 
\begin{figure}
\centering
\includegraphics[width=0.45\textwidth]{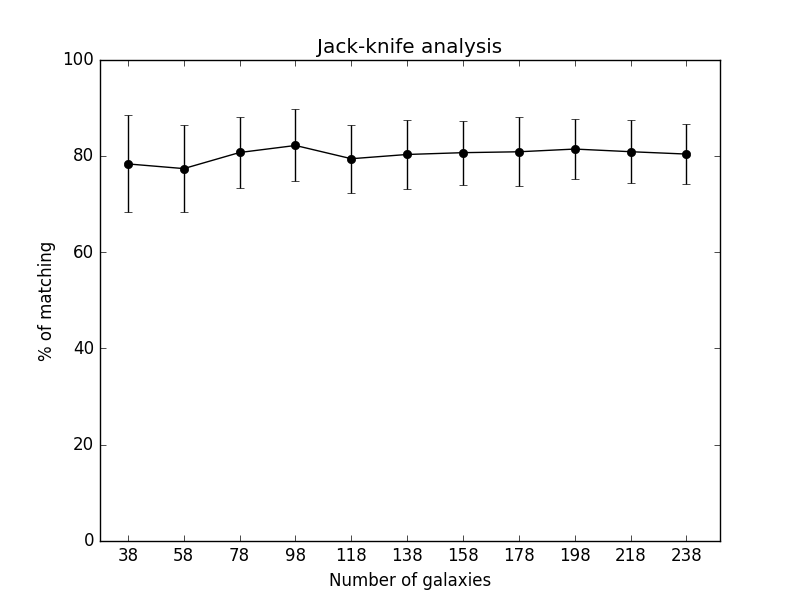}
\caption{The results of a jack-knife technique, which probes the stability of the classification through a cross-validation approach:
we select a sub-sample from the original set and carry out a complete analysis including 
PCA decomposition of the rotation curves and the $k$-means separation into classes. 
On the $x$-axis, we plot the number of the sub-samples drawn from the original set of 238 galaxies, while the $y$-axis corresponds to the 
matching of the original classification in per cent (see Section \ref{S:caveats}).}
\label{fig:jack}
\end{figure}

\emph{(iv) Validation of PCA and $k$-means analyses.}
Both the size and galaxy content of the sample will influence both the PCA representation and any classification of 
the CVCs based on this scheme.  
Since the categories are not clearly defined in the PC space (Fig. \ref{fig:kmeans}), there will naturally be uncertainty in any division into categories.  
We stress that the PCA provides a reduced-dimensionality summary of the CVC data, and the proposed classification is a part of a continuum.

We have assessed the stability of our classification using a cross-validation technique.  We select a sub-sample from the original set and carry out a complete analysis including 
PCA decomposition of the CVCs and the $k$-means separation into classes.  We find that the primary uncertainty in the classification results from the $k$-means algorithm. 
Since the exact solution to the problem requires intensive computation,  heuristic approaches to approximate a near-optimal solution are typically employed.  
These approximate approaches result in a set of galaxies potentially being classified differently on different runs of the algorithm.  
As a first test, we run the algorithm 100 times on the full set of 238 galaxies, which results in the same categorizations for $\sim$80\% of the galaxies on average, 
with the $\sim$ 20 \% representing a systematic error imparted by the $k$-means approach.

Next, we complete the full categorization algorithm on smaller subsets of the original galaxies to assess how sample size and membership affect the classification.  
In Fig. \ref{fig:jack}, we plot the fraction of galaxies that maintain their original classifications as the sample size is changed.  
For each sample size, we draw a sub-sample of that size 20 times from the parent sample, identify the PC representation 
of the CVC and classify the galaxies using $k$-means with $k=4$ groups.  We calculate the fraction of galaxies that retain their original classification.  
Fig. \ref{fig:jack} shows that there remains some uncertainty in the classification, even for numbers approaching that of the original sample.  
We note that the classifications of the SR- and SP-CVCs tend to be more stable in their classification when compared to the 
RP- and FL-CVCs.  

While there are some uncertainties, the stability of the method is encouraging, particularly when applying a coarse machine learning approach to a relatively small sample size.  
The uncertainty appears to be driven primarily by the lack of clear divisions between categories, which reflects the continuous nature of the underlying galactic population.  
While separating the sample into four categories provides a useful means of simply describing the dynamical properties of the galaxy, there may ultimately be more utility in 
using the PCs themselves as a measure of the dynamical properties of the system.

\section{Application}
\label{S:apply}
Our analysis provides promising results in the classification of the galaxies based on their 
total potential using their CVC based on stellar dynamics. 
The sample we used is variable via morphology and large enough to be used as a  
classifier for future studies, especially for a single CVC or small samples of galaxies. 

The practical way to classify a single galaxy is to use the five main PC eigenvectors 
(as shown in the left-hand panel of Fig. \ref{fig:pcaradial}), 
and estimate the new first five PC projections for the new galaxy. 
Note that the CVC decomposition can be done after subtracting the mean circular velocity of our sample 
($\overline V_c$ =196.336 km s$^{-1}$, Section \ref{SS:pca}).
Next, the derived first two main PC projections (PC1 and PC2) are the only values necessary to define the CVC 
class of the new galaxy using the $k$-means clustering of our sample in the right-hand panel of Fig. \ref{fig:kmeans}.
After overplotting the two new PC projections in the (PC1, PC2) parameter space of our sample, 
the minimum distance to the closest $k$-means center will define the CVC class of the new galaxy.
We provide a python script with an example for the application of our PC-classifier on the link \url{https://github.com/Astroua/DynClass}.

\section{Conclusion}
\label{S:summary}
In this study, we derive the total CVCs of 238 (E1--Sdm) CALIFA galaxies. 
The shapes of the CVCs are given by the light profiles of the galaxies (via MGE model), where the amplitudes of the CVCs are given by the value of the dynamical mass-to-light ratio $(M/L)_{\mathrm{dyn}}$, derived through fitting the stellar kinematics and solving the Jeans equations in the axisymmetric case via JAM-MCMC method (Section \ref{S:analysis}).
 
We investigate the maximum circular velocities (i.e., amplitude) and qualitative shape of circular curves through Hubble sequence using PCA and $k$-means clustering technique.  We distinguish four main CVC classes: (1) in the SL class, the circular velocity increases monotonically with the absence of a peak in the centres and mostly represented by late-type spiral galaxies (Sb--Sdm) with the average amplitude of CVCs reaching $\sim$ 150 \kms.  
(2) FL-CVCs have an anaemic peak in the centres or approximately constant circular velocity  within the whole galaxy and include most of the early-type spiral galaxies (Sab--Sbc).  The average amplitude of CVCs for FL galaxies is $\sim$ 210 \kms. 
(3) RP-CVCs have a steeply rising central CVC with a round peak, gradually changing to the flat part of the CVC at radius $\sim 0.5R_e$. This group mostly includes late-type ellipticals (E4--E7), lenticular galaxies (S0--S0a) and early-type spirals (Sa--Sbc) with average amplitude of CVCs $\sim$ 300 \kms.  
Finally, SP-CVCs are steeply increasing in the centre with a sharp peak, which changes abruptly to the flat part of the CVC at radius $\sim 0.2R_e$. This class is mostly represented by early-type (E1--E7), similar to the RP galaxies. The average amplitude of CVCs reaches $\sim$ 400 \kms. 

In general, the SR class is represented by late-type, low-mass, young, faint, metal-poor and disc-dominated galaxies, while 
SP class is represented by early-type, high-mass, old, bright, metal-rich and bulge-dominated galaxies. FL and RP are represented by galaxies with various morphologies and intermediate values of mass, age, luminosity, metallicity and B/D ratio.

CVC-based classifications provide a view on galaxy potential that is complementary to morphological schemes.  Our study provides a step forward in the empirical constraints on current and future theoretical models of galaxy evolution across a wide range of galaxy morphological types and masses.  This approach provides a good way to reduce the complexity present in optical IFU studies of galaxies into a manageable descriptor of galaxy dynamics. 

The following data products of the 238 (E1--Sdm) galaxies will be made available to the 
community at the CALIFA website, \url{http://califa.caha.es/?q=content/science-dataproducts}:
the basic properties of the galaxies from Appendix \ref{A:tables};
the five main PC eigenvectors from Appendix \ref{A:PC_vect};
the five main projection PC$_{i}$ from Appendix \ref{A:PCs}; 
the MGE models from Appendix \ref{A:MGEs};
the CVC models and their uncertainties from the online material.

\section*{Acknowledgements}
This study makes use of the data provided by the Calar Alto Legacy Integral Field Area (CALIFA) survey
(\url{http://califa.caha.es}). Based on observations collected at the Centro Astron{\'o}mico
Hispano Alem{\'a}n (CAHA) at Calar Alto, operated jointly by the Max-Planck-Institut
f{\"u}r Astronomie and the Instituto de Astrof{\'i}sica de Andaluc{\'i}a
(CSIC). CALIFA is the first legacy survey being performed at Calar Alto. The
CALIFA collaboration would like to thank the IAA-CSIC and MPIA-MPG as
major partners of the observatory, and CAHA itself, for the unique access to
telescope time and support in manpower and infrastructures. The CALIFA collaboration
also thanks the CAHA staff for the dedication to this project.

We are grateful to the anonymous referees for their constructive comments, which
helped us to substantially improve the manuscript.
VK acknowledges Glenn van de Ven, Jes{\'u}s Falc{\'o}n-Barroso, and Mariya Lyubenova 
for the relevant discussions.
VK is supported by the Avadh Bhatia Fellowship at the University of Alberta.  
VK, DC and ER are supported by a Discovery Grant from NSERC of Canada.
DC acknowledges support by the Deutsche Forschungsgemeinschaft, DFG through project number SFB956C. 
LG is supported in part by the US National Science 
Foundation under Grant AST-1311862.
LSM thanks support from the Spanish {\em Ministerio de Econom\'ia y Competitividad (MINECO)} via grant AYA2012-31935.
RGD and RGB are supported by the Spanish Ministerio de 31 Ciencia e Innovaci{\'o}n 
under grant AYA2010-15081 and AyA2014-57490-P. 
SFS thanks the CONACYT-125180 and DGAPA-IA100815 projects for providing him support in this
study. EF acknowledges support from MINECO and JA grants, AYA2014-53506-P and FQM-108.
RAM acknowledges support by the Swiss National Science Foundation.
JM-A acknowledge support from the European Research Council
Starting Grant (SEDMorph; PI: V. Wild).

This research has made use of the NASA/IPAC Extragalactic Database (NED), which is operated by the Jet Propulsion Laboratory, California Institute of Technology, under contract with the National Aeronautics and Space Administration

Funding for the Sloan Digital Sky Survey IV has been provided by
the Alfred P. Sloan Foundation, the U.S. Department of Energy Office of
Science, and the Participating Institutions. SDSS-IV acknowledges
support and resources from the Center for High-Performance Computing at
the University of Utah. The SDSS web site is www.sdss.org.

SDSS-IV is managed by the Astrophysical Research Consortium for the 
Participating Institutions of the SDSS Collaboration including the 
Brazilian Participation Group, the Carnegie Institution for Science, 
Carnegie Mellon University, the Chilean Participation Group, the French Participation Group, Harvard-Smithsonian Center for Astrophysics, 
Instituto de Astrof\'isica de Canarias, The Johns Hopkins University, 
Kavli Institute for the Physics and Mathematics of the Universe 
(IPMU)/University of Tokyo, Lawrence Berkeley National Laboratory, 
Leibniz Institut f\"ur Astrophysik Potsdam (AIP),  
Max-Planck-Institut f\"ur Astronomie (MPIA Heidelberg), 
Max-Planck-Institut f\"ur Astrophysik (MPA Garching), 
Max-Planck-Institut f\"ur Extraterrestrische Physik (MPE), 
National Astronomical Observatory of China, New Mexico State University, 
New York University, University of Notre Dame, 
Observat\'ario Nacional/MCTI, the Ohio State University, 
Pennsylvania State University, Shanghai Astronomical Observatory, 
United Kingdom Participation Group,
Universidad Nacional Aut\'onoma de M\'exico, University of Arizona, 
University of Colorado Boulder, University of Oxford, University of Portsmouth, 
University of Utah, University of Virginia, University of Washington, University of Wisconsin, 
Vanderbilt University, and Yale University.

This research made use of $\texttt{ASTROPY}$ (\url{http://www.astropy.org}), a community-developed core $\texttt{PYTHON}$ package for Astronomy (\citealt{AstropyCollaboration2013}).

\footnotesize{
\bibliographystyle{mn2e_new}
\bibliography{BIB_v6.bib}
}

 \appendix
 \label{A:appendix}
\newpage
\section{Second velocity moment maps}
\label{A:maps}
 
For our sample of 238 (E1-Sdm) galaxies, in Figs \ref{fig:maps1}-\ref{fig:maps8}, the first and second columns  
show the second moment maps $\Vrms = \sqrt{V^2+\sigma^2}$ of the data and
the model with overplotted flux and MGE contours, respectively. 
The third columns correspond to the residual field between the observed and modelled $\Vrms$ maps, where 
$RES=|1-(V_{\mathrm{rms,OBS}}/V_{\mathrm{rms,MOD}})|$ and $\overline{RES}$ is the median value of the residual field.
We observed that on average the median of the residual $\overline{RES} \sim 0.10$ for the 238 CALIFA galaxies, 
corresponding to 10 per cent error.

\begin{figure*}
\centering
{\includegraphics[width=0.27\textwidth]{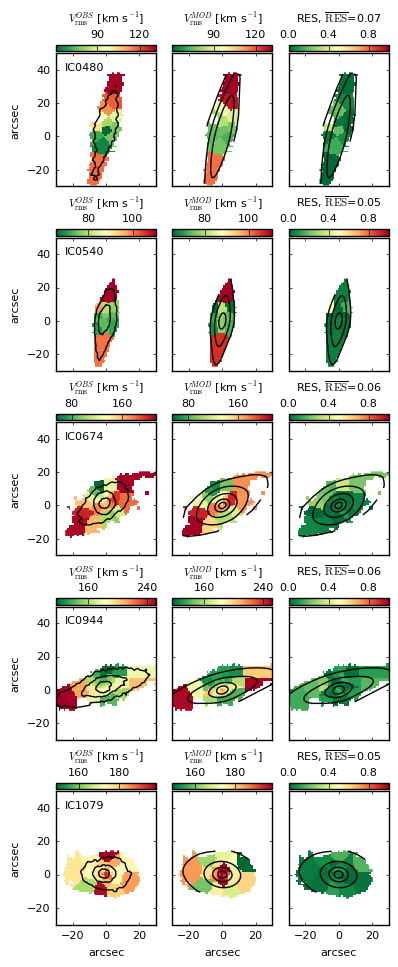}
\includegraphics[width=0.27\textwidth]{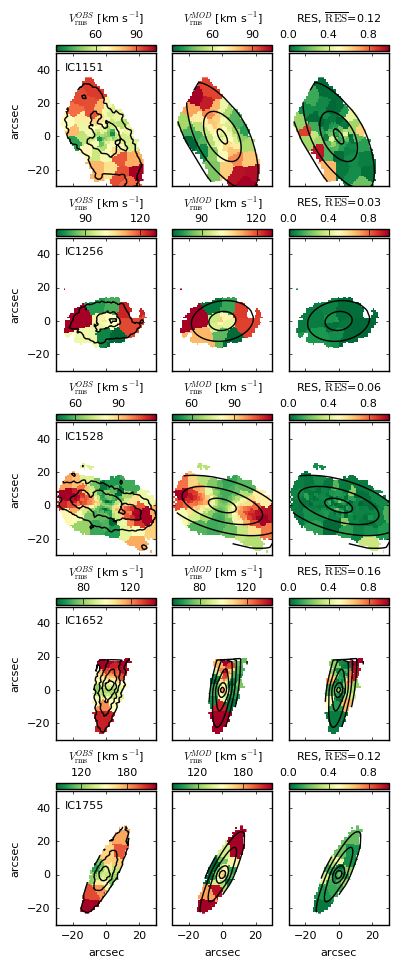}
\includegraphics[width=0.27\textwidth]{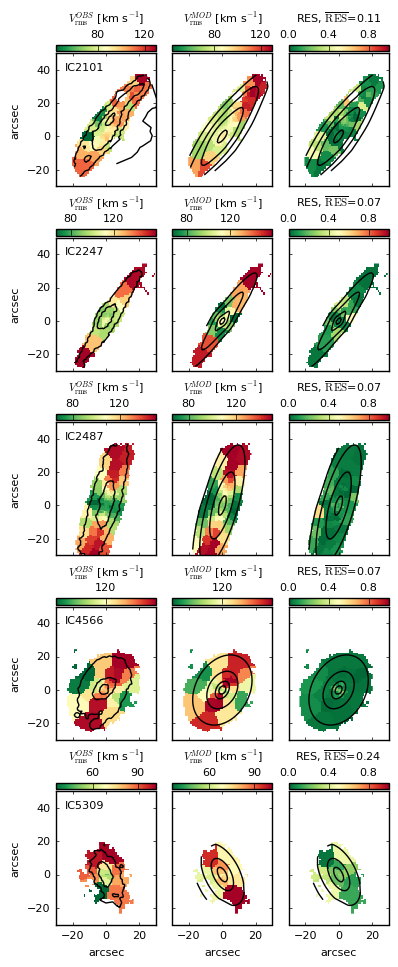}
\includegraphics[width=0.27\textwidth]{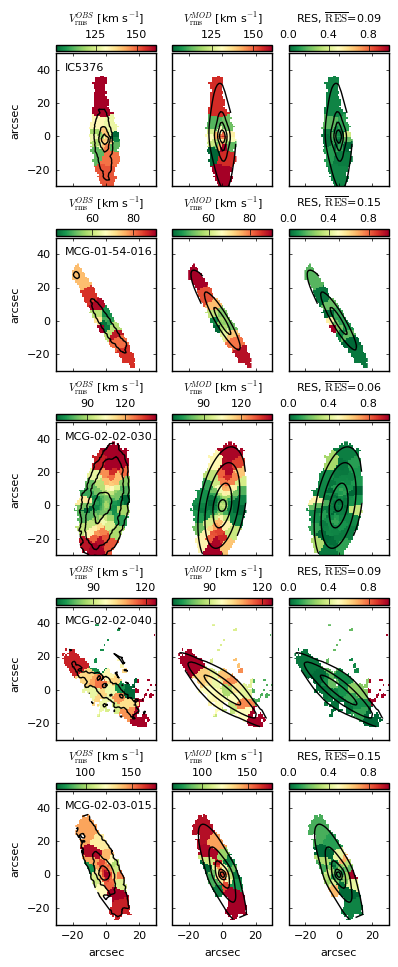}
\includegraphics[width=0.27\textwidth]{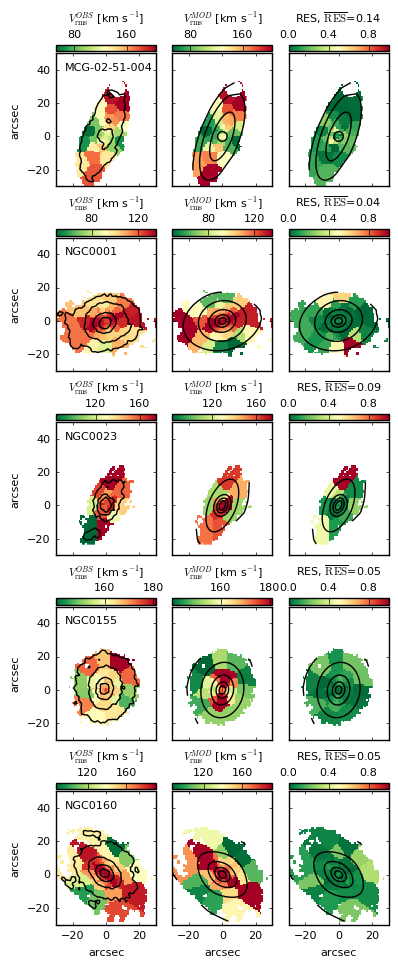}
\includegraphics[width=0.27\textwidth]{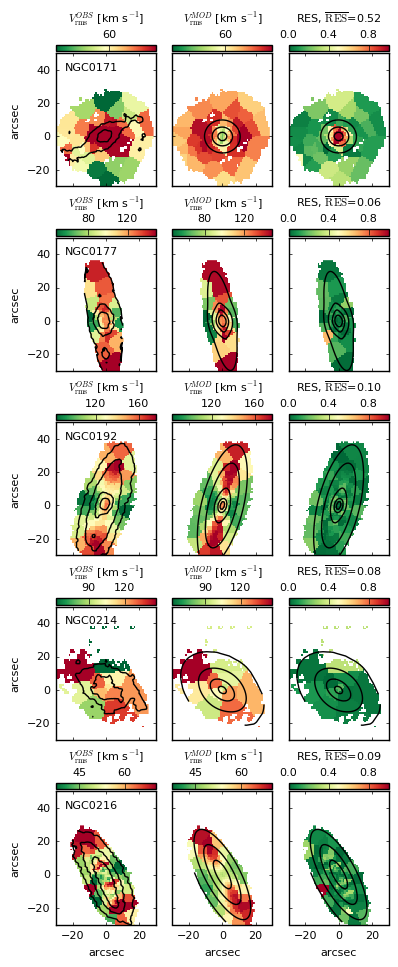}
}
\caption{Second moment maps $\Vrms = \sqrt{V^2+\sigma^2}$ of the data and the model, and residual maps of our sample of 238 (E1--Sdm)
  galaxies.}
\label{fig:maps1}
\end{figure*}

\begin{figure*}
\centering
{\includegraphics[width=0.27\textwidth]{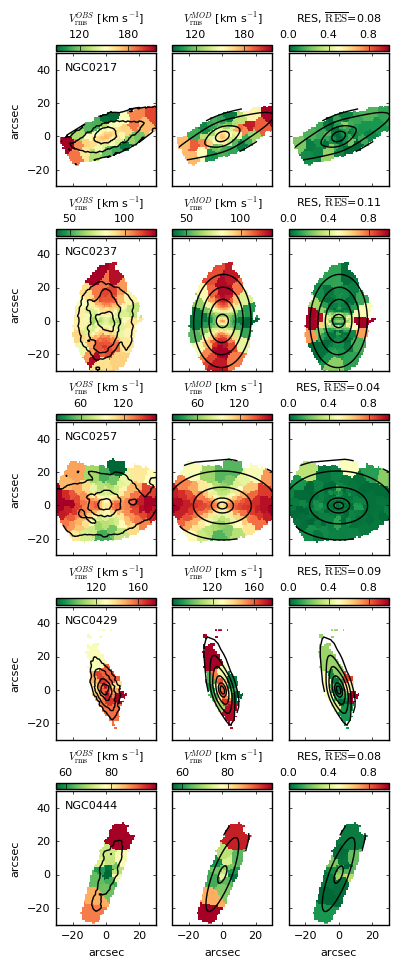}
\includegraphics[width=0.27\textwidth]{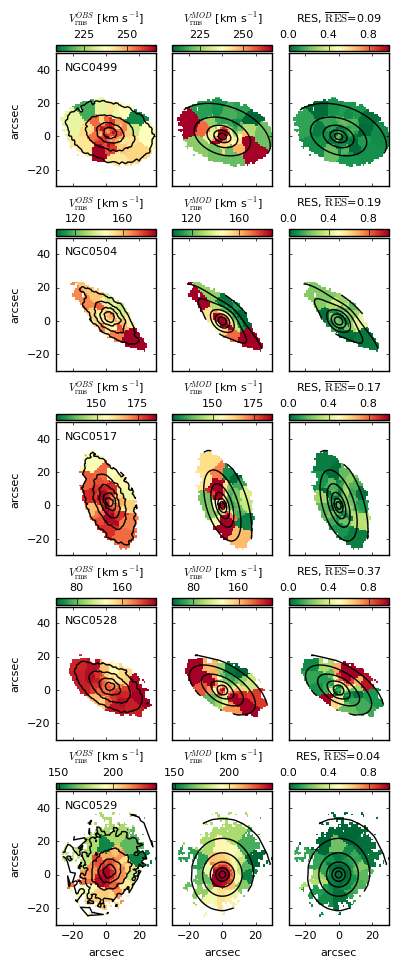}
\includegraphics[width=0.27\textwidth]{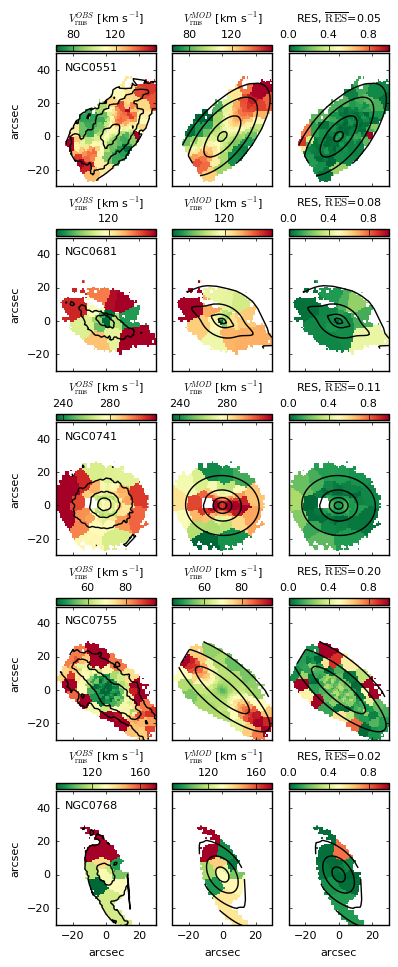}
\includegraphics[width=0.27\textwidth]{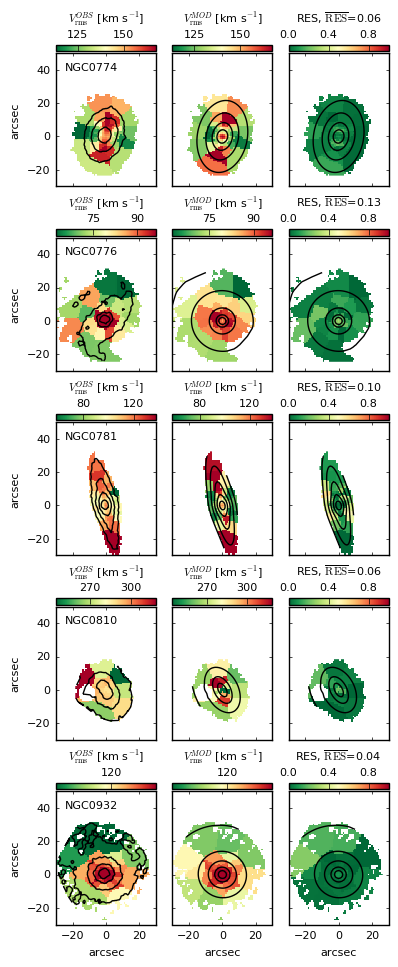}
\includegraphics[width=0.27\textwidth]{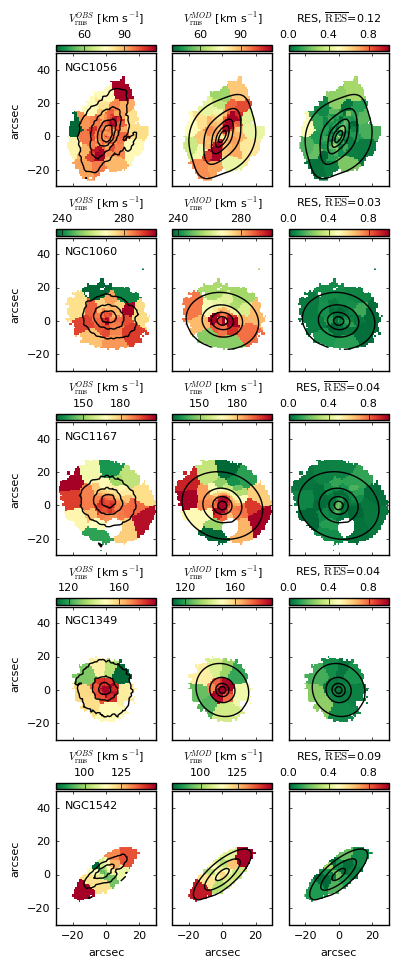}
\includegraphics[width=0.27\textwidth]{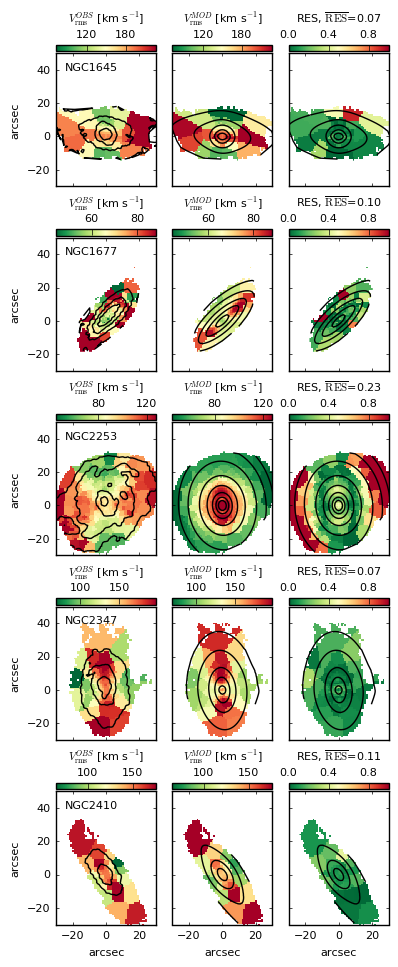}
}
\caption{Second moment maps $\Vrms = \sqrt{V^2+\sigma^2}$ of the data and the model, and residual maps of our sample of 238 (E1--Sdm)
  galaxies.}
\label{fig:maps2}
\end{figure*}

\begin{figure*}
\centering
{\includegraphics[width=0.27\textwidth]{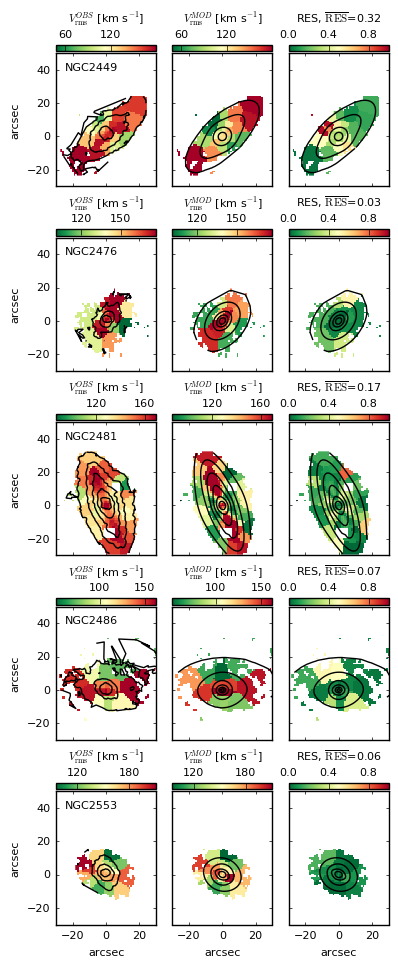}
\includegraphics[width=0.27\textwidth]{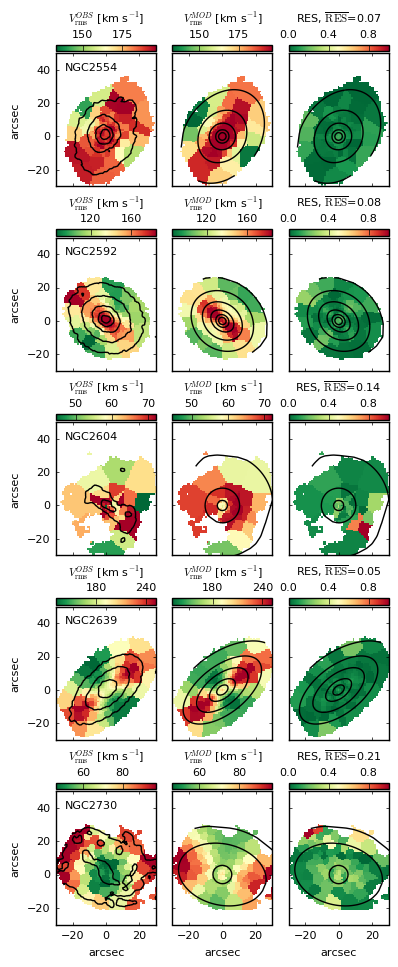}
\includegraphics[width=0.27\textwidth]{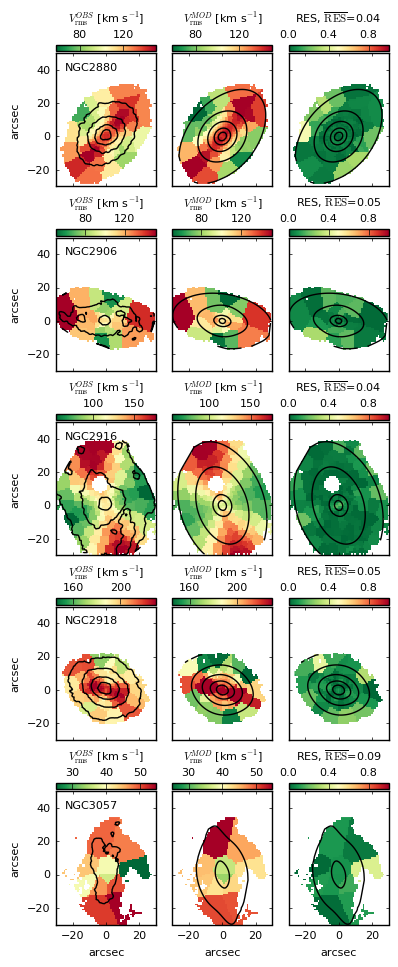}
\includegraphics[width=0.27\textwidth]{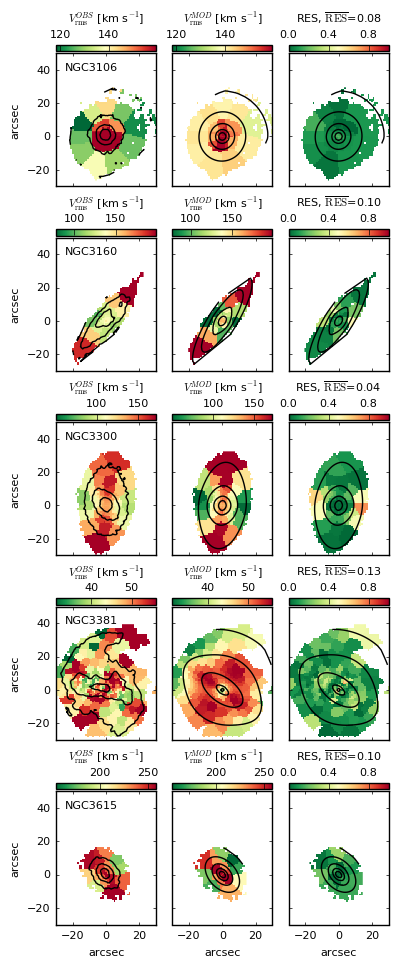}
\includegraphics[width=0.27\textwidth]{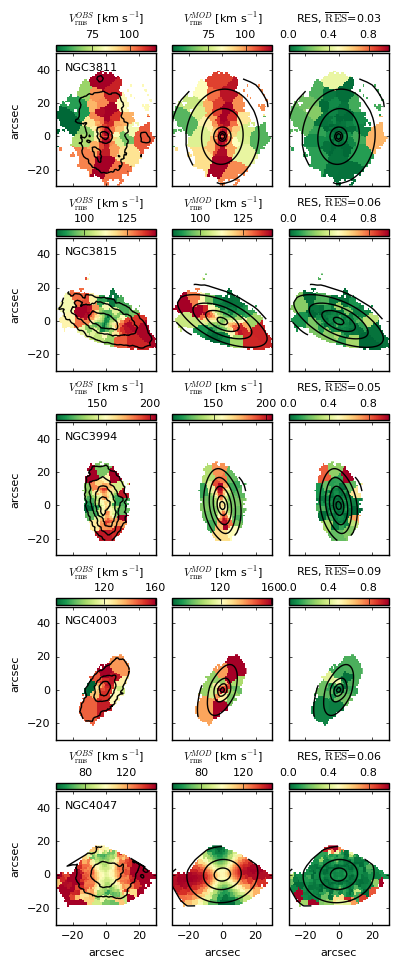}
\includegraphics[width=0.27\textwidth]{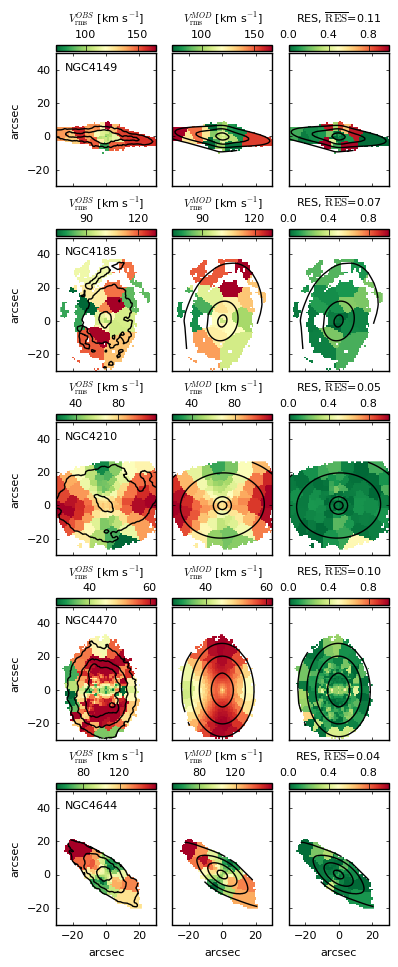}
}
\caption{Second moment maps $\Vrms = \sqrt{V^2+\sigma^2}$ of the data and the model, and residual maps of our sample of 238 (E1--Sdm)
  galaxies.}
\label{fig:maps3}
\end{figure*}

\begin{figure*}
\centering
{\includegraphics[width=0.27\textwidth]{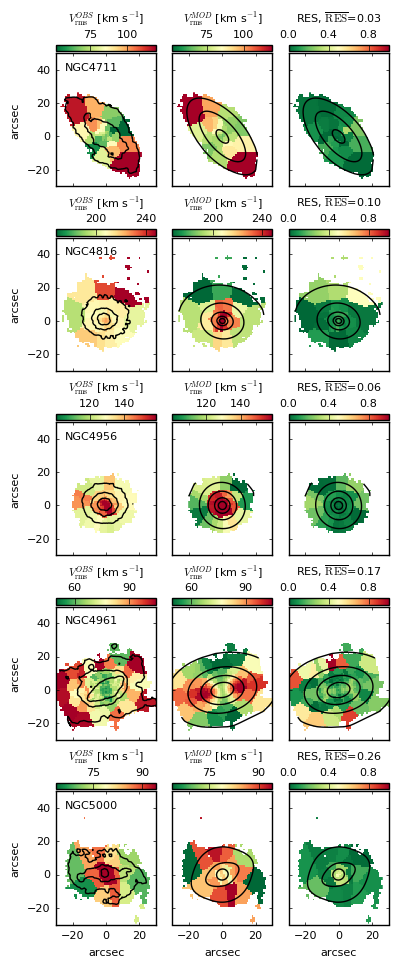}
\includegraphics[width=0.27\textwidth]{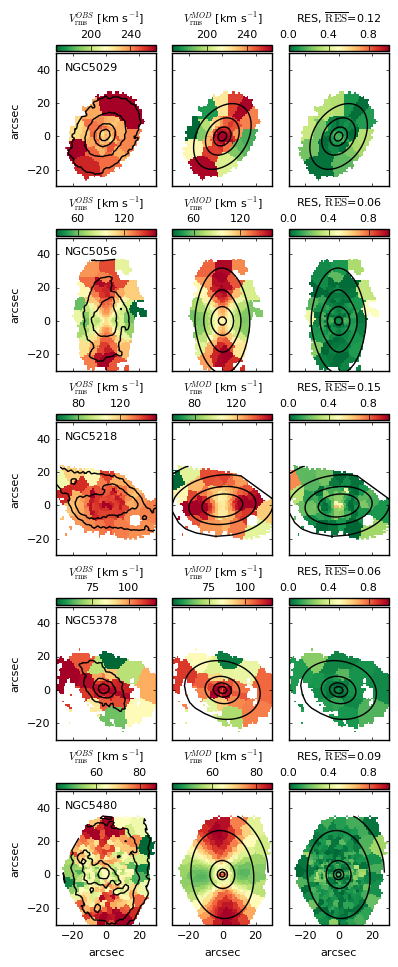}
\includegraphics[width=0.27\textwidth]{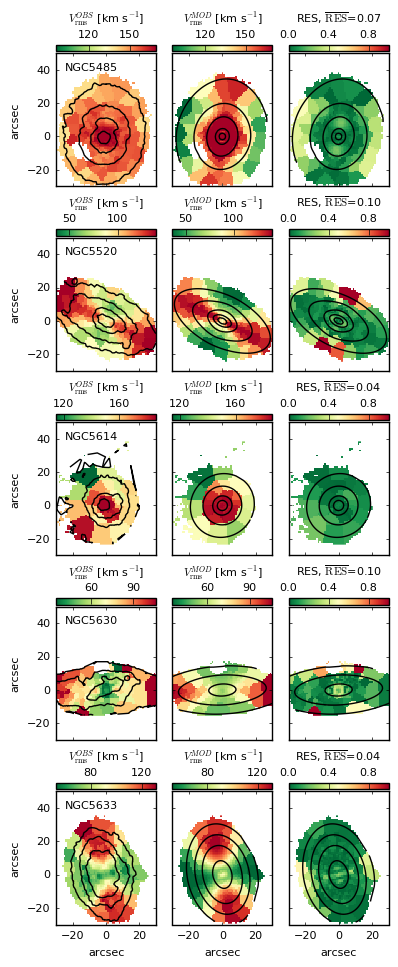}
\includegraphics[width=0.27\textwidth]{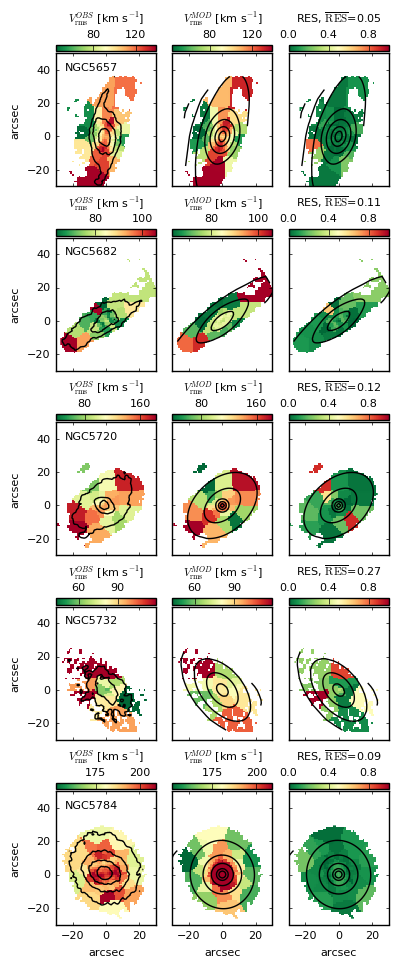}
\includegraphics[width=0.27\textwidth]{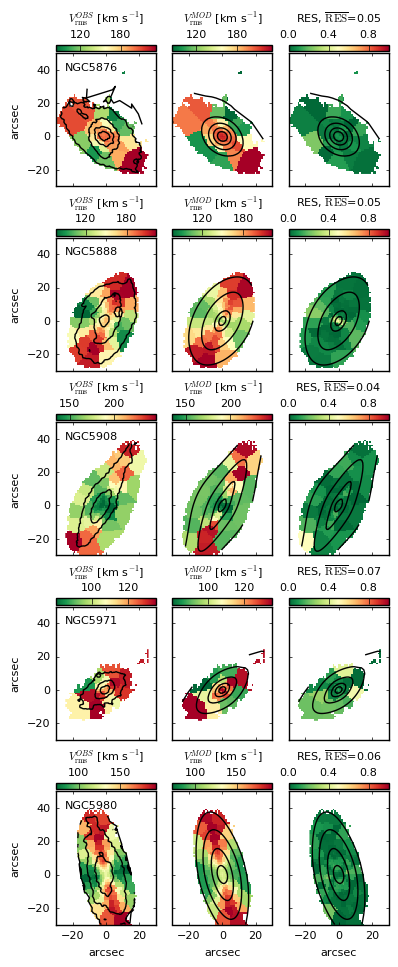}
\includegraphics[width=0.27\textwidth]{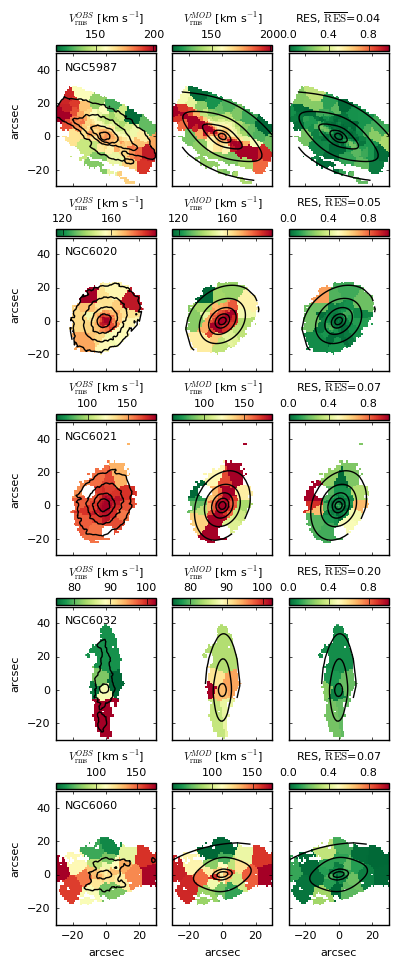}
}
\caption{Second moment maps $\Vrms = \sqrt{V^2+\sigma^2}$ of the data and the model, and residual maps of our sample of 238 (E1--Sdm)
  galaxies.}
\label{fig:maps4}
\end{figure*}

\begin{figure*}
\centering
{\includegraphics[width=0.27\textwidth]{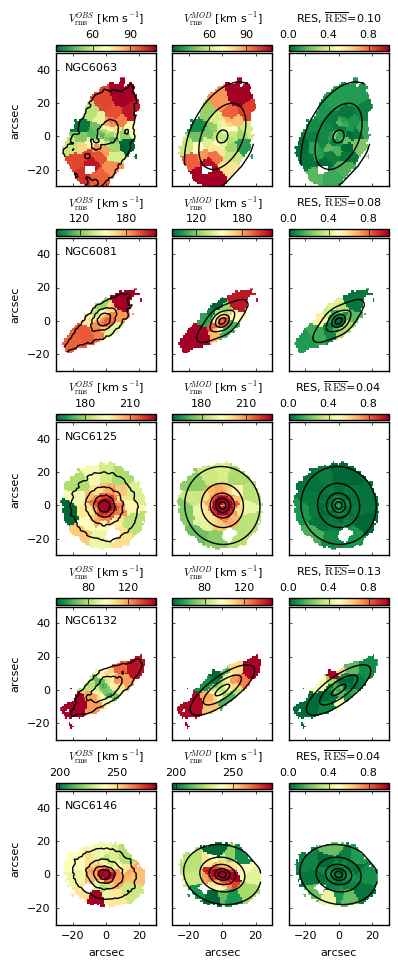}
\includegraphics[width=0.27\textwidth]{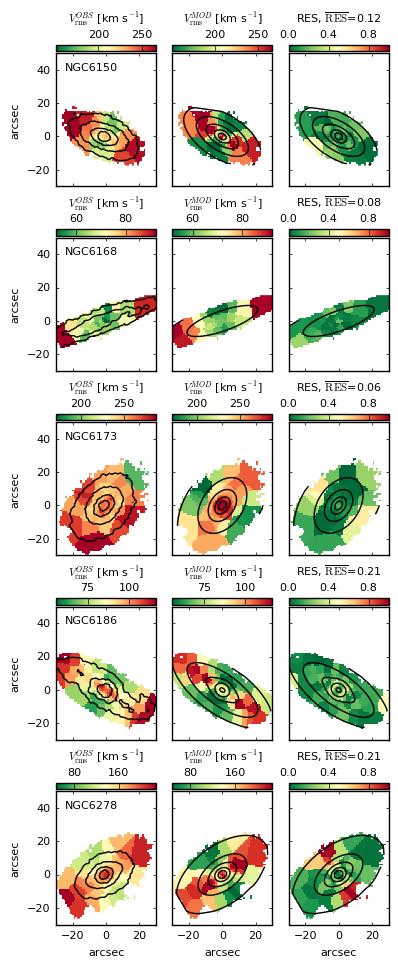}
\includegraphics[width=0.27\textwidth]{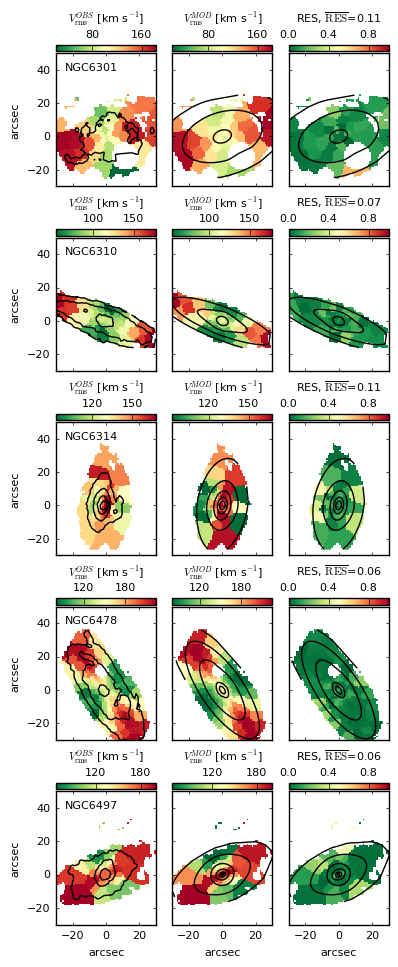}
\includegraphics[width=0.27\textwidth]{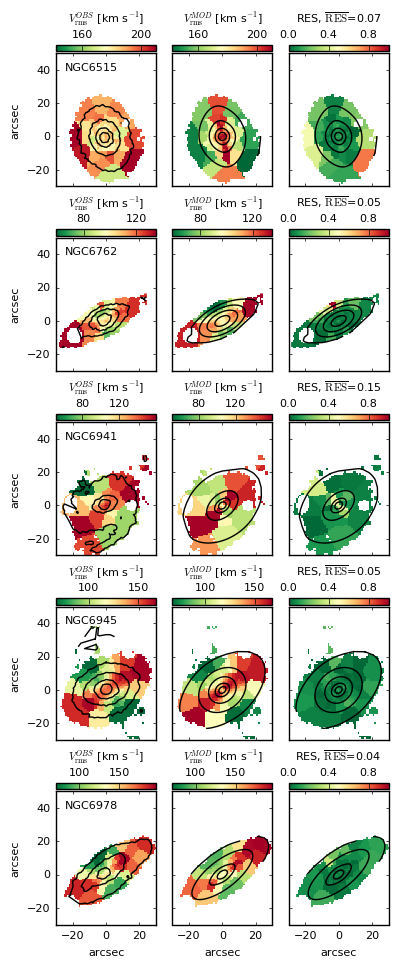}
\includegraphics[width=0.27\textwidth]{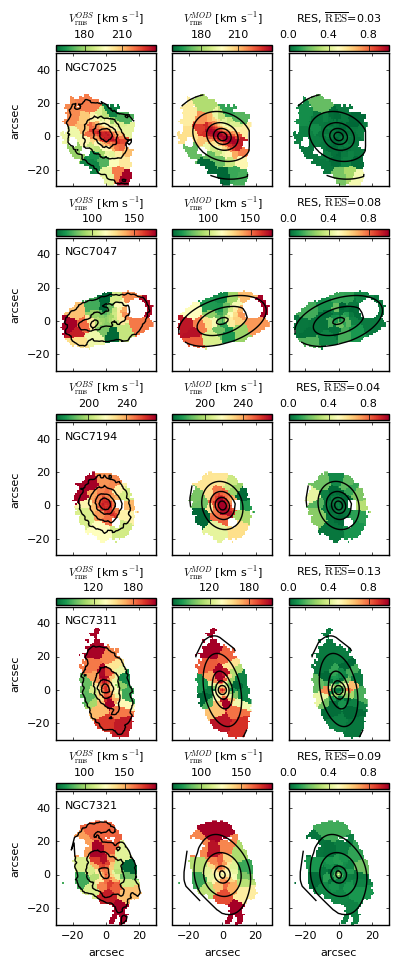}
\includegraphics[width=0.27\textwidth]{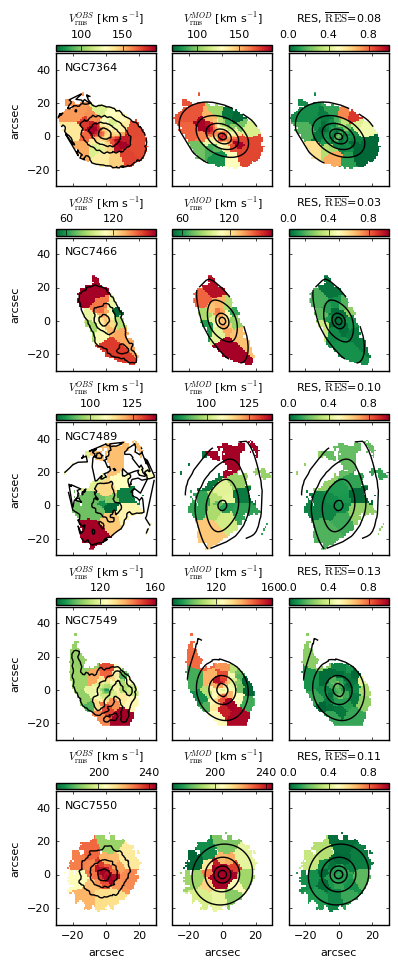}
}
\caption{Second moment maps $\Vrms = \sqrt{V^2+\sigma^2}$ of the data and the model, and residual maps of our sample of 238 (E1--Sdm)
  galaxies.}
\label{fig:maps5}
\end{figure*}

\begin{figure*}
\centering
{\includegraphics[width=0.27\textwidth]{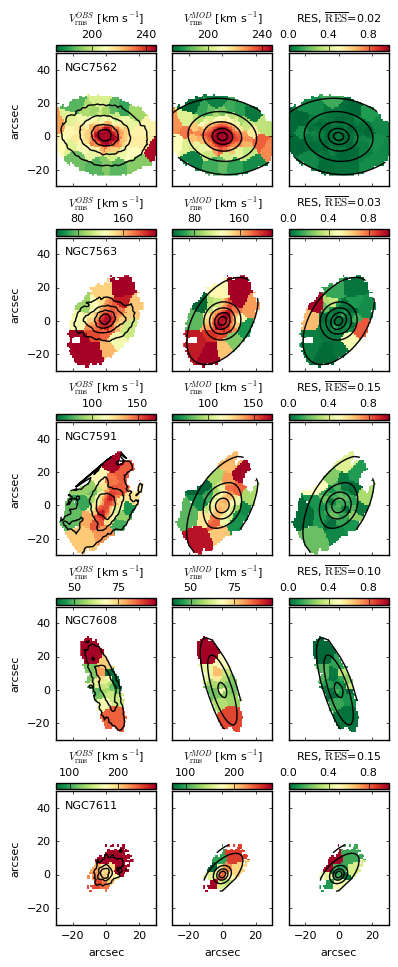}
\includegraphics[width=0.27\textwidth]{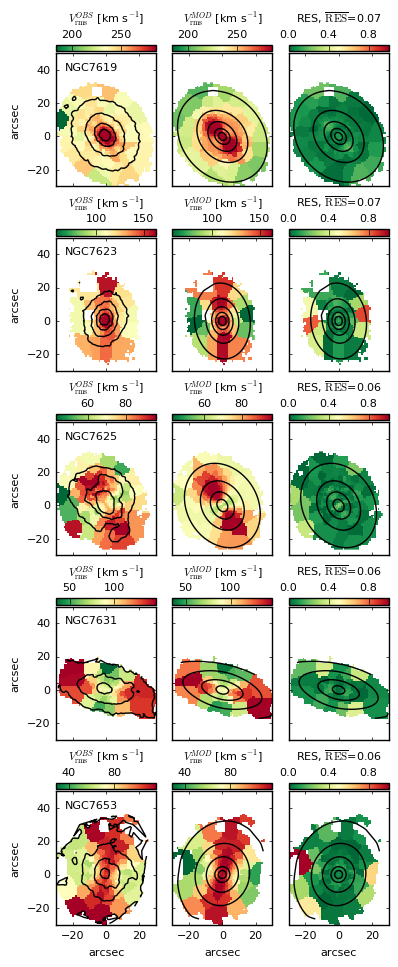}
\includegraphics[width=0.27\textwidth]{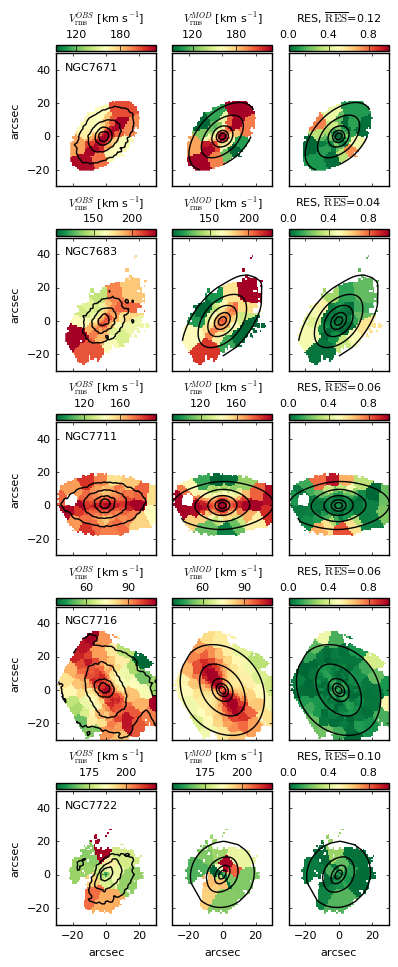}
\includegraphics[width=0.27\textwidth]{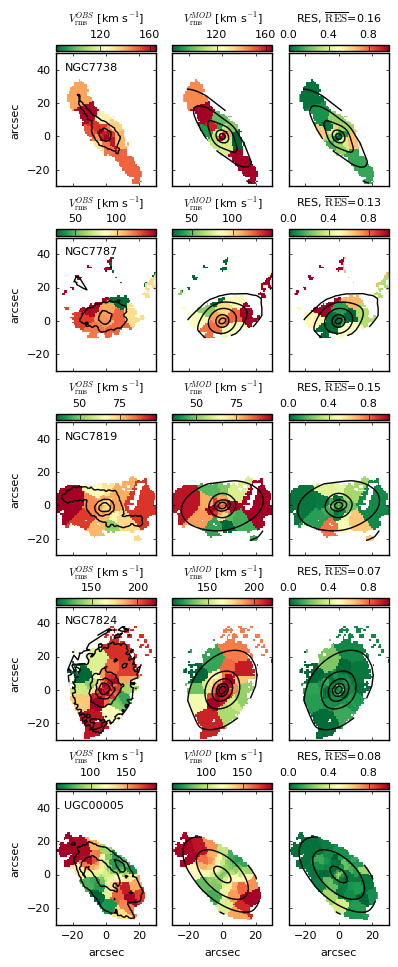}
\includegraphics[width=0.27\textwidth]{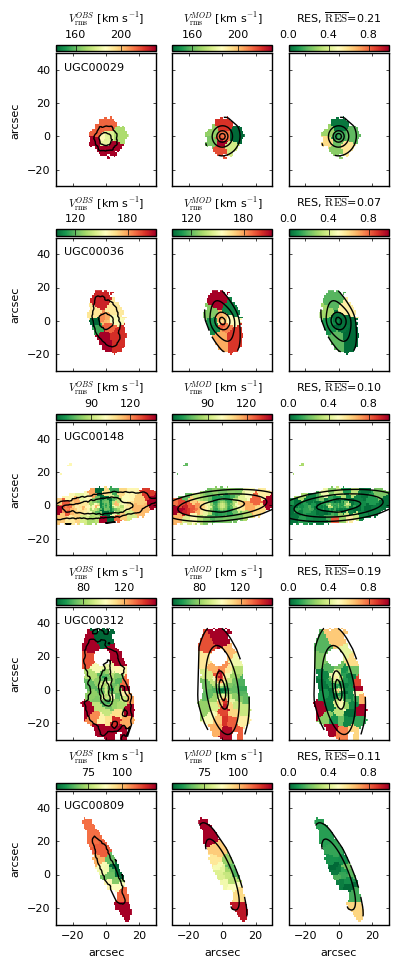}
\includegraphics[width=0.27\textwidth]{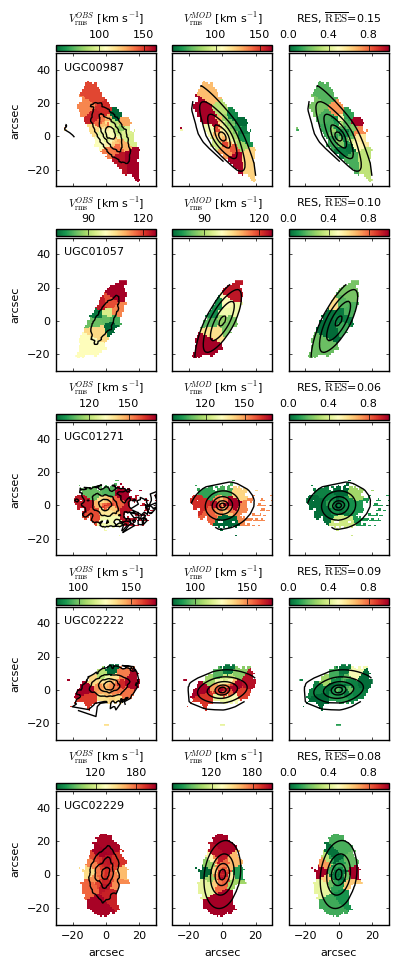}
}
\caption{Second moment maps $\Vrms = \sqrt{V^2+\sigma^2}$ of the data and the model, and residual maps of our sample of 238 (E1--Sdm)
  galaxies.}
\label{fig:maps6}
\end{figure*}

\begin{figure*}
\centering
{\includegraphics[width=0.27\textwidth]{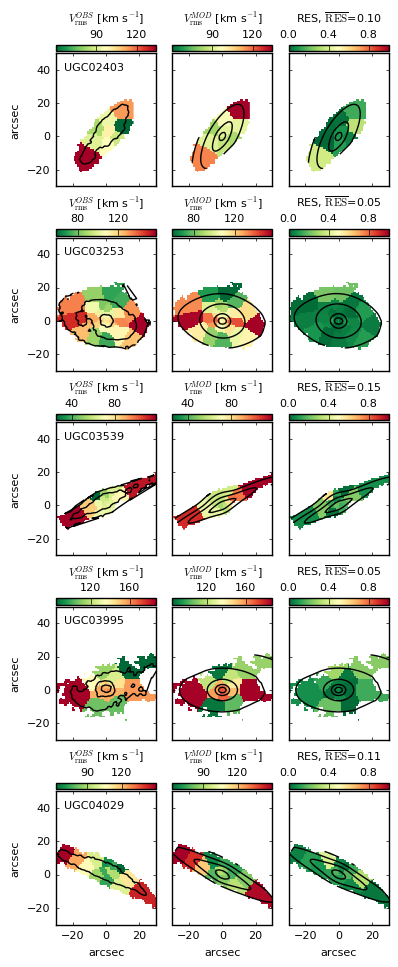}
\includegraphics[width=0.27\textwidth]{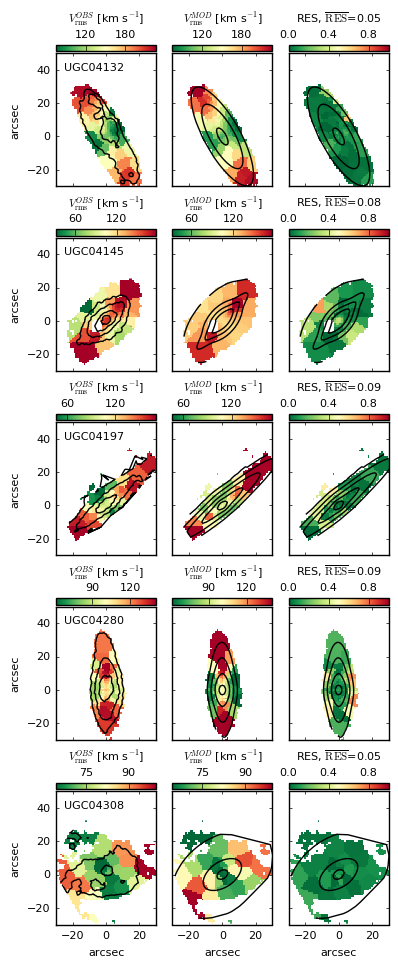}
\includegraphics[width=0.27\textwidth]{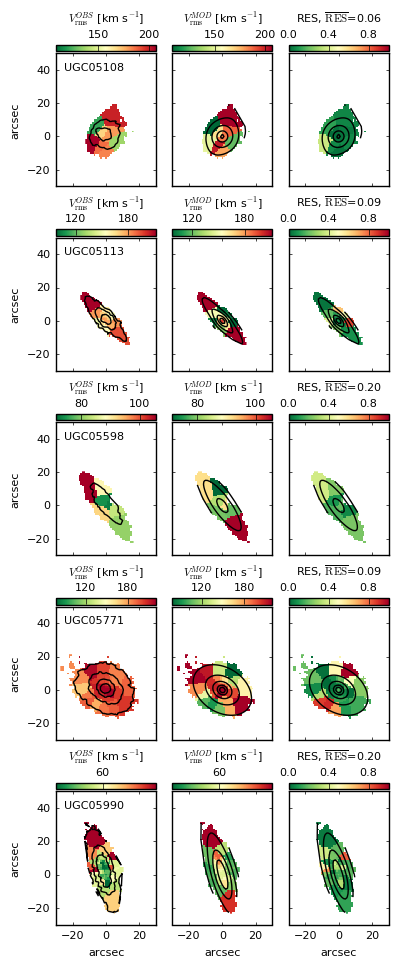}
\includegraphics[width=0.27\textwidth]{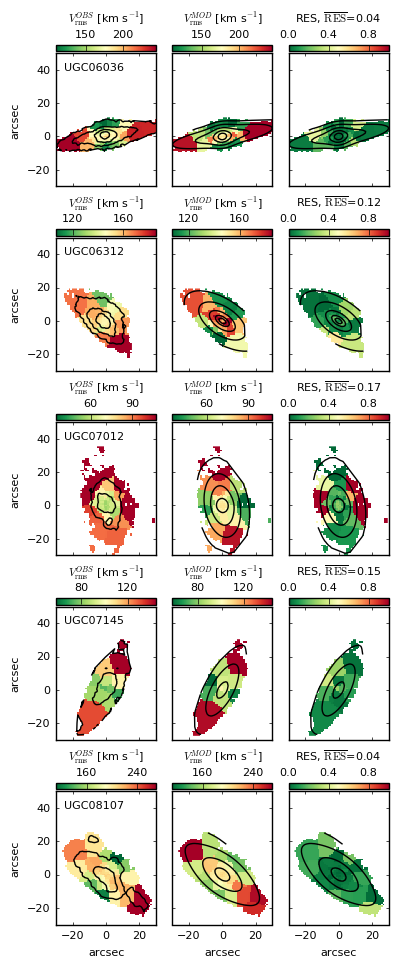}
\includegraphics[width=0.27\textwidth]{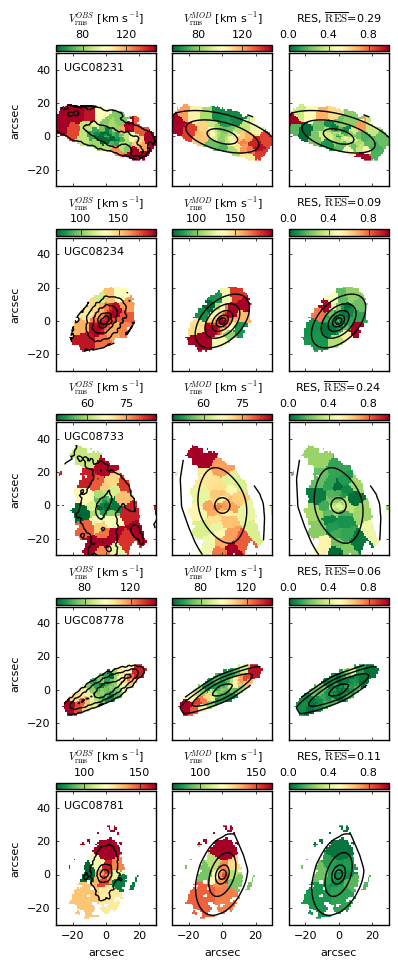}
\includegraphics[width=0.27\textwidth]{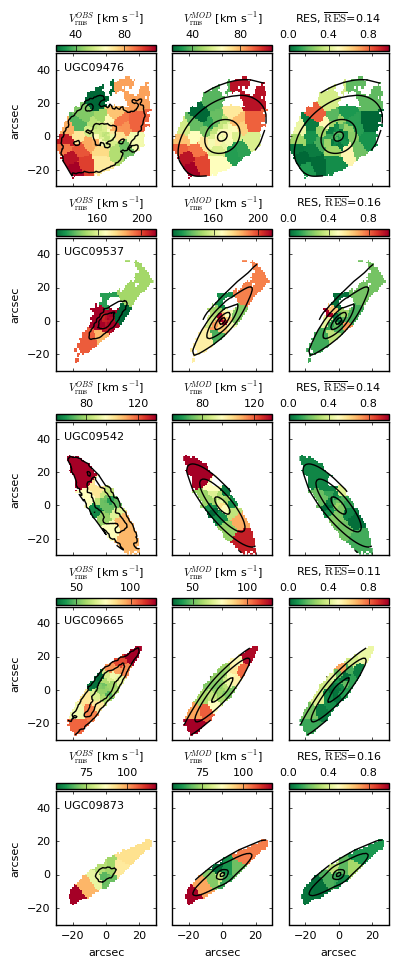}
}
\caption{Second moment maps $\Vrms = \sqrt{V^2+\sigma^2}$ of the data and the model, and residual maps of our sample of 238 (E1--Sdm)
  galaxies.}
\label{fig:maps7}
\end{figure*}

\begin{figure*}
\centering
{\includegraphics[width=0.27\textwidth]{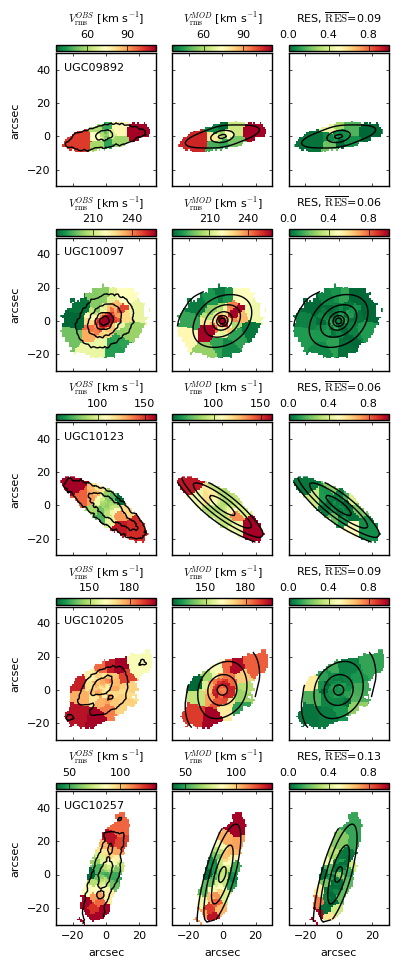}
\includegraphics[width=0.27\textwidth]{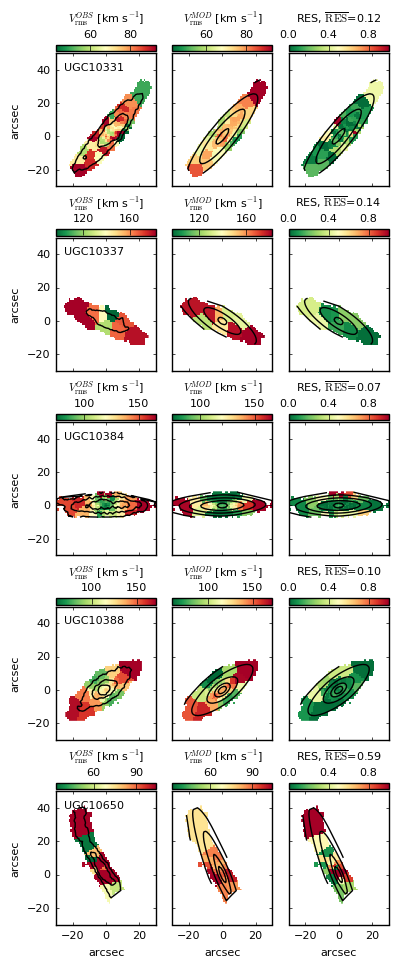}
\includegraphics[width=0.27\textwidth]{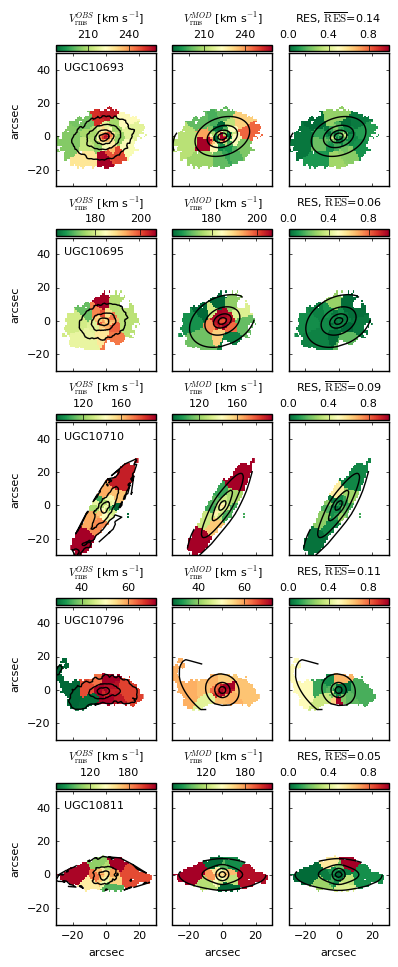}
\includegraphics[width=0.27\textwidth]{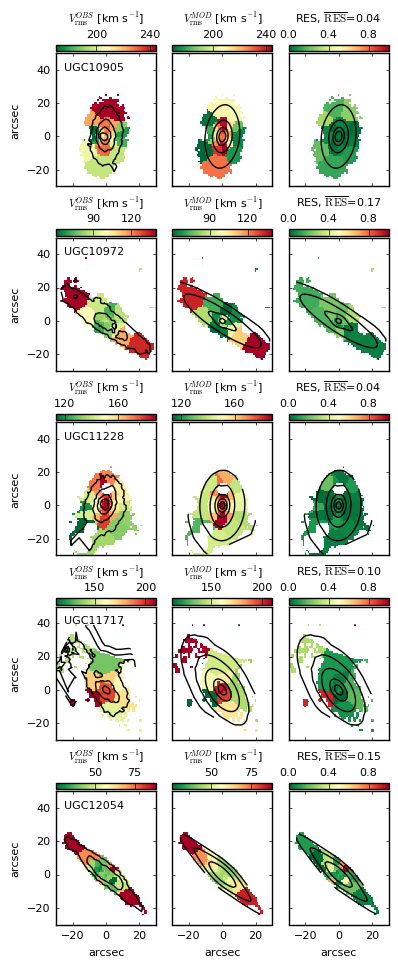}
\includegraphics[width=0.27\textwidth]{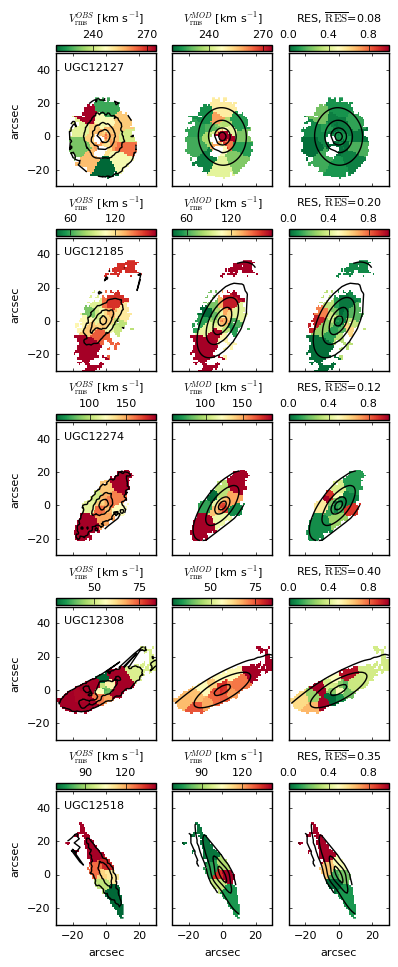}
\includegraphics[width=0.27\textwidth]{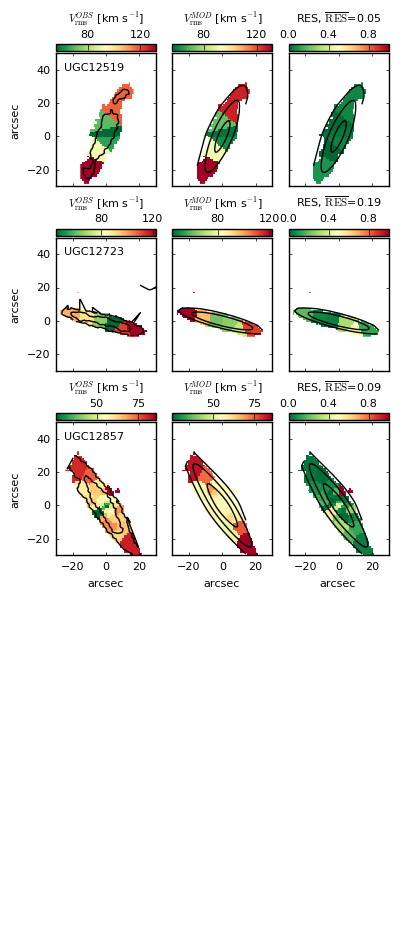}
}
\caption{Second moment maps $\Vrms = \sqrt{V^2+\sigma^2}$ of the data and the model, and residual maps of our sample of 238 (E1--Sdm)
  galaxies.}
\label{fig:maps8}
\end{figure*}

\clearpage
\newpage
\section{Basic properties of the 238 CALIFA galaxies}
\label{A:tables}

In Tables \ref{tab:DS1}--\ref{tab:DS6}, 
we the present basic properties of the 238 (E1--Sdm) CALIFA galaxies from the literature and our own measurements. 
\begin{table*}
\begin{minipage}{180mm}
\caption{Properties of the 238 (E1--Sdm) CALIFA galaxies. 
Columns list (1) galaxy identifier. (2) Hubble type based on by-eye morphological classification from \citet{Walcher2014}. 
(3) Hubble Flow (Galactocentric) distances from NED (using Hubble constant of $H_0$=73 km sec$^{-1}$ Mpc$^{-1}$),  calculated from the weighted mean radial velocities of the radio and optical redshifts, corrected to the Galactic standard of rest (GSR) from the RC3 (Third Reference Catalogue of Bright Galaxies) as described in \citet[][section 3.10c, page 54]{deVaucouleurs1991}. 
(4) and (5) Average photometric position angle and ellipticity from $r$-band SDSS (DR12) images 
 with typical uncertainty of 5$^{\circ}$-10$^{\circ}$ and 5 per cent, respectively,
 using $\texttt{findgalaxy PYTHON}$ procedure of \citealt{Cappellari2002}. 
(6) From the axial ratio of the galaxies ($q=1-\epsilon$, where $\epsilon$ is the average ellipticity of the galaxy,
 determined from $\texttt{findgalaxy}$), we calculated the inclination ($\emph{i}$) assuming $q_o=0.2$ for the intrinsic axial ratio of the galaxies (Section \ref{SS:dyn}).
(7) Effective (half-light) radius measured via growth curve analysis (\citealt{Walcher2014}). 
(8) Radial extent of the stellar kinematic data in arcseconds (\citealt{Falcon-Barroso2017}).
(9) Radial extent of the stellar kinematic data in terms of effective radius.
 (10) Systemic velocity of the galaxies with a typical uncertainty of 5 km~s$^{-1}$, measured as the median value of the 
 velocity field  (Section \ref{SS:sauronifs}).
 (11) and (12) The medians of the posterior distributions of the dynamical mass-to-light ratios and azimuthal velocity anisotropies, respectively,
 from JAM-MCMC model. The uncertainties are estimated as the 25th and 75th percentiles of the data corresponding to the median absolute deviation
 (Section \ref{SS:dyn}).
 (13) CVC (circular velocity curve) class of the 238 (E1--Sdm) galaxies, where 0 -- Slow rising (SR); 1 -- Flat (FL); 
 2 -- Round-peaked (RP); 3 -- Sharp-peaked (SP).} 
\begin{center}
\begin{tabular}{|c|cccccccccccc|}
%
\hline
Galaxy & Type & Dist & PA & $\epsilon$ & $i$ & $R_{\mathrm{e}}$ & $R_{\mathrm{max}}$ & $R_{\mathrm{max}}/R_{\mathrm{e}}$ & $V_{\mathrm{sys}}$ & 
 $(M/L)_{\mathrm{dyn}}$ & $\beta_{\mathrm{z}}$ &  CVC class \\
 &  & $\mathrm{Mpc}$ & ($^{\circ}$) &  & ($^{\circ}$) & (arcsec) & (arcsec) &  & (km s$^{-1}$) &  &  &  \\
(1)    &  (2)  &  (3)  &      (4) &    (5) &    (6) &  (7)   & (8)    &   (9)  & (10) & (11) & (12) & (13) \\  
\hline
IC0480 & Sc & 62 & 167 & 0.80 & 84 & 24 & 38 & 1.6 & 4597 & $11.661_{-0.073}^{+0.069}$ & $0.362_{-0.012}^{+0.009}$ & 0 \\
IC0540 & Sab & 26 & 172 & 0.78 & 81 & 14 & 27 & 1.9 & 2095 & $8.582_{-0.047}^{+0.054}$ & $0.390_{-0.011}^{+0.009}$ & 0 \\
IC0674 & Sab & 103 & 119 & 0.60 & 69 & 9 & 36 & 4.0 & 7459 & $6.439_{-0.022}^{+0.024}$ & $-0.451_{-0.020}^{+0.021}$ & 2 \\
IC0944 & Sab & 96 & 105 & 0.66 & 75 & 19 & 37 & 1.9 & 6974 & $8.054_{-0.021}^{+0.025}$ & $0.440_{-0.005}^{+0.004}$ & 2 \\
IC1079 & E4 & 119 & 83 & 0.50 & 71 & 37 & 26 & 0.7 & 8631 & $6.306_{-0.023}^{+0.025}$ & $0.320_{-0.005}^{+0.004}$ & 1 \\
IC1151 & Scd & 31 & 29 & 0.57 & 68 & 22 & 37 & 1.7 & 2158 & $5.420_{-0.032}^{+0.034}$ & $0.211_{-0.029}^{+0.018}$ & 0 \\
IC1256 & Sb & 67 & 99 & 0.37 & 51 & 17 & 25 & 1.5 & 4691 & $5.852_{-0.038}^{+0.040}$ & $0.598_{-0.010}^{+0.006}$ & 0 \\
IC1528 & Sbc & 53 & 74 & 0.57 & 68 & 23 & 41 & 1.8 & 3741 & $3.582_{-0.014}^{+0.021}$ & $0.402_{-0.015}^{+0.010}$ & 0 \\
IC1652 & S0a & 73 & 171 & 0.78 & 78 & 11 & 26 & 2.4 & 5040 & $2.959_{-0.012}^{+0.013}$ & $0.456_{-0.017}^{+0.009}$ & 1 \\
IC1755 & Sb & 110 & 155 & 0.75 & 78 & 11 & 31 & 2.8 & 7889 & $7.108_{-0.021}^{+0.021}$ & $0.281_{-0.018}^{+0.010}$ & 2 \\
IC2101 & Scd & 61 & 145 & 0.76 & 80 & 25 & 44 & 1.8 & 4445 & $5.773_{-0.039}^{+0.044}$ & $0.601_{-0.008}^{+0.005}$ & 0 \\
IC2247 & Sab & 58 & 145 & 0.77 & 81 & 21 & 41 & 2.0 & 4230 & $7.787_{-0.035}^{+0.035}$ & $0.480_{-0.016}^{+0.011}$ & 1 \\
IC2487 & Sc & 58 & 164 & 0.74 & 76 & 25 & 40 & 1.6 & 4284 & $7.423_{-0.041}^{+0.026}$ & $0.610_{-0.011}^{+0.007}$ & 0 \\
IC4566 & Sb & 81 & 151 & 0.38 & 53 & 15 & 27 & 1.8 & 5523 & $5.556_{-0.016}^{+0.030}$ & $-0.681_{-0.030}^{+0.035}$ & 2 \\
IC5309 & Sc & 60 & 27 & 0.54 & 64 & 17 & 24 & 1.4 & 4153 & $4.163_{-0.037}^{+0.046}$ & $0.420_{-0.019}^{+0.013}$ & 0 \\
IC5376 & Sb & 72 & 3 & 0.73 & 80 & 16 & 36 & 2.2 & 4975 & $7.894_{-0.035}^{+0.041}$ & $0.516_{-0.008}^{+0.006}$ & 2 \\
MCG-01-54-016 & Scd & 42 & 31 & 0.74 & 80 & 24 & 38 & 1.6 & 2915 & $10.861_{-0.186}^{+0.172}$ & $0.818_{-0.015}^{+0.011}$ & 0 \\
MCG-02-02-030 & Sb & 49 & 170 & 0.57 & 69 & 19 & 38 & 2.0 & 3482 & $3.686_{-0.013}^{+0.015}$ & $0.478_{-0.010}^{+0.006}$ & 0 \\
MCG-02-02-040 & Scd & 50 & 51 & 0.63 & 81 & 20 & 34 & 1.7 & 3458 & $7.080_{-0.050}^{+0.064}$ & $0.569_{-0.007}^{+0.005}$ & 0 \\
MCG-02-03-015 & Sab & 80 & 22 & 0.73 & 76 & 12 & 37 & 3.1 & 5773 & $4.886_{-0.029}^{+0.021}$ & $0.450_{-0.015}^{+0.008}$ & 1 \\
MCG-02-51-004 & Sb & 79 & 160 & 0.63 & 69 & 17 & 34 & 2.0 & 5587 & $6.556_{-0.020}^{+0.018}$ & $0.259_{-0.013}^{+0.010}$ & 0 \\
NGC0001 & Sbc & 65 & 105 & 0.31 & 47 & 12 & 30 & 2.5 & 4498 & $3.129_{-0.014}^{+0.020}$ & $0.163_{-0.012}^{+0.012}$ & 0 \\
NGC0023 & Sb & 65 & 157 & 0.33 & 63 & 17 & 26 & 1.5 & 4570 & $1.801_{-0.018}^{+0.034}$ & $0.519_{-0.023}^{+0.011}$ & 1 \\
NGC0155 & E1 & 86 & 170 & 0.18 & 48 & 15 & 25 & 1.7 & 6182 & $4.534_{-0.016}^{+0.018}$ & $0.342_{-0.010}^{+0.006}$ & 2 \\
NGC0160 & Sa & 74 & 51 & 0.47 & 59 & 22 & 35 & 1.6 & 5183 & $5.868_{-0.016}^{+0.034}$ & $0.494_{-0.006}^{+0.004}$ & 3 \\
NGC0171 & Sb & 54 & 84 & 0.04 & 23 & 26 & 32 & 1.2 & 3879 & $1.683_{-0.012}^{+0.034}$ & $0.457_{-0.194}^{+0.035}$ & 1 \\
NGC0177 & Sab & 53 & 190 & 0.75 & 77 & 13 & 37 & 2.8 & 3810 & $5.498_{-0.026}^{+0.023}$ & $0.079_{-0.013}^{+0.011}$ & 2 \\
NGC0192 & Sab & 58 & 166 & 0.57 & 72 & 22 & 39 & 1.8 & 4192 & $3.791_{-0.009}^{+0.011}$ & $0.507_{-0.011}^{+0.005}$ & 1 \\
NGC0214 & Sbc & 64 & 52 & 0.29 & 56 & 18 & 31 & 1.7 & 4501 & $2.608_{-0.013}^{+0.015}$ & $-0.030_{-0.019}^{+0.024}$ & 0 \\
NGC0216 & Sd & 21 & 208 & 0.65 & 73 & 20 & 35 & 1.8 & 1544 & $3.301_{-0.039}^{+0.039}$ & $0.559_{-0.017}^{+0.014}$ & 0 \\
NGC0217 & Sa & 55 & 113 & 0.76 & 79 & 23 & 41 & 1.8 & 3989 & $5.593_{-0.014}^{+0.016}$ & $0.349_{-0.006}^{+0.004}$ & 2 \\
NGC0237 & Sc & 58 & 178 & 0.38 & 52 & 15 & 35 & 2.3 & 4125 & $2.732_{-0.015}^{+0.017}$ & $0.595_{-0.024}^{+0.009}$ & 0 \\
NGC0257 & Sc & 74 & 89 & 0.38 & 52 & 21 & 40 & 1.9 & 5273 & $3.448_{-0.010}^{+0.012}$ & $0.056_{-0.030}^{+0.016}$ & 1 \\
NGC0429 & Sa & 78 & 14 & 0.72 & 78 & 6 & 29 & 4.8 & 5650 & $3.948_{-0.018}^{+0.029}$ & $0.432_{-0.008}^{+0.006}$ & 2 \\
NGC0444 & Scd & 68 & 159 & 0.76 & 78 & 23 & 32 & 1.4 & 4770 & $6.327_{-0.052}^{+0.065}$ & $0.694_{-0.017}^{+0.009}$ & 0 \\
NGC0499 & E5 & 62 & 71 & 0.34 & 54 & 21 & 31 & 1.5 & 4359 & $6.672_{-0.029}^{+0.100}$ & $0.727_{-0.013}^{+0.002}$ & 3 \\
NGC0504 & S0 & 60 & 48 & 0.66 & 76 & 8 & 30 & 3.8 & 4236 & $7.243_{-0.020}^{+0.020}$ & $0.122_{-0.007}^{+0.006}$ & 2 \\
NGC0517 & S0 & 60 & 17 & 0.52 & 66 & 10 & 34 & 3.4 & 4122 & $5.136_{-0.023}^{+0.042}$ & $0.712_{-0.006}^{+0.002}$ & 2 \\
NGC0528 & S0 & 68 & 58 & 0.49 & 61 & 12 & 28 & 2.3 & 4773 & $5.408_{-0.015}^{+0.017}$ & $0.517_{-0.012}^{+0.005}$ & 2 \\
NGC0529 & E4 & 68 & 179 & 0.18 & 35 & 12 & 37 & 3.1 & 4791 & $5.811_{-0.014}^{+0.022}$ & $0.063_{-0.014}^{+0.013}$ & 2 \\
NGC0551 & Sbc & 73 & 140 & 0.62 & 67 & 19 & 44 & 2.3 & 5124 & $4.714_{-0.021}^{+0.032}$ & $0.388_{-0.023}^{+0.007}$ & 0 \\
\hline
\end{tabular}
\label{tab:DS1}
\end{center}
\end{minipage}
\end{table*}

\begin{table*}
\begin{minipage}{180mm}
\caption{Properties of the 238 (E1--Sdm) CALIFA galaxies. 
Columns list (1) galaxy identifier. (2) Hubble type based on by-eye morphological classification from \citet{Walcher2014}. 
(3) Hubble Flow (Galactocentric) distances from NED (using Hubble constant of $H_0$=73 km sec$^{-1}$ Mpc$^{-1}$),  calculated from the weighted mean radial velocities of the radio and optical redshifts, corrected to the Galactic standard of rest (GSR) from the RC3 (Third Reference Catalogue of Bright Galaxies) as described in \citet[][section 3.10c, page 54]{deVaucouleurs1991}. 
(4) and (5) Average photometric position angle and ellipticity from $r$-band SDSS (DR12) images 
 with typical uncertainty of 5$^{\circ}$-10$^{\circ}$ and 5 per cent, respectively,
 using $\texttt{findgalaxy PYTHON}$ procedure of \citealt{Cappellari2002}. 
(6) From the axial ratio of the galaxies ($q=1-\epsilon$, where $\epsilon$ is the average ellipticity of the galaxy,
 determined from $\texttt{findgalaxy}$), we calculated the inclination ($\emph{i}$) assuming $q_o=0.2$ for the intrinsic axial ratio of the galaxies (Section \ref{SS:dyn}).
(7) Effective (half-light) radius measured via growth curve analysis (\citealt{Walcher2014}). 
(8) Radial extent of the stellar kinematic data in arcseconds (\citealt{Falcon-Barroso2017}).
(9) Radial extent of the stellar kinematic data in terms of effective radius.
 (10) Systemic velocity of the galaxies with a typical uncertainty of 5 km~s$^{-1}$, measured as the median value of the 
 velocity field  (Section \ref{SS:sauronifs}).
 (11) and (12) The medians of the posterior distributions of the dynamical mass-to-light ratios and azimuthal velocity anisotropies, respectively,
 from JAM-MCMC model. The uncertainties are estimated as the 25th and 75th percentiles of the data corresponding to the median absolute deviation
 (Section \ref{SS:dyn}).
 (13) CVC (circular velocity curve) class of the 238 (E1--Sdm) galaxies, where 0 -- Slow rising (SR); 1 -- Flat (FL); 
 2 -- Round-peaked (RP); 3 -- Sharp-peaked (SP).} 
\begin{center}
\begin{tabular}{|c|cccccccccccc|}
%
\hline
Galaxy & Type & Dist & PA & $\epsilon$ & $i$ & $R_{\mathrm{e}}$ & $R_{\mathrm{max}}$ & $R_{\mathrm{max}}/R_{\mathrm{e}}$ & $V_{\mathrm{sys}}$ & 
 $(M/L)_{\mathrm{dyn}}$ & $\beta_{\mathrm{z}}$ &  CVC class \\
 &  & $\mathrm{Mpc}$ & ($^{\circ}$) &  & ($^{\circ}$) & (arcsec) & (arcsec) &  & (km s$^{-1}$) &  &  &  \\
(1)    &  (2)  &  (3)  &      (4) &    (5) &    (6) &  (7)   & (8)    &   (9)  & (10) & (11) & (12) & (13) \\ 
\hline
NGC0681 & Sa & 24 & 67 & 0.32 & 86 & 30 & 37 & 1.2 & 1743 & $6.745_{-0.041}^{+0.050}$ & $-0.010_{-0.011}^{+0.010}$ & 1 \\
NGC0741 & E1 & 77 & 86 & 0.21 & 38 & 35 & 32 & 0.9 & 5524 & $8.999_{-0.021}^{+0.024}$ & $0.265_{-0.009}^{+0.007}$ & 3 \\
NGC0755 & Scd & 23 & 49 & 0.67 & 76 & 28 & 39 & 1.4 & 1655 & $4.933_{-0.042}^{+0.044}$ & $0.447_{-0.011}^{+0.009}$ & 0 \\
NGC0768 & Sc & 97 & 35 & 0.55 & 67 & 15 & 34 & 2.3 & 6898 & $7.081_{-0.036}^{+0.041}$ & $0.093_{-0.014}^{+0.014}$ & 2 \\
NGC0774 & S0 & 64 & 169 & 0.18 & 49 & 12 & 26 & 2.2 & 4595 & $4.691_{-0.013}^{+0.027}$ & $0.554_{-0.008}^{+0.005}$ & 2 \\
NGC0776 & Sb & 69 & 67 & 0.05 & 29 & 19 & 32 & 1.7 & 4866 & $1.976_{-0.015}^{+0.020}$ & $-0.334_{-0.089}^{+0.084}$ & 1 \\
NGC0781 & Sa & 49 & 12 & 0.71 & 79 & 8 & 32 & 4.0 & 3482 & $2.738_{-0.010}^{+0.011}$ & $0.413_{-0.009}^{+0.005}$ & 0 \\
NGC0810 & E5 & 106 & 28 & 0.32 & 61 & 17 & 20 & 1.2 & 7716 & $12.168_{-0.058}^{+0.047}$ & $0.275_{-0.008}^{+0.007}$ & 2 \\
NGC0932 & S0a & 57 & 71 & 0.10 & 29 & 18 & 33 & 1.8 & 4066 & $3.314_{-0.010}^{+0.014}$ & $0.036_{-0.034}^{+0.026}$ & 1 \\
NGC1056 & Sa & 22 & 154 & 0.32 & 70 & 14 & 37 & 2.6 & 1556 & $3.115_{-0.018}^{+0.029}$ & $0.252_{-0.016}^{+0.008}$ & 0 \\
NGC1060 & E3 & 72 & 81 & 0.17 & 42 & 27 & 26 & 1.0 & 5132 & $5.797_{-0.024}^{+0.043}$ & $0.788_{-0.008}^{+0.002}$ & 3 \\
NGC1167 & S0 & 69 & 73 & 0.18 & 38 & 24 & 30 & 1.2 & 4885 & $5.198_{-0.015}^{+0.016}$ & $0.591_{-0.008}^{+0.003}$ & 3 \\
NGC1349 & E6 & 90 & 43 & 0.14 & 32 & 17 & 21 & 1.2 & 6525 & $5.395_{-0.034}^{+0.031}$ & $0.440_{-0.032}^{+0.012}$ & 3 \\
NGC1542 & Sab & 50 & 130 & 0.54 & 70 & 15 & 23 & 1.5 & 3687 & $5.259_{-0.033}^{+0.046}$ & $0.776_{-0.007}^{+0.004}$ & 0 \\
NGC1645 & S0a & 66 & 85 & 0.59 & 66 & 13 & 39 & 3.0 & 4749 & $6.579_{-0.015}^{+0.017}$ & $-0.050_{-0.010}^{+0.007}$ & 2 \\
NGC1677 & Scd & 36 & 133 & 0.73 & 77 & 12 & 29 & 2.4 & 2725 & $4.317_{-0.053}^{+0.060}$ & $0.465_{-0.025}^{+0.015}$ & 0 \\
NGC2253 & Sbc & 50 & 5 & 0.25 & 41 & 15 & 36 & 2.4 & 3527 & $9.701_{-0.043}^{+0.067}$ & $0.815_{-0.001}^{+0.001}$ & 2 \\
NGC2347 & Sbc & 62 & 5 & 0.38 & 56 & 18 & 42 & 2.3 & 4408 & $4.306_{-0.032}^{+0.021}$ & $0.599_{-0.024}^{+0.005}$ & 2 \\
NGC2410 & Sb & 64 & 34 & 0.71 & 76 & 21 & 37 & 1.8 & 4638 & $5.487_{-0.016}^{+0.018}$ & $0.205_{-0.012}^{+0.010}$ & 2 \\
NGC2449 & Sab & 66 & 134 & 0.56 & 66 & 16 & 33 & 2.1 & 4842 & $4.078_{-0.011}^{+0.014}$ & $-0.095_{-0.018}^{+0.013}$ & 0 \\
NGC2476 & E6 & 51 & 140 & 0.28 & 51 & 9 & 22 & 2.4 & 3628 & $4.097_{-0.023}^{+0.026}$ & $0.398_{-0.021}^{+0.009}$ & 2 \\
NGC2481 & S0 & 29 & 20 & 0.50 & 74 & 9 & 34 & 3.8 & 2147 & $4.241_{-0.013}^{+0.016}$ & $0.447_{-0.035}^{+0.004}$ & 2 \\
NGC2486 & Sab & 63 & 93 & 0.42 & 56 & 15 & 29 & 1.9 & 4602 & $5.837_{-0.032}^{+0.034}$ & $-0.019_{-0.031}^{+0.028}$ & 1 \\
NGC2553 & Sb & 64 & 65 & 0.46 & 61 & 9 & 19 & 2.1 & 4622 & $7.403_{-0.035}^{+0.035}$ & $-0.264_{-0.016}^{+0.017}$ & 2 \\
NGC2554 & S0a & 56 & 146 & 0.35 & 49 & 19 & 38 & 2.0 & 4064 & $5.097_{-0.010}^{+0.013}$ & $-0.029_{-0.008}^{+0.006}$ & 3 \\
NGC2592 & E4 & 26 & 49 & 0.20 & 51 & 9 & 28 & 3.1 & 1937 & $6.868_{-0.013}^{+0.013}$ & $-0.029_{-0.007}^{+0.006}$ & 2 \\
NGC2604 & Sd & 28 & 179 & 0.08 & 22 & 26 & 36 & 1.4 & 2048 & $4.619_{-0.044}^{+0.047}$ & $0.552_{-0.034}^{+0.024}$ & 0 \\
NGC2639 & Sa & 46 & 128 & 0.38 & 61 & 17 & 38 & 2.2 & 3149 & $4.334_{-0.023}^{+0.011}$ & $0.579_{-0.020}^{+0.003}$ & 2 \\
NGC2730 & Scd & 51 & 77 & 0.23 & 49 & 24 & 39 & 1.6 & 3762 & $4.926_{-0.031}^{+0.045}$ & $0.377_{-0.021}^{+0.015}$ & 0 \\
NGC2880 & E7 & 22 & 140 & 0.36 & 54 & 18 & 36 & 2.0 & 1553 & $4.365_{-0.008}^{+0.010}$ & $0.032_{-0.008}^{+0.006}$ & 1 \\
NGC2906 & Sbc & 28 & 81 & 0.45 & 58 & 19 & 33 & 1.7 & 2110 & $5.364_{-0.018}^{+0.021}$ & $0.390_{-0.014}^{+0.010}$ & 1 \\
NGC2916 & Sbc & 50 & 18 & 0.36 & 56 & 26 & 40 & 1.5 & 3678 & $4.480_{-0.009}^{+0.010}$ & $-0.403_{-0.019}^{+0.022}$ & 1 \\
NGC2918 & E6 & 93 & 74 & 0.34 & 53 & 12 & 28 & 2.3 & 6653 & $6.008_{-0.015}^{+0.018}$ & $0.424_{-0.006}^{+0.003}$ & 2 \\
NGC3057 & Sdm & 23 & 191 & 0.38 & 81 & 32 & 34 & 1.1 & 1508 & $6.614_{-0.101}^{+0.107}$ & $-0.307_{-0.046}^{+0.036}$ & 0 \\
NGC3106 & Sab & 84 & 169 & 0.01 & 25 & 21 & 32 & 1.5 & 6111 & $4.023_{-0.014}^{+0.014}$ & $0.017_{-0.034}^{+0.033}$ & 1 \\
NGC3160 & Sab & 95 & 148 & 0.68 & 75 & 15 & 36 & 2.4 & 6777 & $8.517_{-0.046}^{+0.039}$ & $0.488_{-0.010}^{+0.007}$ & 2 \\
NGC3300 & S0a & 41 & 174 & 0.45 & 59 & 13 & 32 & 2.5 & 2989 & $4.728_{-0.013}^{+0.021}$ & $0.027_{-0.014}^{+0.010}$ & 2 \\
NGC3381 & Sd & 22 & 54 & 0.13 & 63 & 24 & 42 & 1.8 & 1594 & $2.305_{-0.027}^{+0.034}$ & $0.350_{-0.012}^{+0.010}$ & 0 \\
NGC3615 & E5 & 91 & 42 & 0.40 & 53 & 15 & 18 & 1.2 & 6581 & $5.669_{-0.028}^{+0.039}$ & $0.572_{-0.013}^{+0.006}$ & 3 \\
NGC3811 & Sbc & 43 & 174 & 0.29 & 48 & 21 & 39 & 1.9 & 3064 & $3.817_{-0.012}^{+0.014}$ & $0.120_{-0.013}^{+0.013}$ & 1 \\
\hline
\end{tabular}
\label{tab:DS2}
\end{center}
\end{minipage}
\end{table*}

\begin{table*}
\begin{minipage}{180mm}
\caption{Properties of the 238 (E1--Sdm) CALIFA galaxies. 
Columns list (1) galaxy identifier. (2) Hubble type based on by-eye morphological classification from \citet{Walcher2014}. 
(3) Hubble Flow (Galactocentric) distances from NED (using Hubble constant of $H_0$=73 km sec$^{-1}$ Mpc$^{-1}$),  calculated from the weighted mean radial velocities of the radio and optical redshifts, corrected to the Galactic standard of rest (GSR) from the RC3 (Third Reference Catalogue of Bright Galaxies) as described in \citet[][section 3.10c, page 54]{deVaucouleurs1991}. 
(4) and (5) Average photometric position angle and ellipticity from $r$-band SDSS (DR12) images 
 with typical uncertainty of 5$^{\circ}$-10$^{\circ}$ and 5 per cent, respectively,
 using $\texttt{findgalaxy PYTHON}$ procedure of \citealt{Cappellari2002}. 
(6) From the axial ratio of the galaxies ($q=1-\epsilon$, where $\epsilon$ is the average ellipticity of the galaxy,
 determined from $\texttt{findgalaxy}$), we calculated the inclination ($\emph{i}$) assuming $q_o=0.2$ for the intrinsic axial ratio of the galaxies (Section \ref{SS:dyn}).
(7) Effective (half-light) radius measured via growth curve analysis (\citealt{Walcher2014}). 
(8) Radial extent of the stellar kinematic data in arcseconds (\citealt{Falcon-Barroso2017}).
(9) Radial extent of the stellar kinematic data in terms of effective radius.
 (10) Systemic velocity of the galaxies with a typical uncertainty of 5 km~s$^{-1}$, measured as the median value of the 
 velocity field  (Section \ref{SS:sauronifs}).
 (11) and (12) The medians of the posterior distributions of the dynamical mass-to-light ratios and azimuthal velocity anisotropies, respectively,
 from JAM-MCMC model. The uncertainties are estimated as the 25th and 75th percentiles of the data corresponding to the median absolute deviation
 (Section \ref{SS:dyn}).
 (13) CVC (circular velocity curve) class of the 238 (E1--Sdm) galaxies, where 0 -- Slow rising (SR); 1 -- Flat (FL); 
 2 -- Round-peaked (RP); 3 -- Sharp-peaked (SP).} 
\begin{center}
\begin{tabular}{|c|cccccccccccc|}
%
\hline
Galaxy & Type & Dist & PA & $\epsilon$ & $i$ & $R_{\mathrm{e}}$ & $R_{\mathrm{max}}$ & $R_{\mathrm{max}}/R_{\mathrm{e}}$ & $V_{\mathrm{sys}}$ & 
 $(M/L)_{\mathrm{dyn}}$ & $\beta_{\mathrm{z}}$ &  CVC class \\
 &  & $\mathrm{Mpc}$ & ($^{\circ}$) &  & ($^{\circ}$) & (arcsec) & (arcsec) &  & (km s$^{-1}$) &  &  &  \\
(1)    &  (2)  &  (3)  &      (4) &    (5) &    (6) &  (7)   & (8)    &   (9)  & (10) & (11) & (12) & (13) \\  
\hline
NGC3815 & Sbc & 50 & 65 & 0.57 & 71 & 14 & 34 & 2.4 & 3655 & $5.177_{-0.019}^{+0.021}$ & $0.257_{-0.009}^{+0.007}$ & 0 \\
NGC3994 & Sbc & 42 & 8 & 0.44 & 68 & 9 & 26 & 2.9 & 3115 & $5.358_{-0.027}^{+0.034}$ & $0.441_{-0.010}^{+0.006}$ & 2 \\
NGC4003 & S0a & 89 & 159 & 0.36 & 64 & 14 & 22 & 1.6 & 6499 & $4.037_{-0.021}^{+0.025}$ & $0.557_{-0.031}^{+0.008}$ & 1 \\
NGC4047 & Sbc & 48 & 97 & 0.19 & 38 & 16 & 33 & 2.1 & 3381 & $3.801_{-0.017}^{+0.025}$ & $0.699_{-0.019}^{+0.006}$ & 2 \\
NGC4149 & Sa & 43 & 84 & 0.73 & 81 & 18 & 36 & 2.0 & 3030 & $6.325_{-0.029}^{+0.033}$ & $0.405_{-0.017}^{+0.010}$ & 0 \\
NGC4185 & Sbc & 53 & 164 & 0.34 & 52 & 30 & 38 & 1.3 & 3831 & $5.304_{-0.020}^{+0.033}$ & $0.398_{-0.016}^{+0.009}$ & 0 \\
NGC4210 & Sb & 39 & 95 & 0.26 & 43 & 21 & 36 & 1.7 & 2694 & $4.544_{-0.014}^{+0.016}$ & $-0.447_{-0.046}^{+0.044}$ & 0 \\
NGC4470 & Sc & 31 & 1 & 0.32 & 71 & 15 & 33 & 2.2 & 2309 & $1.636_{-0.015}^{+0.024}$ & $0.001_{-0.021}^{+0.019}$ & 0 \\
NGC4644 & Sb & 69 & 56 & 0.67 & 76 & 12 & 29 & 2.4 & 4933 & $6.180_{-0.021}^{+0.021}$ & $-0.680_{-0.033}^{+0.037}$ & 0 \\
NGC4711 & Sbc & 56 & 42 & 0.47 & 67 & 17 & 33 & 1.9 & 4027 & $4.202_{-0.018}^{+0.019}$ & $0.128_{-0.014}^{+0.010}$ & 0 \\
NGC4816 & E1 & 95 & 79 & 0.27 & 52 & 30 & 30 & 1.0 & 6813 & $9.178_{-0.024}^{+0.024}$ & $0.430_{-0.004}^{+0.004}$ & 3 \\
NGC4956 & E1 & 66 & 38 & 0.12 & 30 & 9 & 21 & 2.3 & 4683 & $2.730_{-0.009}^{+0.011}$ & $-0.138_{-0.024}^{+0.025}$ & 2 \\
NGC4961 & Scd & 35 & 100 & 0.33 & 49 & 15 & 33 & 2.2 & 2549 & $5.201_{-0.035}^{+0.036}$ & $0.632_{-0.010}^{+0.007}$ & 0 \\
NGC5000 & Sbc & 77 & 115 & 0.29 & 63 & 16 & 30 & 1.9 & 5586 & $2.499_{-0.014}^{+0.015}$ & $0.365_{-0.012}^{+0.008}$ & 0 \\
NGC5029 & E6 & 121 & 146 & 0.37 & 51 & 25 & 28 & 1.1 & 8563 & $9.458_{-0.025}^{+0.026}$ & $0.326_{-0.005}^{+0.005}$ & 3 \\
NGC5056 & Sc & 77 & 181 & 0.48 & 63 & 15 & 38 & 2.5 & 5534 & $4.376_{-0.012}^{+0.016}$ & $-0.182_{-0.019}^{+0.017}$ & 0 \\
NGC5218 & Sab & 42 & 95 & 0.33 & 57 & 18 & 36 & 2.0 & 2906 & $4.542_{-0.018}^{+0.042}$ & $0.768_{-0.006}^{+0.002}$ & 2 \\
NGC5378 & Sb & 43 & 77 & 0.25 & 42 & 24 & 34 & 1.4 & 2925 & $5.601_{-0.025}^{+0.028}$ & $0.612_{-0.013}^{+0.004}$ & 1 \\
NGC5480 & Scd & 27 & 186 & 0.19 & 46 & 25 & 41 & 1.6 & 1906 & $2.780_{-0.018}^{+0.025}$ & $0.430_{-0.022}^{+0.014}$ & 0 \\
NGC5485 & E5 & 28 & 171 & 0.24 & 45 & 31 & 38 & 1.2 & 1902 & $6.688_{-0.019}^{+0.017}$ & $0.522_{-0.010}^{+0.003}$ & 3 \\
NGC5520 & Sbc & 27 & 64 & 0.50 & 61 & 12 & 34 & 2.8 & 1879 & $4.934_{-0.019}^{+0.023}$ & $0.330_{-0.014}^{+0.009}$ & 0 \\
NGC5614 & Sa & 54 & 133 & 0.19 & 36 & 18 & 35 & 1.9 & 3818 & $3.792_{-0.013}^{+0.015}$ & $0.008_{-0.012}^{+0.013}$ & 2 \\
NGC5630 & Sdm & 38 & 93 & 0.66 & 75 & 22 & 38 & 1.7 & 2642 & $5.805_{-0.038}^{+0.043}$ & $0.198_{-0.026}^{+0.023}$ & 0 \\
NGC5633 & Sbc & 34 & 193 & 0.33 & 49 & 13 & 35 & 2.7 & 2333 & $3.157_{-0.027}^{+0.021}$ & $0.596_{-0.084}^{+0.010}$ & 0 \\
NGC5657 & Sbc & 54 & 165 & 0.62 & 71 & 10 & 39 & 3.9 & 3881 & $4.961_{-0.016}^{+0.017}$ & $-0.177_{-0.014}^{+0.014}$ & 2 \\
NGC5682 & Scd & 33 & 129 & 0.65 & 77 & 26 & 38 & 1.5 & 2278 & $13.554_{-0.142}^{+0.158}$ & $0.501_{-0.020}^{+0.014}$ & 0 \\
NGC5720 & Sbc & 108 & 131 & 0.31 & 49 & 16 & 27 & 1.7 & 7687 & $5.211_{-0.019}^{+0.017}$ & $0.059_{-0.013}^{+0.011}$ & 1 \\
NGC5732 & Sbc & 53 & 39 & 0.47 & 60 & 14 & 32 & 2.3 & 3768 & $5.972_{-0.055}^{+0.064}$ & $0.317_{-0.020}^{+0.017}$ & 0 \\
NGC5784 & S0 & 75 & 190 & 0.31 & 59 & 13 & 29 & 2.2 & 5418 & $6.000_{-0.011}^{+0.013}$ & $-0.035_{-0.005}^{+0.004}$ & 2 \\
NGC5876 & S0a & 47 & 50 & 0.58 & 67 & 12 & 31 & 2.6 & 3274 & $9.370_{-0.025}^{+0.023}$ & $0.094_{-0.008}^{+0.005}$ & 2 \\
NGC5888 & Sb & 121 & 152 & 0.39 & 68 & 16 & 31 & 1.9 & 8583 & $5.749_{-0.011}^{+0.013}$ & $-0.888_{-0.023}^{+0.060}$ & 2 \\
NGC5908 & Sa & 47 & 154 & 0.56 & 81 & 34 & 42 & 1.2 & 3327 & $8.192_{-0.016}^{+0.020}$ & $-0.026_{-0.005}^{+0.006}$ & 2 \\
NGC5971 & Sb & 61 & 131 & 0.55 & 68 & 12 & 27 & 2.2 & 3385 & $6.655_{-0.033}^{+0.047}$ & $0.429_{-0.008}^{+0.006}$ & 0 \\
NGC5980 & Sbc & 57 & 193 & 0.62 & 71 & 17 & 40 & 2.4 & 4045 & $4.474_{-0.014}^{+0.022}$ & $0.401_{-0.012}^{+0.008}$ & 2 \\
NGC5987 & Sa & 44 & 61 & 0.61 & 75 & 33 & 37 & 1.1 & 2994 & $6.452_{-0.014}^{+0.024}$ & $0.308_{-0.006}^{+0.004}$ & 1 \\
NGC6020 & E4 & 60 & 133 & 0.29 & 50 & 19 & 25 & 1.3 & 4272 & $6.520_{-0.027}^{+0.037}$ & $0.417_{-0.008}^{+0.005}$ & 3 \\
NGC6021 & E5 & 66 & 156 & 0.30 & 48 & 9 & 27 & 3.0 & 4698 & $6.203_{-0.025}^{+0.029}$ & $0.475_{-0.005}^{+0.003}$ & 2 \\
NGC6032 & Sbc & 60 & 179 & 0.48 & 76 & 27 & 39 & 1.4 & 4331 & $3.198_{-0.028}^{+0.063}$ & $0.757_{-0.018}^{+0.004}$ & 0 \\
NGC6060 & Sb & 62 & 100 & 0.58 & 65 & 28 & 35 & 1.2 & 4407 & $4.937_{-0.017}^{+0.020}$ & $0.325_{-0.021}^{+0.010}$ & 1 \\
NGC6063 & Sbc & 40 & 155 & 0.45 & 59 & 20 & 36 & 1.8 & 2853 & $5.875_{-0.025}^{+0.025}$ & $-0.073_{-0.031}^{+0.025}$ & 0 \\
\hline
\end{tabular}
\label{tab:DS3}
\end{center}
\end{minipage}
\end{table*}

\begin{table*}
\begin{minipage}{180mm}
\caption{Properties of the 238 (E1--Sdm) CALIFA galaxies. 
Columns list (1) galaxy identifier. (2) Hubble type based on by-eye morphological classification from \citet{Walcher2014}. 
(3) Hubble Flow (Galactocentric) distances from NED (using Hubble constant of $H_0$=73 km sec$^{-1}$ Mpc$^{-1}$),  calculated from the weighted mean radial velocities of the radio and optical redshifts, corrected to the Galactic standard of rest (GSR) from the RC3 (Third Reference Catalogue of Bright Galaxies) as described in \citet[][section 3.10c, page 54]{deVaucouleurs1991}. 
(4) and (5) Average photometric position angle and ellipticity from $r$-band SDSS (DR12) images 
 with typical uncertainty of 5$^{\circ}$-10$^{\circ}$ and 5 per cent, respectively,
 using $\texttt{findgalaxy PYTHON}$ procedure of \citealt{Cappellari2002}. 
(6) From the axial ratio of the galaxies ($q=1-\epsilon$, where $\epsilon$ is the average ellipticity of the galaxy,
 determined from $\texttt{findgalaxy}$), we calculated the inclination ($\emph{i}$) assuming $q_o=0.2$ for the intrinsic axial ratio of the galaxies (Section \ref{SS:dyn}).
(7) Effective (half-light) radius measured via growth curve analysis (\citealt{Walcher2014}). 
(8) Radial extent of the stellar kinematic data in arcseconds (\citealt{Falcon-Barroso2017}).
(9) Radial extent of the stellar kinematic data in terms of effective radius.
 (10) Systemic velocity of the galaxies with a typical uncertainty of 5 km~s$^{-1}$, measured as the median value of the 
 velocity field  (Section \ref{SS:sauronifs}).
 (11) and (12) The medians of the posterior distributions of the dynamical mass-to-light ratios and azimuthal velocity anisotropies, respectively,
 from JAM-MCMC model. The uncertainties are estimated as the 25th and 75th percentiles of the data corresponding to the median absolute deviation
 (Section \ref{SS:dyn}).
 (13) CVC (circular velocity curve) class of the 238 (E1--Sdm) galaxies, where 0 -- Slow rising (SR); 1 -- Flat (FL); 
 2 -- Round-peaked (RP); 3 -- Sharp-peaked (SP).} 
\begin{center}
\begin{tabular}{|c|cccccccccccc|}
%
\hline
Galaxy & Type & Dist & PA & $\epsilon$ & $i$ & $R_{\mathrm{e}}$ & $R_{\mathrm{max}}$ & $R_{\mathrm{max}}/R_{\mathrm{e}}$ & $V_{\mathrm{sys}}$ & 
 $(M/L)_{\mathrm{dyn}}$ & $\beta_{\mathrm{z}}$ &  CVC class \\
 &  & $\mathrm{Mpc}$ & ($^{\circ}$) &  & ($^{\circ}$) & (arcsec) & (arcsec) &  & (km s$^{-1}$) &  &  &  \\
(1)    &  (2)  &  (3)  &      (4) &    (5) &    (6) &  (7)   & (8)    &   (9)  & (10) & (11) & (12) & (13) \\  
\hline
NGC6081 & S0a & 72 & 129 & 0.60 & 71 & 12 & 30 & 2.5 & 5073 & $6.046_{-0.022}^{+0.025}$ & $0.197_{-0.012}^{+0.008}$ & 2 \\
NGC6125 & E1 & 69 & 199 & 0.04 & 26 & 21 & 28 & 1.3 & 4678 & $6.507_{-0.012}^{+0.018}$ & $0.389_{-0.008}^{+0.006}$ & 3 \\
NGC6132 & Sbc & 69 & 126 & 0.66 & 72 & 14 & 31 & 2.2 & 4911 & $5.884_{-0.027}^{+0.025}$ & $0.384_{-0.018}^{+0.011}$ & 0 \\
NGC6146 & E5 & 123 & 79 & 0.35 & 49 & 15 & 26 & 1.7 & 8700 & $5.968_{-0.018}^{+0.024}$ & $0.270_{-0.006}^{+0.005}$ & 2 \\
NGC6150 & E7 & 122 & 58 & 0.46 & 61 & 11 & 29 & 2.6 & 8645 & $6.897_{-0.019}^{+0.019}$ & $0.633_{-0.005}^{+0.002}$ & 2 \\
NGC6168 & Sc & 36 & 109 & 0.77 & 77 & 26 & 34 & 1.3 & 2539 & $7.706_{-0.070}^{+0.078}$ & $0.548_{-0.017}^{+0.011}$ & 0 \\
NGC6173 & E6 & 122 & 145 & 0.39 & 53 & 38 & 32 & 0.8 & 8694 & $6.146_{-0.017}^{+0.017}$ & $0.544_{-0.004}^{+0.002}$ & 3 \\
NGC6186 & Sb & 42 & 55 & 0.36 & 69 & 20 & 35 & 1.8 & 2937 & $2.341_{-0.012}^{+0.014}$ & $0.218_{-0.019}^{+0.013}$ & 1 \\
NGC6278 & S0a & 41 & 126 & 0.45 & 57 & 11 & 33 & 3.0 & 2869 & $5.284_{-0.017}^{+0.015}$ & $0.517_{-0.008}^{+0.003}$ & 2 \\
NGC6301 & Sbc & 117 & 108 & 0.42 & 57 & 24 & 39 & 1.6 & 8219 & $5.787_{-0.014}^{+0.017}$ & $-0.119_{-0.017}^{+0.014}$ & 0 \\
NGC6310 & Sb & 50 & 248 & 0.72 & 76 & 23 & 33 & 1.4 & 3393 & $6.152_{-0.022}^{+0.022}$ & $0.347_{-0.026}^{+0.012}$ & 0 \\
NGC6314 & Sab & 93 & 171 & 0.44 & 60 & 12 & 37 & 3.1 & 6548 & $3.258_{-0.013}^{+0.020}$ & $0.712_{-0.006}^{+0.003}$ & 2 \\
NGC6478 & Sc & 96 & 215 & 0.58 & 69 & 23 & 38 & 1.7 & 6701 & $5.026_{-0.014}^{+0.012}$ & $0.612_{-0.012}^{+0.005}$ & 1 \\
NGC6497 & Sab & 87 & 114 & 0.45 & 61 & 13 & 34 & 2.6 & 5981 & $5.385_{-0.013}^{+0.012}$ & $-0.199_{-0.013}^{+0.010}$ & 3 \\
NGC6515 & E3 & 97 & 193 & 0.33 & 62 & 19 & 28 & 1.5 & 6764 & $5.769_{-0.019}^{+0.021}$ & $0.163_{-0.006}^{+0.004}$ & 3 \\
NGC6762 & Sab & 43 & 118 & 0.70 & 78 & 9 & 31 & 3.4 & 2932 & $4.733_{-0.021}^{+0.016}$ & $0.120_{-0.011}^{+0.008}$ & 0 \\
NGC6941 & Sb & 87 & 131 & 0.28 & 53 & 20 & 32 & 1.6 & 6171 & $4.154_{-0.010}^{+0.013}$ & $-0.270_{-0.022}^{+0.020}$ & 1 \\
NGC6945 & S0 & 54 & 129 & 0.33 & 55 & 13 & 31 & 2.4 & 3750 & $3.622_{-0.008}^{+0.011}$ & $-0.326_{-0.014}^{+0.013}$ & 2 \\
NGC6978 & Sb & 84 & 128 & 0.62 & 70 & 18 & 34 & 1.9 & 5876 & $5.003_{-0.014}^{+0.011}$ & $0.380_{-0.008}^{+0.007}$ & 2 \\
NGC7025 & S0a & 71 & 66 & 0.31 & 46 & 13 & 31 & 2.4 & 4916 & $5.381_{-0.010}^{+0.014}$ & $0.033_{-0.011}^{+0.007}$ & 2 \\
NGC7047 & Sbc & 82 & 108 & 0.52 & 61 & 18 & 29 & 1.6 & 5729 & $4.719_{-0.022}^{+0.021}$ & $0.230_{-0.023}^{+0.013}$ & 0 \\
NGC7194 & E3 & 113 & 20 & 0.31 & 47 & 17 & 22 & 1.3 & 8037 & $7.426_{-0.027}^{+0.036}$ & $0.117_{-0.012}^{+0.007}$ & 2 \\
NGC7311 & Sa & 64 & 11 & 0.46 & 61 & 12 & 37 & 3.1 & 4474 & $2.752_{-0.011}^{+0.011}$ & $0.294_{-0.028}^{+0.011}$ & 2 \\
NGC7321 & Sbc & 100 & 15 & 0.34 & 56 & 15 & 32 & 2.1 & 7060 & $4.272_{-0.010}^{+0.015}$ & $-0.634_{-0.020}^{+0.031}$ & 0 \\
NGC7364 & Sab & 68 & 62 & 0.34 & 49 & 12 & 32 & 2.7 & 4797 & $3.906_{-0.016}^{+0.013}$ & $0.758_{-0.013}^{+0.003}$ & 2 \\
NGC7466 & Sbc & 106 & 206 & 0.58 & 67 & 13 & 31 & 2.4 & 7417 & $5.280_{-0.017}^{+0.016}$ & $0.087_{-0.017}^{+0.015}$ & 0 \\
NGC7489 & Sbc & 88 & 169 & 0.43 & 67 & 20 & 39 & 2.0 & 6170 & $3.516_{-0.016}^{+0.022}$ & $0.134_{-0.018}^{+0.011}$ & 0 \\
NGC7549 & Sbc & 67 & 196 & 0.58 & 72 & 20 & 34 & 1.7 & 4616 & $7.854_{-0.038}^{+0.042}$ & $-0.514_{-0.015}^{+0.019}$ & 2 \\
NGC7550 & E4 & 72 & 138 & 0.08 & 22 & 24 & 25 & 1.0 & 5029 & $7.122_{-0.023}^{+0.024}$ & $0.318_{-0.016}^{+0.014}$ & 3 \\
NGC7562 & E4 & 51 & 83 & 0.35 & 66 & 20 & 36 & 1.8 & 3588 & $5.285_{-0.012}^{+0.028}$ & $0.220_{-0.005}^{+0.003}$ & 2 \\
NGC7563 & Sa & 59 & 146 & 0.40 & 55 & 9 & 31 & 3.4 & 4310 & $6.571_{-0.015}^{+0.015}$ & $0.048_{-0.006}^{+0.005}$ & 2 \\
NGC7591 & Sbc & 70 & 151 & 0.49 & 62 & 16 & 33 & 2.1 & 4950 & $3.785_{-0.012}^{+0.012}$ & $-0.048_{-0.011}^{+0.011}$ & 2 \\
NGC7608 & Sbc & 50 & 18 & 0.71 & 75 & 20 & 33 & 1.6 & 3522 & $4.043_{-0.039}^{+0.033}$ & $0.734_{-0.014}^{+0.007}$ & 0 \\
NGC7611 & S0 & 47 & 136 & 0.49 & 63 & 11 & 21 & 1.9 & 3204 & $7.129_{-0.062}^{+0.068}$ & $0.086_{-0.019}^{+0.017}$ & 2 \\
NGC7619 & E3 & 54 & 42 & 0.16 & 47 & 35 & 34 & 1.0 & 3732 & $5.679_{-0.013}^{+0.016}$ & $0.366_{-0.008}^{+0.003}$ & 3 \\
NGC7623 & S0 & 53 & 7 & 0.27 & 44 & 10 & 31 & 3.1 & 3660 & $4.314_{-0.030}^{+0.013}$ & $0.714_{-0.008}^{+0.002}$ & 2 \\
NGC7625 & Sa & 25 & 211 & 0.10 & 41 & 14 & 34 & 2.4 & 1615 & $2.273_{-0.017}^{+0.020}$ & $0.510_{-0.109}^{+0.011}$ & 0 \\
NGC7631 & Sb & 54 & 76 & 0.60 & 68 & 17 & 33 & 1.9 & 3706 & $4.150_{-0.010}^{+0.012}$ & $0.179_{-0.018}^{+0.015}$ & 0 \\
NGC7653 & Sb & 61 & 167 & 0.17 & 35 & 12 & 38 & 3.2 & 4239 & $2.832_{-0.011}^{+0.019}$ & $-0.730_{-0.068}^{+0.071}$ & 0 \\
NGC7671 & S0 & 59 & 136 & 0.38 & 56 & 11 & 26 & 2.4 & 3885 & $6.004_{-0.014}^{+0.017}$ & $0.311_{-0.009}^{+0.006}$ & 2 \\
\hline
\end{tabular}
\label{tab:DS4}
\end{center}
\end{minipage}
\end{table*}

\begin{table*}
\begin{minipage}{180mm}
\caption{Properties of the 238 (E1--Sdm) CALIFA galaxies. 
Columns list (1) galaxy identifier. (2) Hubble type based on by-eye morphological classification from \citet{Walcher2014}. 
(3) Hubble Flow (Galactocentric) distances from NED (using Hubble constant of $H_0$=73 km sec$^{-1}$ Mpc$^{-1}$),  calculated from the weighted mean radial velocities of the radio and optical redshifts, corrected to the Galactic standard of rest (GSR) from the RC3 (Third Reference Catalogue of Bright Galaxies) as described in \citet[][section 3.10c, page 54]{deVaucouleurs1991}. 
(4) and (5) Average photometric position angle and ellipticity from $r$-band SDSS (DR12) images 
 with typical uncertainty of 5$^{\circ}$-10$^{\circ}$ and 5 per cent, respectively,
 using $\texttt{findgalaxy PYTHON}$ procedure of \citealt{Cappellari2002}. 
(6) From the axial ratio of the galaxies ($q=1-\epsilon$, where $\epsilon$ is the average ellipticity of the galaxy,
 determined from $\texttt{findgalaxy}$), we calculated the inclination ($\emph{i}$) assuming $q_o=0.2$ for the intrinsic axial ratio of the galaxies (Section \ref{SS:dyn}).
(7) Effective (half-light) radius measured via growth curve analysis (\citealt{Walcher2014}). 
(8) Radial extent of the stellar kinematic data in arcseconds (\citealt{Falcon-Barroso2017}).
(9) Radial extent of the stellar kinematic data in terms of effective radius.
 (10) Systemic velocity of the galaxies with a typical uncertainty of 5 km~s$^{-1}$, measured as the median value of the 
 velocity field  (Section \ref{SS:sauronifs}).
 (11) and (12) The medians of the posterior distributions of the dynamical mass-to-light ratios and azimuthal velocity anisotropies, respectively,
 from JAM-MCMC model. The uncertainties are estimated as the 25th and 75th percentiles of the data corresponding to the median absolute deviation
 (Section \ref{SS:dyn}).
 (13) CVC (circular velocity curve) class of the 238 (E1--Sdm) galaxies, where 0 -- Slow rising (SR); 1 -- Flat (FL); 
 2 -- Round-peaked (RP); 3 -- Sharp-peaked (SP).} 
\begin{center}
\begin{tabular}{|c|cccccccccccc|}
%
\hline
Galaxy & Type & Dist & PA & $\epsilon$ & $i$ & $R_{\mathrm{e}}$ & $R_{\mathrm{max}}$ & $R_{\mathrm{max}}/R_{\mathrm{e}}$ & $V_{\mathrm{sys}}$ & 
 $(M/L)_{\mathrm{dyn}}$ & $\beta_{\mathrm{z}}$ &  CVC class \\
 &  & $\mathrm{Mpc}$ & ($^{\circ}$) &  & ($^{\circ}$) & (arcsec) & (arcsec) &  & (km s$^{-1}$) &  &  &  \\
(1)    &  (2)  &  (3)  &      (4) &    (5) &    (6) &  (7)   & (8)    &   (9)  & (10) & (11) & (12) & (13) \\   
\hline
NGC7683 & S0 & 53 & 141 & 0.48 & 62 & 14 & 33 & 2.4 & 3725 & $5.646_{-0.021}^{+0.018}$ & $0.309_{-0.018}^{+0.006}$ & 2 \\
NGC7711 & E7 & 58 & 92 & 0.58 & 65 & 15 & 42 & 2.8 & 4049 & $4.564_{-0.011}^{+0.019}$ & $0.345_{-0.013}^{+0.004}$ & 2 \\
NGC7716 & Sb & 37 & 35 & 0.19 & 55 & 21 & 38 & 1.8 & 2565 & $2.485_{-0.007}^{+0.007}$ & $-0.176_{-0.019}^{+0.012}$ & 1 \\
NGC7722 & Sab & 57 & 148 & 0.19 & 54 & 21 & 24 & 1.1 & 4024 & $6.660_{-0.030}^{+0.033}$ & $0.102_{-0.012}^{+0.010}$ & 3 \\
NGC7738 & Sb & 94 & 43 & 0.37 & 72 & 14 & 37 & 2.6 & 6708 & $3.685_{-0.020}^{+0.019}$ & $0.508_{-0.041}^{+0.008}$ & 1 \\
NGC7787 & Sab & 93 & 115 & 0.48 & 62 & 11 & 23 & 2.1 & 6622 & $5.480_{-0.048}^{+0.048}$ & $-0.100_{-0.021}^{+0.022}$ & 0 \\
NGC7819 & Sc & 70 & 100 & 0.40 & 54 & 23 & 37 & 1.6 & 4908 & $3.754_{-0.019}^{+0.021}$ & $0.175_{-0.025}^{+0.023}$ & 1 \\
NGC7824 & Sab & 86 & 146 & 0.35 & 54 & 11 & 38 & 3.5 & 6071 & $6.052_{-0.027}^{+0.031}$ & $0.250_{-0.017}^{+0.011}$ & 2 \\
UGC00005 & Sbc & 101 & 227 & 0.49 & 65 & 16 & 33 & 2.1 & 7166 & $5.291_{-0.014}^{+0.015}$ & $-0.626_{-0.020}^{+0.032}$ & 2 \\
UGC00029 & E1 & 123 & 172 & 0.26 & 45 & 17 & 13 & 0.8 & 8687 & $7.690_{-0.066}^{+0.075}$ & $-0.504_{-0.056}^{+0.065}$ & 3 \\
UGC00036 & Sab & 88 & 17 & 0.59 & 66 & 10 & 20 & 2.0 & 6249 & $7.877_{-0.032}^{+0.029}$ & $-0.099_{-0.013}^{+0.013}$ & 2 \\
UGC00148 & Sc & 60 & 95 & 0.66 & 76 & 20 & 36 & 1.8 & 4148 & $4.755_{-0.035}^{+0.041}$ & $0.705_{-0.016}^{+0.006}$ & 0 \\
UGC00312 & Sd & 61 & 8 & 0.56 & 75 & 20 & 38 & 1.9 & 4288 & $8.364_{-0.057}^{+0.064}$ & $-0.357_{-0.035}^{+0.030}$ & 0 \\
UGC00809 & Scd & 60 & 22 & 0.79 & 78 & 20 & 36 & 1.8 & 4154 & $13.184_{-0.098}^{+0.101}$ & $0.532_{-0.010}^{+0.008}$ & 0 \\
UGC00987 & Sa & 66 & 31 & 0.69 & 75 & 12 & 34 & 2.8 & 4615 & $4.650_{-0.017}^{+0.028}$ & $0.262_{-0.016}^{+0.011}$ & 2 \\
UGC01057 & Sc & 89 & 152 & 0.69 & 77 & 14 & 27 & 1.9 & 6308 & $5.074_{-0.038}^{+0.034}$ & $0.420_{-0.012}^{+0.009}$ & 0 \\
UGC01271 & S0a & 71 & 101 & 0.42 & 60 & 9 & 29 & 3.2 & 5017 & $4.959_{-0.030}^{+0.035}$ & $-0.423_{-0.028}^{+0.031}$ & 2 \\
UGC02222 & S0a & 71 & 98 & 0.56 & 66 & 10 & 23 & 2.3 & 4949 & $5.060_{-0.019}^{+0.024}$ & $0.314_{-0.011}^{+0.007}$ & 2 \\
UGC02229 & S0a & 100 & 172 & 0.38 & 59 & 19 & 25 & 1.3 & 7219 & $8.181_{-0.027}^{+0.031}$ & $0.597_{-0.005}^{+0.004}$ & 2 \\
UGC02403 & Sb & 57 & 150 & 0.62 & 70 & 19 & 26 & 1.4 & 4144 & $4.777_{-0.037}^{+0.040}$ & $0.569_{-0.010}^{+0.008}$ & 0 \\
UGC03253 & Sb & 59 & 84 & 0.36 & 62 & 15 & 33 & 2.2 & 4093 & $5.915_{-0.019}^{+0.022}$ & $-0.455_{-0.020}^{+0.024}$ & 0 \\
UGC03539 & Sc & 47 & 117 & 0.78 & 84 & 20 & 38 & 1.9 & 3257 & $11.785_{-0.090}^{+0.090}$ & $-0.555_{-0.060}^{+0.058}$ & 0 \\
UGC03995 & Sb & 64 & 90 & 0.53 & 62 & 25 & 39 & 1.6 & 4746 & $5.642_{-0.025}^{+0.023}$ & $0.345_{-0.009}^{+0.007}$ & 1 \\
UGC04029 & Sc & 60 & 63 & 0.80 & 81 & 26 & 37 & 1.4 & 4411 & $7.837_{-0.037}^{+0.045}$ & $0.394_{-0.018}^{+0.009}$ & 0 \\
UGC04132 & Sbc & 71 & 29 & 0.71 & 76 & 22 & 35 & 1.6 & 5111 & $7.697_{-0.019}^{+0.029}$ & $0.024_{-0.016}^{+0.011}$ & 2 \\
UGC04145 & Sa & 62 & 138 & 0.48 & 81 & 9 & 29 & 3.2 & 4609 & $7.279_{-0.030}^{+0.028}$ & $0.077_{-0.007}^{+0.009}$ & 0 \\
UGC04197 & Sab & 61 & 131 & 0.74 & 82 & 18 & 41 & 2.3 & 4490 & $7.807_{-0.022}^{+0.026}$ & $0.303_{-0.013}^{+0.012}$ & 0 \\
UGC04280 & Sb & 49 & 1 & 0.68 & 75 & 11 & 36 & 3.3 & 3512 & $5.942_{-0.029}^{+0.023}$ & $0.291_{-0.010}^{+0.009}$ & 0 \\
UGC04308 & Sc & 48 & 125 & 0.23 & 58 & 24 & 33 & 1.4 & 3538 & $4.002_{-0.030}^{+0.030}$ & $-0.404_{-0.037}^{+0.040}$ & 0 \\
UGC05108 & Sb & 110 & 151 & 0.56 & 64 & 9 & 19 & 2.1 & 8001 & $7.372_{-0.029}^{+0.039}$ & $-0.337_{-0.016}^{+0.015}$ & 1 \\
UGC05113 & S0a & 95 & 41 & 0.71 & 78 & 8 & 22 & 2.8 & 6752 & $5.324_{-0.024}^{+0.024}$ & $0.510_{-0.008}^{+0.006}$ & 2 \\
UGC05598 & Sb & 76 & 36 & 0.73 & 76 & 15 & 27 & 1.8 & 5597 & $5.574_{-0.057}^{+0.071}$ & $0.713_{-0.012}^{+0.007}$ & 0 \\
UGC05771 & E6 & 102 & 57 & 0.31 & 47 & 12 & 27 & 2.2 & 7360 & $7.771_{-0.032}^{+0.034}$ & $0.501_{-0.009}^{+0.004}$ & 3 \\
UGC05990 & Sc & 21 & 16 & 0.70 & 78 & 12 & 33 & 2.8 & 1566 & $8.605_{-0.114}^{+0.137}$ & $0.656_{-0.011}^{+0.010}$ & 0 \\
UGC06036 & Sa & 89 & 100 & 0.74 & 81 & 11 & 38 & 3.5 & 6492 & $7.644_{-0.017}^{+0.017}$ & $0.005_{-0.007}^{+0.008}$ & 2 \\
UGC06312 & Sab & 85 & 56 & 0.62 & 68 & 13 & 29 & 2.2 & 6295 & $7.055_{-0.025}^{+0.028}$ & $0.370_{-0.006}^{+0.005}$ & 2 \\
UGC07012 & Scd & 42 & 10 & 0.47 & 58 & 14 & 30 & 2.1 & 3070 & $6.506_{-0.052}^{+0.061}$ & $0.289_{-0.027}^{+0.020}$ & 0 \\
UGC07145 & Sbc & 91 & 152 & 0.62 & 69 & 16 & 32 & 2.0 & 6562 & $7.018_{-0.051}^{+0.059}$ & $0.744_{-0.010}^{+0.004}$ & 0 \\
UGC08107 & Sa & 115 & 52 & 0.62 & 72 & 16 & 33 & 2.1 & 8220 & $12.324_{-0.047}^{+0.039}$ & $0.390_{-0.007}^{+0.006}$ & 2 \\
UGC08231 & Sd & 35 & 75 & 0.63 & 71 & 19 & 33 & 1.7 & 2482 & $19.564_{-0.170}^{+0.171}$ & $0.249_{-0.023}^{+0.017}$ & 0 \\
\hline
\end{tabular}
\label{tab:DS5}
\end{center}
\end{minipage}
\end{table*}

\begin{table*}
\begin{minipage}{180mm}
\caption{Properties of the 238 (E1--Sdm) CALIFA galaxies. 
Columns list (1) galaxy identifier. (2) Hubble type based on by-eye morphological classification from \citet{Walcher2014}. 
(3) Hubble Flow (Galactocentric) distances from NED (using Hubble constant of $H_0$=73 km sec$^{-1}$ Mpc$^{-1}$),  calculated from the weighted mean radial velocities of the radio and optical redshifts, corrected to the Galactic standard of rest (GSR) from the RC3 (Third Reference Catalogue of Bright Galaxies) as described in \citet[][section 3.10c, page 54]{deVaucouleurs1991}. 
(4) and (5) Average photometric position angle and ellipticity from $r$-band SDSS (DR12) images 
 with typical uncertainty of 5$^{\circ}$-10$^{\circ}$ and 5 per cent, respectively,
 using $\texttt{findgalaxy PYTHON}$ procedure of \citealt{Cappellari2002}. 
(6) From the axial ratio of the galaxies ($q=1-\epsilon$, where $\epsilon$ is the average ellipticity of the galaxy,
 determined from $\texttt{findgalaxy}$), we calculated the inclination ($\emph{i}$) assuming $q_o=0.2$ for the intrinsic axial ratio of the galaxies (Section \ref{SS:dyn}).
(7) Effective (half-light) radius measured via growth curve analysis (\citealt{Walcher2014}). 
(8) Radial extent of the stellar kinematic data in arcseconds (\citealt{Falcon-Barroso2017}).
(9) Radial extent of the stellar kinematic data in terms of effective radius.
 (10) Systemic velocity of the galaxies with a typical uncertainty of 5 km~s$^{-1}$, measured as the median value of the 
 velocity field  (Section \ref{SS:sauronifs}).
 (11) and (12) The medians of the posterior distributions of the dynamical mass-to-light ratios and azimuthal velocity anisotropies, respectively,
 from JAM-MCMC model. The uncertainties are estimated as the 25th and 75th percentiles of the data corresponding to the median absolute deviation
 (Section \ref{SS:dyn}).
 (13) CVC (circular velocity curve) class of the 238 (E1--Sdm) galaxies, where 0 -- Slow rising (SR); 1 -- Flat (FL); 
 2 -- Round-peaked (RP); 3 -- Sharp-peaked (SP).} 
\begin{center}
\begin{tabular}{|c|cccccccccccc|}
%
\hline
Galaxy & Type & Dist & PA & $\epsilon$ & $i$ & $R_{\mathrm{e}}$ & $R_{\mathrm{max}}$ & $R_{\mathrm{max}}/R_{\mathrm{e}}$ & $V_{\mathrm{sys}}$ & 
 $(M/L)_{\mathrm{dyn}}$ & $\beta_{\mathrm{z}}$ &  CVC class \\
 &  & $\mathrm{Mpc}$ & ($^{\circ}$) &  & ($^{\circ}$) & (arcsec) & (arcsec) &  & (km s$^{-1}$) &  &  &  \\
(1)    &  (2)  &  (3)  &      (4) &    (5) &    (6) &  (7)   & (8)    &   (9)  & (10) & (11) & (12) & (13) \\   
\hline
UGC08234 & S0 & 113 & 134 & 0.40 & 55 & 8 & 24 & 3.0 & 8043 & $3.166_{-0.013}^{+0.021}$ & $0.565_{-0.020}^{+0.004}$ & 2 \\
UGC08733 & Sdm & 33 & 195 & 0.42 & 64 & 30 & 40 & 1.3 & 2329 & $9.708_{-0.075}^{+0.080}$ & $0.528_{-0.010}^{+0.009}$ & 0 \\
UGC08778 & Sb & 46 & 117 & 0.74 & 78 & 15 & 27 & 1.8 & 3235 & $5.848_{-0.029}^{+0.036}$ & $0.362_{-0.015}^{+0.013}$ & 0 \\
UGC08781 & Sb & 104 & 161 & 0.42 & 59 & 15 & 29 & 1.9 & 7554 & $4.789_{-0.027}^{+0.036}$ & $0.289_{-0.010}^{+0.008}$ & 2 \\
UGC09476 & Sbc & 46 & 130 & 0.34 & 49 & 21 & 40 & 1.9 & 3247 & $3.754_{-0.018}^{+0.016}$ & $0.111_{-0.036}^{+0.033}$ & 0 \\
UGC09537 & Sb & 122 & 139 & 0.76 & 82 & 20 & 40 & 2.0 & 8778 & $6.438_{-0.029}^{+0.040}$ & $0.659_{-0.005}^{+0.003}$ & 1 \\
UGC09542 & Sc & 76 & 40 & 0.69 & 73 & 21 & 37 & 1.8 & 5412 & $6.245_{-0.037}^{+0.040}$ & $0.788_{-0.008}^{+0.004}$ & 0 \\
UGC09665 & Sb & 37 & 140 & 0.76 & 80 & 18 & 33 & 1.8 & 2524 & $6.302_{-0.038}^{+0.050}$ & $0.285_{-0.031}^{+0.025}$ & 0 \\
UGC09873 & Sb & 79 & 125 & 0.77 & 78 & 21 & 33 & 1.6 & 5587 & $7.666_{-0.054}^{+0.056}$ & $0.427_{-0.019}^{+0.020}$ & 1 \\
UGC09892 & Sbc & 80 & 101 & 0.74 & 77 & 16 & 26 & 1.6 & 5632 & $4.904_{-0.028}^{+0.037}$ & $-0.083_{-0.048}^{+0.046}$ & 0 \\
UGC10097 & E5 & 84 & 125 & 0.14 & 51 & 14 & 27 & 1.9 & 5912 & $6.390_{-0.015}^{+0.016}$ & $0.415_{-0.005}^{+0.004}$ & 3 \\
UGC10123 & Sab & 54 & 53 & 0.72 & 79 & 18 & 31 & 1.7 & 3723 & $8.337_{-0.041}^{+0.037}$ & $0.476_{-0.011}^{+0.008}$ & 0 \\
UGC10205 & S0a & 92 & 152 & 0.41 & 54 & 19 & 35 & 1.8 & 6500 & $8.828_{-0.023}^{+0.019}$ & $0.055_{-0.008}^{+0.008}$ & 2 \\
UGC10257 & Sbc & 54 & 164 & 0.76 & 77 & 20 & 38 & 1.9 & 3826 & $6.514_{-0.037}^{+0.040}$ & $0.531_{-0.024}^{+0.016}$ & 0 \\
UGC10331 & Sc & 64 & 141 & 0.77 & 78 & 19 & 41 & 2.2 & 4447 & $4.990_{-0.103}^{+0.073}$ & $0.776_{-0.038}^{+0.006}$ & 0 \\
UGC10337 & Sb & 121 & 56 & 0.54 & 72 & 17 & 26 & 1.5 & 8685 & $6.969_{-0.044}^{+0.027}$ & $0.869_{-0.011}^{+0.001}$ & 0 \\
UGC10384 & Sb & 69 & 91 & 0.76 & 79 & 11 & 35 & 3.2 & 4935 & $6.230_{-0.038}^{+0.042}$ & $0.505_{-0.019}^{+0.013}$ & 0 \\
UGC10388 & Sa & 65 & 129 & 0.68 & 73 & 11 & 28 & 2.5 & 4600 & $6.253_{-0.019}^{+0.024}$ & $0.292_{-0.009}^{+0.008}$ & 2 \\
UGC10650 & Scd & 42 & 202 & 0.77 & 83 & 23 & 43 & 1.9 & 2999 & $13.753_{-0.281}^{+0.306}$ & $0.798_{-0.007}^{+0.008}$ & 0 \\
UGC10693 & E7 & 117 & 109 & 0.34 & 49 & 22 & 31 & 1.4 & 8259 & $5.434_{-0.024}^{+0.028}$ & $0.774_{-0.008}^{+0.002}$ & 3 \\
UGC10695 & E5 & 116 & 116 & 0.42 & 55 & 24 & 27 & 1.1 & 8156 & $7.652_{-0.026}^{+0.028}$ & $0.390_{-0.007}^{+0.005}$ & 3 \\
UGC10710 & Sb & 117 & 148 & 0.74 & 78 & 20 & 36 & 1.8 & 8239 & $8.040_{-0.028}^{+0.035}$ & $0.568_{-0.011}^{+0.007}$ & 0 \\
UGC10796 & Scd & 45 & 236 & 0.44 & 57 & 20 & 32 & 1.6 & 3060 & $6.207_{-0.109}^{+0.098}$ & $0.293_{-0.028}^{+0.021}$ & 0 \\
UGC10811 & Sb & 122 & 90 & 0.65 & 71 & 12 & 29 & 2.4 & 8581 & $8.154_{-0.043}^{+0.046}$ & $-0.313_{-0.026}^{+0.023}$ & 2 \\
UGC10905 & S0a & 110 & 173 & 0.47 & 69 & 15 & 25 & 1.7 & 7679 & $6.456_{-0.020}^{+0.017}$ & $0.203_{-0.004}^{+0.005}$ & 2 \\
UGC10972 & Sbc & 66 & 57 & 0.62 & 77 & 24 & 34 & 1.4 & 4617 & $5.231_{-0.023}^{+0.028}$ & $0.612_{-0.016}^{+0.006}$ & 1 \\
UGC11228 & S0 & 82 & 178 & 0.31 & 59 & 12 & 33 & 2.8 & 5745 & $4.673_{-0.016}^{+0.031}$ & $0.439_{-0.009}^{+0.004}$ & 2 \\
UGC11717 & Sab & 89 & 213 & 0.57 & 65 & 17 & 39 & 2.3 & 6181 & $9.592_{-0.066}^{+0.080}$ & $0.607_{-0.006}^{+0.005}$ & 2 \\
UGC12054 & Sc & 31 & 47 & 0.78 & 83 & 15 & 33 & 2.2 & 2039 & $7.214_{-0.128}^{+0.106}$ & $0.140_{-0.041}^{+0.036}$ & 0 \\
UGC12127 & E1 & 116 & 185 & 0.15 & 44 & 36 & 25 & 0.7 & 8169 & $7.714_{-0.033}^{+0.027}$ & $0.444_{-0.005}^{+0.005}$ & 3 \\
UGC12185 & Sb & 94 & 153 & 0.48 & 60 & 12 & 33 & 2.8 & 6518 & $4.967_{-0.027}^{+0.027}$ & $0.584_{-0.017}^{+0.006}$ & 2 \\
UGC12274 & Sa & 108 & 142 & 0.58 & 66 & 17 & 27 & 1.6 & 7553 & $6.030_{-0.032}^{+0.033}$ & $0.529_{-0.015}^{+0.007}$ & 2 \\
UGC12308 & Scd & 33 & 119 & 0.80 & 79 & 27 & 38 & 1.4 & 2204 & $12.393_{-0.119}^{+0.139}$ & $0.655_{-0.010}^{+0.008}$ & 0 \\
UGC12518 & Sb & 40 & 25 & 0.75 & 80 & 17 & 34 & 2.0 & 3740 & $3.869_{-0.127}^{+0.952}$ & $0.905_{-0.024}^{+0.004}$ & 0 \\
UGC12519 & Sc & 62 & 157 & 0.70 & 81 & 21 & 34 & 1.6 & 4325 & $5.886_{-0.047}^{+0.049}$ & $-0.188_{-0.029}^{+0.025}$ & 0 \\
UGC12723 & Sc & 77 & 76 & 0.83 & 82 & 17 & 27 & 1.6 & 5400 & $9.874_{-0.090}^{+0.081}$ & $0.588_{-0.023}^{+0.016}$ & 0 \\
UGC12857 & Sbc & 36 & 34 & 0.74 & 82 & 19 & 36 & 1.9 & 2466 & $4.287_{-0.049}^{+0.051}$ & $0.554_{-0.031}^{+0.014}$ & 0 \\
\hline
\end{tabular}
\label{tab:DS6}
\end{center}
\end{minipage}
\end{table*}
\clearpage
\section{Main PC Eigenvectors of the 238 CVCs }
\label{A:PC_vect} 
In Table \ref{tab:PCvectors}, we present the five main PC eigenvectors as a common basis of the 238 CVCs (columns: 2, 3, 4, 5 and 6); see Fig. \ref{fig:pcaradial} and Section \ref{SS:pca}. We normalize the radius $R$ of the CVCs to the effective radius of the galaxies, $R_{\mathrm{e}}$ (column 1). We use 50 points spanning $R/R_{\mathrm{e}}$ $\in$ [0,1.5]. We logarithmically sample the normalized radius since the CVCs vary much more in the centre than in their outer parts.
\begin{table*}
\begin{minipage}{180mm}
\caption{Main PC eigenvectors of the 238 CVCs}
\begin{center}
\begin{tabular}{|c|ccccc|}
\hline
\multicolumn{1}{|c|}{Normalized PC radius}  & \multicolumn{5}{c|}{Main PC eigenvectors of the 238 CVCs }  \\ 
\hline
$R/R_{\mathrm{e}}$ & $\textbf{\textit{u}}_1$ & $\textbf{\textit{u}}_2$ & $\textbf{\textit{u}}_3$ & $\textbf{\textit{u}}_4$ 
& $\textbf{\textit{u}}_5$ \\
(1)    &  (2)  &  (3)  &      (4) &    (5)  & (6) \\ 
\hline
0.05 & 64.76932 & 37.259213 & 11.21853 & 8.9489543 & 4.6411178 \\
0.054 & 68.055877 & 37.676725 & 10.948696 & 8.2453302 & 3.9447158 \\
0.057 & 71.360914 & 37.860693 & 10.536666 & 7.3751072 & 3.1373666 \\
0.062 & 74.666903 & 37.79002 & 9.9738811 & 6.3419901 & 2.2366909 \\
0.066 & 77.955658 & 37.448003 & 9.255452 & 5.1579907 & 1.2700509 \\
0.071 & 81.208198 & 36.823327 & 8.3807879 & 3.8442673 & 0.27449214 \\
0.076 & 84.404509 & 35.910691 & 7.3539918 & 2.4312438 & -0.70464477 \\
0.081 & 87.523333 & 34.710926 & 6.1840008 & 0.95782335 & -1.6166277 \\
0.087 & 90.542206 & 33.230662 & 4.8844873 & -0.53034653 & -2.4096168 \\
0.093 & 93.437834 & 31.481733 & 3.4735645 & -1.9834188 & -3.0357776 \\
0.1 & 96.186956 & 29.480676 & 1.9734466 & -3.3503051 & -3.4563487 \\
0.107 & 98.767639 & 27.248632 & 0.41017445 & -4.5815536 & -3.6456969 \\
0.115 & 101.16079 & 24.81173 & -1.1865335 & -5.6317402 & -3.5936475 \\
0.123 & 103.35157 & 22.201673 & -2.7834851 & -6.4613244 & -3.3059167 \\
0.132 & 105.33025 & 19.455889 & -4.3446576 & -7.0383221 & -2.8030829 \\
0.142 & 107.09223 & 16.616691 & -5.83257 & -7.3401554 & -2.1187458 \\
0.152 & 108.63717 & 13.729172 & -7.2103017 & -7.3556303 & -1.2973089 \\
0.163 & 109.96751 & 10.838253 & -8.443638 & -7.0864419 & -0.39138065 \\
0.174 & 111.08682 & 7.9855522 & -9.5028233 & -6.5475253 & 0.54139922 \\
0.187 & 111.9985 & 5.2068615 & -10.36366 & -5.7659212 & 1.4420876 \\
0.2 & 112.70507 & 2.5307238 & -11.008105 & -4.7784258 & 2.2543876 \\
0.215 & 113.20818 & -0.021777731 & -11.424611 & -3.628653 & 2.9282843 \\
0.23 & 113.50901 & -2.4363144 & -11.608392 & -2.3641805 & 3.4229664 \\
0.247 & 113.60883 & -4.7042019 & -11.561535 & -1.0341945 & 3.7089523 \\
0.265 & 113.50953 & -6.8213952 & -11.292886 & 0.31222026 & 3.769558 \\
0.284 & 113.21381 & -8.7878239 & -10.817487 & 1.627416 & 3.6018536 \\
0.304 & 112.7253 & -10.607081 & -10.155462 & 2.8664942 & 3.2171157 \\
0.326 & 112.04845 & -12.28619 & -9.3303486 & 3.9889544 & 2.640534 \\
0.349 & 111.18832 & -13.835095 & -8.3670289 & 4.9605188 & 1.9099085 \\
0.374 & 110.15019 & -15.265495 & -7.2896241 & 5.75468 & 1.0731298 \\
0.401 & 108.93919 & -16.588888 & -6.1197481 & 6.3532541 & 0.18453767 \\
0.43 & 107.5604 & -17.814046 & -4.8753713 & 6.7456375 & -0.69945797 \\
0.461 & 106.01947 & -18.944648 & -3.5702978 & 6.9270234 & -1.5248958 \\
0.494 & 104.32389 & -19.977969 & -2.2141963 & 6.8964496 & -2.2442446 \\
0.53 & 102.48434 & -20.905308 & -0.81308321 & 6.6555928 & -2.8193531 \\
0.568 & 100.51531 & -21.714176 & 0.62961878 & 6.2088247 & -3.2231187 \\
0.608 & 98.434589 & -22.391575 & 2.1114059 & 5.5643904 & -3.4398877 \\
0.652 & 96.261432 & -22.927287 & 3.6284619 & 4.7359681 & -3.4648218 \\
0.699 & 94.014357 & -23.316129 & 5.1733235 & 3.7437964 & -3.3026104 \\
0.749 & 91.709379 & -23.558792 & 6.7330028 & 2.6147314 & -2.9658773 \\
0.803 & 89.359544 & -23.661338 & 8.288097 & 1.3811375 & -2.4736442 \\
0.861 & 86.975896 & -23.633893 & 9.8130658 & 0.078952947 & -1.849921 \\
0.923 & 84.569475 & -23.489183 & 11.277512 & -1.2544409 & -1.1224072 \\
0.989 & 82.153758 & -23.241346 & 12.648154 & -2.5820091 & -0.32111982 \\
1.06 & 79.746783 & -22.905296 & 13.891131 & -3.8682639 & 0.52307182 \\
1.136 & 77.372457 & -22.496554 & 14.974439 & -5.0799294 & 1.3799602 \\
1.218 & 75.060533 & -22.031309 & 15.870277 & -6.1863094 & 2.2212881 \\
1.306 & 72.845036 & -21.526336 & 16.557231 & -7.1595417 & 3.0216525 \\
1.399 & 70.761173 & -20.998587 & 17.022091 & -7.974969 & 3.7589679 \\
1.5 & 68.841272 & -20.464343 & 17.261184 & -8.6117751 & 4.4146419 \\
\hline
\end{tabular}
\label{tab:PCvectors}
\end{center}
\end{minipage}
\end{table*}
\clearpage
\section{The five main projections PC$_{i}$ of the 238 CVCs}
\label{A:PCs} 
In Tables \ref{tab:PCs1}--\ref{tab:PCs5}, 
we present the five main projections PC$_{i}$ of the CVCs 
for each of the 238 (E1--Sdm) CALIFA galaxies as described in Section \ref{SS:pca}. 
Using the linear combination between those five projections PC$_{i}$ and the five main PC vectors, plus 
adding the mean value $\overline{V_c}$ of the 238 CVCs, we are able to reconstruct well all CVCs of our 
sample (see also Fig. \ref{fig:califa} and equation \ref{eq:eigen}).
\begin{table*}
\begin{minipage}{180mm}
\begin{center}
\caption{The five main projections PC$_{i}$ of the 238 CVCs.}
\begin{tabular}{|c|ccccc|}
\hline
Galaxy & PC1 & PC2 & PC3 & PC4 & PC5 \\
(1)    &  (2)  &  (3)  &      (4) &    (5)  & (6) \\ 
\hline
IC0480 & -1.20926 & -2.3276 & 0.79536 & -1.35533 & -1.45758 \\
IC0540 & -1.0277 & -1.65598 & -1.3821 & -1.15281 & -0.449284 \\
IC0674 & 0.409293 & -2.77205 & -0.482785 & -0.747374 & -0.628182 \\
IC0944 & 0.723189 & -2.14871 & -1.0831 & -2.88036 & 0.103113 \\
IC1079 & 0.841509 & 0.195445 & -0.612319 & -0.398093 & -1.05088 \\
IC1151 & -1.09668 & -1.18971 & -1.06528 & -1.23835 & -1.00047 \\
IC1256 & -0.780024 & -2.31828 & 1.0241 & -2.00657 & -1.61986 \\
IC1528 & -1.06868 & -1.71063 & -0.832056 & -1.55337 & -0.656684 \\
IC1652 & -0.399075 & -0.587136 & -0.819743 & 0.0887954 & -0.519594 \\
IC1755 & 0.319741 & -2.12241 & -1.78489 & -1.30644 & -0.330496 \\
IC2101 & -1.10349 & -2.14584 & -0.356404 & -0.668677 & -1.34662 \\
IC2247 & -0.326579 & -0.569163 & -0.998491 & -1.3598 & -0.65643 \\
IC2487 & -0.7437 & -1.56494 & 0.786716 & -0.642283 & -1.65642 \\
IC4566 & 0.143066 & -1.65649 & -1.52814 & -1.70792 & -0.975006 \\
IC5309 & -0.861252 & -1.4714 & -1.22377 & -0.535366 & -2.40962 \\
IC5376 & -0.00336622 & -1.90825 & -2.0792 & -0.546939 & -1.85613 \\
MCG-01-54-016 & -1.26462 & -1.16168 & -0.966791 & 0.0883945 & -1.88919 \\
MCG-02-02-030 & -0.74409 & -1.35248 & -0.449203 & -1.70031 & -0.128108 \\
MCG-02-02-040 & -1.12141 & -1.81402 & 0.506846 & -1.54051 & -1.59858 \\
MCG-02-03-015 & 0.284227 & -0.810432 & -2.37091 & -2.84387 & -1.05553 \\
MCG-02-51-004 & -0.222262 & -1.64179 & 0.0107271 & -4.22379 & 1.09131 \\
NGC0001 & 0.0590133 & -1.663 & -1.97393 & -1.40448 & -1.40828 \\
NGC0023 & 0.443446 & -1.15225 & -3.70625 & -2.50189 & 1.17438 \\
NGC0155 & 0.806298 & -1.76249 & -1.64517 & -1.54686 & -1.36864 \\
NGC0160 & 0.997055 & -0.64756 & -1.72468 & -1.26951 & -2.04502 \\
NGC0171 & -0.600168 & -0.213682 & -1.90794 & -1.14194 & 0.0927998 \\
NGC0177 & -0.242677 & -2.18961 & -2.42122 & 0.320728 & -1.67213 \\
NGC0192 & 0.166642 & -0.132587 & -1.08111 & -3.16354 & -1.02631 \\
NGC0214 & -0.226703 & -1.54425 & -0.0834138 & -1.19997 & -1.8465 \\
NGC0216 & -1.38734 & -1.25915 & -1.52472 & -0.727423 & -1.86062 \\
NGC0217 & 0.274476 & -1.81879 & -1.11533 & -1.805 & -0.435183 \\
NGC0237 & -0.84227 & -1.88684 & -0.205547 & -1.45908 & -1.28688 \\
NGC0257 & -0.101976 & -1.1933 & -0.705063 & -2.51945 & -0.676655 \\
NGC0429 & -0.149136 & -3.44591 & -0.393745 & 1.22482 & -2.22765 \\
NGC0444 & -1.14733 & -1.08421 & -1.1132 & -1.31008 & -0.482442 \\
NGC0499 & 2.25356 & -1.22882 & -0.392556 & -1.36232 & -1.31672 \\
NGC0504 & 0.391501 & -3.77708 & -0.293322 & 0.772278 & -1.78642 \\
NGC0517 & 0.696317 & -2.71926 & -1.73005 & -1.03803 & -0.50969 \\
NGC0528 & 0.421842 & -3.37503 & -0.0295673 & -0.865415 & -0.820105 \\
NGC0529 & 1.5928 & -2.69623 & -0.639182 & -2.30655 & -0.826763 \\
NGC0551 & -0.650087 & -1.50068 & -0.0254303 & -2.48497 & -1.04241 \\
NGC0681 & -0.211422 & -0.785398 & 0.0452183 & -0.881701 & -0.394939 \\
NGC0741 & 3.31104 & -0.0460114 & 1.97253 & -0.162111 & 0.0313921 \\
NGC0755 & -1.33352 & -1.65394 & -0.661884 & -1.17364 & -1.2099 \\
NGC0768 & -0.319662 & -2.77593 & -0.338327 & -1.08984 & 0.167925 \\
NGC0774 & -0.0371416 & -3.26697 & -0.103042 & -0.429948 & -0.599349 \\
NGC0776 & -0.488312 & -0.018084 & -1.80124 & -1.94228 & -2.27739 \\
NGC0781 & -0.625485 & -1.97724 & -1.47397 & -0.650459 & -1.15739 \\
NGC0810 & 2.12353 & -4.44166 & -0.30524 & -1.2928 & 0.255498 \\
NGC0932 & 0.403977 & -0.725801 & -1.6333 & -1.51453 & -2.38971 \\
NGC1056 & -0.597191 & -1.61349 & -1.48216 & 0.935024 & -1.71221 \\
\hline
\end{tabular}
\label{tab:PCs1}
\end{center}
\end{minipage}
\end{table*}
\begin{table*}
\begin{minipage}{180mm}
\begin{center}
\caption{The five main projections PC$_{i}$ of the 238 CVCs.}
\begin{tabular}{|c|ccccc|}
\hline
Galaxy & PC1 & PC2 & PC3 & PC4 & PC5 \\
(1)    &  (2)  &  (3)  &      (4) &    (5)  & (6) \\ 
\hline
NGC1060 & 3.14852 & -1.20864 & -0.33093 & -1.89527 & -1.45218 \\
NGC1167 & 1.48641 & -1.59929 & 0.173107 & -1.84505 & -1.09924 \\
NGC1349 & 1.03684 & -1.20366 & -0.461765 & -2.77004 & -0.905643 \\
NGC1542 & -0.65718 & -2.18568 & -0.857595 & -1.31328 & 0.0514257 \\
NGC1645 & 0.924748 & -2.59453 & -2.65253 & -0.797754 & 0.236382 \\
NGC1677 & -1.22476 & -1.33862 & -1.1884 & -1.34635 & -1.34623 \\
NGC2253 & 0.391909 & -1.69758 & -3.62714 & -0.686867 & -0.722437 \\
NGC2347 & 0.338657 & -2.0277 & 0.204938 & 0.373939 & -1.96271 \\
NGC2410 & 0.015673 & -1.83271 & -1.69969 & -1.35327 & -1.68101 \\
NGC2449 & -0.115256 & -1.48962 & -0.970933 & -1.87502 & -0.689973 \\
NGC2476 & 0.536627 & -2.68001 & -1.84769 & -0.360344 & -1.18392 \\
NGC2481 & -0.0299444 & -2.54517 & -1.2607 & -0.630674 & -0.649462 \\
NGC2486 & 0.555538 & 1.29881 & -0.790794 & 1.03536 & 2.4756 \\
NGC2553 & 0.539209 & -2.96772 & -1.27083 & -0.0851655 & -1.39875 \\
NGC2554 & 1.30283 & -0.962678 & -1.24524 & -3.05613 & -2.01115 \\
NGC2592 & 0.47325 & -3.18959 & -1.15446 & -0.233936 & -0.468901 \\
NGC2604 & -1.26277 & -0.829645 & -1.26636 & -1.42856 & -0.984058 \\
NGC2639 & 0.701131 & -3.0644 & 0.65683 & -1.61775 & -0.78759 \\
NGC2730 & -1.11375 & -1.53485 & -0.571618 & -2.2206 & -0.294456 \\
NGC2880 & 0.133457 & -1.30168 & -1.88764 & -1.27788 & -1.83529 \\
NGC2906 & -0.277558 & -0.754825 & -0.109797 & -1.86315 & -0.298498 \\
NGC2916 & -0.123477 & -1.11497 & 0.222431 & -2.99827 & -0.370833 \\
NGC2918 & 1.14822 & -4.45705 & 0.632056 & -0.365575 & -0.788331 \\
NGC3057 & -1.47475 & -0.88222 & -1.62413 & -1.05265 & -1.33162 \\
NGC3106 & 0.826744 & 0.126182 & -1.4728 & -0.776833 & -2.55434 \\
NGC3160 & 0.013251 & -1.90649 & -0.225168 & -1.72557 & -0.713456 \\
NGC3300 & -0.0698678 & -2.06551 & -1.7395 & -1.19746 & -0.229852 \\
NGC3381 & -1.38387 & -0.545028 & -1.58721 & -0.354488 & -1.11364 \\
NGC3615 & 2.23241 & -1.07155 & -1.68654 & -2.25966 & -2.55642 \\
NGC3811 & -0.0807657 & 0.628409 & -0.128984 & -1.84056 & -2.31545 \\
NGC3815 & -0.589177 & -1.82769 & -0.33038 & -1.99619 & -1.65024 \\
NGC3994 & 0.169504 & -2.80224 & -0.56651 & -2.01865 & 1.21709 \\
NGC4003 & 0.179141 & -0.770381 & -0.59803 & 0.53172 & 0.849872 \\
NGC4047 & -0.166514 & -2.55476 & 1.13155 & -0.475375 & -2.25634 \\
NGC4149 & -0.55653 & -1.64648 & -1.06817 & -1.91999 & -0.49263 \\
NGC4185 & -0.506522 & -1.28953 & 0.422828 & -2.26548 & -1.29101 \\
NGC4210 & -0.696877 & -0.899789 & -0.19917 & -3.12556 & -0.607097 \\
NGC4470 & -1.52534 & -1.14175 & -0.878827 & -0.833444 & -1.07501 \\
NGC4644 & -0.598155 & -2.0802 & -0.894709 & -1.59717 & -0.155306 \\
NGC4711 & -1.01796 & -1.50683 & -0.208537 & -2.09663 & -1.06802 \\
NGC4816 & 1.70659 & 1.2395 & 1.04209 & -0.0435526 & 0.425543 \\
NGC4956 & 0.125409 & -2.92302 & -0.834996 & -0.822844 & -0.114177 \\
NGC4961 & -0.909017 & -1.93505 & -0.539806 & -0.566348 & -2.9985 \\
NGC5000 & -1.04415 & -1.28577 & -1.398 & -1.16529 & -1.13803 \\
NGC5029 & 2.06655 & -1.41662 & 0.193125 & -1.91201 & -2.84009 \\
NGC5056 & -0.48565 & -1.52834 & -0.327255 & -2.1092 & -2.05446 \\
NGC5218 & -0.675768 & -3.15847 & 1.10512 & -0.314519 & -2.44868 \\
NGC5378 & -0.0223568 & -0.17787 & -1.38833 & -1.219 & -1.85529 \\
NGC5480 & -0.887018 & -0.477911 & -0.572379 & -1.34536 & -1.1238 \\
NGC5485 & 0.885995 & -0.863803 & 0.00398722 & 0.660348 & -1.04989 \\
NGC5520 & -0.415538 & -1.58961 & -1.2535 & 1.08169 & -1.62206 \\
\hline
\end{tabular}
\label{tab:PCs2}
\end{center}
\end{minipage}
\end{table*}
\begin{table*}
\begin{minipage}{180mm}
\caption{The five main projections PC$_{i}$ of the 238 CVCs.}
\begin{center}
\begin{tabular}{|c|ccccc|}
\hline
Galaxy & PC1 & PC2 & PC3 & PC4 & PC5 \\
(1)    &  (2)  &  (3)  &      (4) &    (5)  & (6) \\ 
\hline
NGC5614 & 0.626909 & -2.15989 & -1.37688 & -1.95352 & -0.759416 \\
NGC5630 & -1.22364 & -1.30351 & -0.805675 & -1.03281 & -1.16507 \\
NGC5633 & -1.00172 & -2.15113 & 0.692719 & -2.12723 & -0.488874 \\
NGC5657 & -0.492301 & -2.42364 & -1.23416 & -0.124995 & -1.24796 \\
NGC5682 & -1.01703 & -1.64883 & -1.74058 & -0.593311 & -0.870584 \\
NGC5720 & 0.572892 & 0.602459 & 0.918431 & -0.638802 & 2.77394 \\
NGC5732 & -0.898305 & -1.78692 & -0.79812 & -1.45943 & -1.18247 \\
NGC5784 & 1.44617 & -1.98282 & -2.07935 & -2.70938 & -0.751516 \\
NGC5876 & 0.827159 & -3.63918 & -1.33785 & 0.622526 & -1.45295 \\
NGC5888 & 0.345202 & -1.96351 & -0.0116483 & -2.9603 & 0.530924 \\
NGC5908 & 0.580897 & -1.70357 & 1.86103 & -0.450364 & -1.59016 \\
NGC5971 & -0.234269 & -1.60258 & -1.80396 & -1.33797 & -1.32671 \\
NGC5980 & -0.379456 & -2.42495 & -0.066242 & -1.72009 & -0.246283 \\
NGC5987 & 0.825058 & 0.202409 & -0.0197113 & -0.4315 & -1.11461 \\
NGC6020 & 1.10619 & -0.863381 & -1.32647 & -1.47519 & -2.44128 \\
NGC6021 & 0.75056 & -3.57711 & -0.735847 & -0.294582 & -0.526381 \\
NGC6032 & -1.12095 & -0.870351 & -1.0976 & -1.59065 & -1.18814 \\
NGC6060 & 0.0652768 & -0.454215 & -0.62023 & -3.71057 & -0.0419281 \\
NGC6063 & -1.08955 & -1.39203 & -0.0150701 & -1.73812 & -0.67609 \\
NGC6081 & 0.744095 & -2.13503 & -1.59998 & -0.954489 & -1.89919 \\
NGC6125 & 2.51513 & 0.889342 & -0.308561 & -1.95412 & -2.97736 \\
NGC6132 & -0.798896 & -1.56309 & 0.830146 & -0.408422 & -1.58356 \\
NGC6146 & 2.47239 & -2.32306 & -1.56287 & -2.38165 & -1.90229 \\
NGC6150 & 1.29296 & -3.50879 & 0.223116 & -1.72499 & -0.444164 \\
NGC6168 & -1.37067 & -1.72106 & -0.618762 & -0.917962 & -1.35513 \\
NGC6173 & 2.4895 & 0.951474 & 0.349885 & -0.38986 & -0.578864 \\
NGC6186 & -0.70412 & -0.523137 & -1.20044 & -0.0438859 & 0.47021 \\
NGC6278 & 0.5913 & -3.06759 & -1.21877 & -0.675217 & -0.233788 \\
NGC6301 & -0.437515 & -2.14835 & 1.90313 & -2.93975 & -0.336462 \\
NGC6310 & -0.445764 & -1.19929 & -0.581098 & -2.46027 & -0.404885 \\
NGC6314 & 0.377287 & -2.12739 & -2.17955 & -0.871077 & -1.38105 \\
NGC6478 & 0.322259 & -0.705531 & 0.971574 & -4.24043 & -1.54252 \\
NGC6497 & 1.1269 & 1.56045 & 1.6568 & 3.56648 & 3.21273 \\
NGC6515 & 0.991123 & -1.3043 & -1.24659 & -1.94864 & -2.06186 \\
NGC6762 & -0.881306 & -2.50781 & -0.0972395 & -0.401164 & -1.40666 \\
NGC6941 & 0.177489 & -0.718485 & -1.39284 & -1.9978 & -0.00824516 \\
NGC6945 & 0.255424 & -1.61691 & -1.66164 & -1.68237 & -1.11491 \\
NGC6978 & 0.0227149 & -1.89752 & -0.397437 & -1.88286 & -0.497865 \\
NGC7025 & 1.39389 & -3.44742 & -0.763846 & -0.86445 & -0.589905 \\
NGC7047 & -0.48258 & -1.66649 & 0.320467 & -3.86304 & 0.447648 \\
NGC7194 & 2.10062 & -2.30269 & -1.84827 & -2.45428 & -1.22815 \\
NGC7311 & 0.518819 & -2.80461 & -1.20811 & -0.928716 & -0.716494 \\
NGC7321 & 0.12649 & -1.53494 & 0.490853 & -1.31074 & 1.86042 \\
NGC7364 & 0.557072 & -2.31453 & -0.741403 & -2.14695 & -2.00103 \\
NGC7466 & -0.110419 & -1.54346 & -1.5577 & -2.56754 & -0.394559 \\
NGC7489 & -0.777748 & -1.51366 & 0.584255 & -1.49091 & -2.2348 \\
NGC7549 & -0.013092 & -2.17648 & -0.984777 & -1.818 & -0.692386 \\
NGC7550 & 2.18375 & -1.6648 & 1.29361 & -1.71546 & -3.90838 \\
NGC7562 & 1.60219 & -2.20663 & -0.782445 & -2.15979 & -1.66792 \\
NGC7563 & 0.785775 & -3.98466 & 0.287908 & 0.102194 & -2.1711 \\
NGC7591 & -0.264197 & -2.432 & -1.26894 & -0.0198076 & -1.88376 \\
\hline
\end{tabular}
\label{tab:PCs3}
\end{center}
\end{minipage}
\end{table*}
\begin{table*}
\begin{minipage}{180mm}
\begin{center}
\caption{The five main projections PC$_{i}$ of the 238 CVCs.}
\begin{tabular}{|c|ccccc|}
\hline
Galaxy & PC1 & PC2 & PC3 & PC4 & PC5 \\
(1)    &  (2)  &  (3)  &      (4) &    (5)  & (6) \\ 
\hline
NGC7608 & -1.25382 & -0.974813 & -0.655114 & -1.49076 & -1.4333 \\
NGC7611 & 1.38192 & -2.60431 & -2.99434 & -1.35975 & 0.980652 \\
NGC7619 & 2.67635 & 0.96161 & -0.541645 & -0.394273 & -0.367046 \\
NGC7623 & 0.506492 & -3.29185 & -0.991882 & -0.0626366 & -1.09721 \\
NGC7625 & -0.753266 & -1.65591 & -1.07167 & 0.326613 & -1.41829 \\
NGC7631 & -0.613771 & -1.7269 & -0.741237 & -1.8912 & -1.29132 \\
NGC7653 & -0.304786 & -1.92099 & -1.19279 & -1.4258 & -0.709507 \\
NGC7671 & 1.28804 & -2.50376 & -1.92126 & -2.40655 & 0.843059 \\
NGC7683 & 0.799171 & -2.59421 & -1.01781 & -1.41289 & -1.15274 \\
NGC7711 & 0.925026 & -1.60574 & -1.65224 & -1.84335 & -1.23785 \\
NGC7716 & -0.308364 & -0.226569 & -2.2927 & -1.94909 & -1.69735 \\
NGC7722 & 1.06193 & -0.979749 & -1.10416 & -1.88796 & -1.93734 \\
NGC7738 & 0.0336331 & -0.937297 & -1.37609 & -2.24552 & -2.13936 \\
NGC7787 & -0.305826 & -1.43139 & -1.9494 & -1.84802 & -0.788726 \\
NGC7819 & -0.58144 & -0.269941 & -2.16629 & -1.80617 & -0.450127 \\
NGC7824 & 1.44191 & -2.97776 & -2.5445 & -1.4094 & 0.947523 \\
UGC00005 & -0.562864 & -2.44482 & 0.882002 & -2.67218 & 0.0905813 \\
UGC00029 & 1.38817 & -0.632478 & -0.581561 & -1.58943 & -1.60693 \\
UGC00036 & 0.500905 & -2.48737 & 0.309335 & -1.82194 & -1.5261 \\
UGC00148 & -1.06812 & -1.63918 & 0.283095 & -1.6438 & -0.533773 \\
UGC00312 & -0.606696 & -1.87013 & -0.923356 & 0.20359 & -0.303226 \\
UGC00809 & -1.30326 & -2.15657 & 0.749096 & -1.68255 & -0.988628 \\
UGC00987 & -0.172349 & -2.13615 & -1.10264 & -1.58724 & -0.493098 \\
UGC01057 & -0.833112 & -1.47802 & -0.24579 & -2.42167 & -1.58075 \\
UGC01271 & 0.477361 & -1.48872 & -0.726871 & -0.213719 & -2.42679 \\
UGC02222 & 0.347309 & -2.5012 & -1.49554 & -1.26473 & 0.286401 \\
UGC02229 & 0.897063 & -1.80358 & -0.994325 & -1.68955 & -3.05358 \\
UGC02403 & -0.645573 & -1.11054 & -0.25598 & -1.9803 & -0.730478 \\
UGC03253 & -0.213947 & -1.31896 & -1.37795 & -1.93616 & -0.0537322 \\
UGC03539 & -0.89508 & -1.80346 & -1.14041 & -0.867211 & -1.33255 \\
UGC03995 & 0.474327 & -0.0702206 & -0.401241 & -1.71494 & -0.648783 \\
UGC04029 & -0.634351 & -1.19482 & -0.418676 & -0.561495 & -0.391212 \\
UGC04132 & -0.298852 & -2.41825 & 1.06062 & -3.52727 & 0.247902 \\
UGC04145 & -0.88732 & -3.18494 & 2.47724 & -1.82675 & -1.16889 \\
UGC04197 & -0.367152 & -1.11075 & -0.20997 & -2.05196 & -1.89764 \\
UGC04280 & -0.683564 & -1.62057 & -0.910266 & -1.71303 & -1.13216 \\
UGC04308 & -0.962203 & -0.988778 & -0.864225 & -2.10756 & -0.887249 \\
UGC05108 & 0.748679 & -0.619859 & 2.47141 & 0.946725 & -1.62914 \\
UGC05113 & 0.466305 & -2.22344 & -1.84701 & -1.91937 & -0.173713 \\
UGC05598 & -1.03932 & -1.49279 & -1.03799 & -2.03769 & -0.157719 \\
UGC05771 & 1.76954 & -1.34517 & -0.850302 & -2.17789 & -2.30426 \\
UGC05990 & -1.44234 & -1.54689 & -1.00317 & -0.398566 & -2.17121 \\
UGC06036 & 0.736569 & -2.58025 & -1.63428 & -0.715293 & -0.74499 \\
UGC06312 & 0.543326 & -1.69868 & -1.90135 & -1.93157 & -1.23111 \\
UGC07012 & -1.00155 & -1.63583 & -1.13029 & -1.21017 & -1.1714 \\
UGC07145 & -0.801511 & -1.86991 & -0.293473 & -1.97158 & -0.953033 \\
UGC08107 & 0.456131 & -3.70777 & 0.798667 & -1.80536 & 1.23986 \\
UGC08231 & -0.902553 & -2.17499 & -0.227133 & -0.914167 & -1.36939 \\
UGC08234 & 0.697236 & -3.53237 & -0.906581 & -0.396873 & 0.101844 \\
UGC08733 & -1.19376 & -1.08321 & -0.968352 & -1.39974 & -1.24759 \\
UGC08778 & -0.972854 & -1.6397 & -0.986926 & -1.82459 & -0.198743 \\
\hline
\end{tabular}
\label{tab:PCs4}
\end{center}
\end{minipage}
\end{table*}
\begin{table*}
\begin{minipage}{180mm}
\caption{The five main projections PC$_{i}$ of the 238 CVCs.}
\begin{center}
\begin{tabular}{|c|ccccc|}
\hline
Galaxy & PC1 & PC2 & PC3 & PC4 & PC5 \\
(1)    &  (2)  &  (3)  &      (4) &    (5)  & (6) \\ 
\hline
UGC08781 & 0.129011 & -2.03356 & -1.45934 & -1.8092 & -0.53345 \\
UGC09476 & -1.00204 & -1.04906 & -0.557631 & -2.02659 & -1.29871 \\
UGC09537 & 0.594893 & -0.83711 & -0.833453 & -1.22542 & -4.1369 \\
UGC09542 & -0.917297 & -1.56885 & -0.119025 & -1.19161 & -1.37776 \\
UGC09665 & -1.13236 & -1.43426 & -1.02951 & -1.87287 & -0.529737 \\
UGC09873 & -0.798433 & -0.205172 & -0.87993 & -1.47617 & -0.474682 \\
UGC09892 & -0.936655 & -0.803293 & -1.59138 & -1.25113 & -0.994022 \\
UGC10097 & 2.17048 & -1.59701 & -1.1101 & -2.46573 & -1.84765 \\
UGC10123 & -0.78853 & -2.49877 & -0.0889768 & -0.994331 & -0.338527 \\
UGC10205 & 0.510202 & -2.4934 & 1.08186 & -2.61422 & -2.32096 \\
UGC10257 & -0.941415 & -1.24607 & -0.773566 & -1.30932 & -0.527751 \\
UGC10331 & -1.20751 & -1.30369 & -1.48972 & -1.66741 & -0.283607 \\
UGC10337 & -0.112275 & -1.78647 & 0.948613 & -3.40711 & -1.52422 \\
UGC10384 & -0.703524 & -1.34828 & -0.203839 & -1.9664 & 1.02627 \\
UGC10388 & -0.114117 & -2.28869 & -1.98973 & -0.645969 & -0.264282 \\
UGC10650 & -0.939656 & -1.16788 & -2.19008 & -1.47155 & -0.6418 \\
UGC10693 & 1.80447 & -0.818887 & -1.51373 & -1.93755 & -1.92207 \\
UGC10695 & 1.04318 & -0.807674 & -0.337421 & -1.302 & -1.72437 \\
UGC10710 & 0.0822857 & -1.38522 & -0.341625 & -1.96588 & -1.92577 \\
UGC10796 & -1.03984 & -0.550743 & -1.87062 & -0.87959 & -1.26422 \\
UGC10811 & 0.441968 & -1.40624 & -1.71174 & -1.82401 & -0.21878 \\
UGC10905 & 1.26279 & -2.96128 & -2.51871 & -1.11793 & 0.87601 \\
UGC10972 & -0.702804 & -0.4874 & 0.351719 & -1.62141 & -0.957344 \\
UGC11228 & 0.763098 & -1.89596 & -1.9166 & -1.05657 & -1.48065 \\
UGC11717 & 0.545557 & -2.48664 & -0.819629 & -2.37346 & -0.673287 \\
UGC12054 & -1.41809 & -1.24747 & -1.28259 & -1.00902 & -1.82222 \\
UGC12127 & 2.31114 & 0.814716 & 0.00545966 & 0.478953 & -0.782018 \\
UGC12185 & -0.269127 & -2.35725 & -0.41775 & -1.81632 & -0.350056 \\
UGC12274 & 0.343704 & -1.75987 & -1.38584 & -2.2932 & 0.0435052 \\
UGC12308 & -1.09323 & -1.07904 & -1.7085 & -1.34418 & -1.93308 \\
UGC12518 & -1.2169 & -1.24586 & -2.26026 & -0.472193 & -1.30585 \\
UGC12519 & -0.99504 & -1.74992 & 0.0775947 & -1.602 & -1.56866 \\
UGC12723 & -1.37479 & -1.9558 & 0.318592 & -1.50003 & -1.16429 \\
UGC12857 & -1.4187 & -1.41638 & -1.12355 & -0.794839 & -1.60526 \\
\hline
\end{tabular}
\label{tab:PCs5}
\end{center}
\end{minipage}
\end{table*}

\clearpage
\section{Multi-Gaussian expansion (MGE) models of the 238 CALIFA galaxies}
\label{A:MGEs} 

We present tables with the parameters of the multi-Gaussian expansion (MGE) models
for each galaxy from our sample of 238 (E1--Sdm) CALIFA targets, where $I_{0,j}$ is the central surface brightness (SB), $\xi_{j}'$
is the dispersion along the major $x'$-axis, and $q'$ is the flattening of the each Gaussian.
\newpage
\clearpage

\begin{table*}
\caption{Multi-Gaussian Expansion (MGE) models of the 238 (E1-Sdm) galaxies.}
\begin{tabular}{|c|ccc|ccc|ccc|ccc|}
\hline
&\multicolumn{3}{|c|}{IC0480}&\multicolumn{3}{|c|}{IC0540}&\multicolumn{3}{|c|}{IC0674}&\multicolumn{3}{|c|}{IC0944}\\ 
\hline
$j$ & $I_{0,j}$ & $\xi_{j}'$ & $q'$ & $I_{0,j}$ & $\xi_{j}'$ & $q'$ & $I_{0,j}$ & $\xi_{j}'$ & $q'$ & $I_{0,j}$ & $\xi_{j}'$ & $q'$ \\
\hline
&(L$_{\odot}$\,pc$^{-2}$)&(arcsec)&&(L$_{\odot}$\,pc$^{-2}$)&(arcsec)&&(L$_{\odot}$\,pc$^{-2}$)&(arcsec)&&(L$_{\odot}$\,pc$^{-2}$)&(arcsec) & \\ 
\hline
1 & 62.93 & 13.82 & 0.317 & 404.89 & 2.25 & 0.161 & 416.95 & 0.39 & 0.360 & 1213.04 & 1.90 & 0.664 \\
2 & 79.19 & 18.30 & 0.118 & 418.59 & 2.33 & 0.501 & 1345.07 & 1.07 & 0.743 & 272.39 & 4.43 & 0.664 \\
3 & 11.40 & 33.62 & 0.317 & 144.61 & 7.35 & 0.501 & 35.47 & 1.78 & 0.360 & 95.99 & 10.72 & 0.265 \\
4 & 32.52 & 33.62 & 0.118 & 50.26 & 11.47 & 0.501 & 707.17 & 2.55 & 0.743 & 13.65 & 14.47 & 0.664 \\
5 & ... & ... & ... & 173.52 & 15.37 & 0.161 & 201.76 & 6.79 & 0.743 & 113.91 & 17.93 & 0.265 \\
\hline
\end{tabular}
\end{table*}

\clearpage
\newpage

\begin{table*}
\caption{Multi-Gaussian Expansion (MGE) models of the 238 (E1-Sdm) galaxies.}
\begin{tabular}{|c|ccc|ccc|ccc|ccc|}
\hline
&\multicolumn{3}{|c|}{MCG-01-54-016}&\multicolumn{3}{|c|}{MCG-02-02-030}&\multicolumn{3}{|c|}{MCG-02-02-040}&\multicolumn{3}{|c|}{MCG-02-03-015}\\ 
\hline
$j$ & $I_{0,j}$ & $\xi_{j}'$ & $q'$ & $I_{0,j}$ & $\xi_{j}'$ & $q'$ & $I_{0,j}$ & $\xi_{j}'$ & $q'$ & $I_{0,j}$ & $\xi_{j}'$ & $q'$ \\
\hline
&(L$_{\odot}$\,pc$^{-2}$)&(arcsec)&&(L$_{\odot}$\,pc$^{-2}$)&(arcsec)&&(L$_{\odot}$\,pc$^{-2}$)&(arcsec)&&(L$_{\odot}$\,pc$^{-2}$)&(arcsec) & \\ 
\hline
1 & 273.14 & 0.94 & 0.191 & 741.18 & 0.85 & 0.674 & 278.37 & 1.34 & 0.167 & 2041.22 & 0.94 & 0.737 \\
2 & 123.02 & 8.03 & 0.187 & 621.77 & 2.68 & 0.674 & 96.56 & 6.28 & 0.167 & 2611.65 & 1.28 & 0.244 \\
3 & 57.84 & 22.98 & 0.187 & 19.15 & 7.47 & 0.674 & 50.90 & 13.21 & 0.508 & 536.10 & 2.24 & 0.737 \\
4 & ... & ... & ... & 289.54 & 14.97 & 0.355 & 203.01 & 16.77 & 0.167 & 56.42 & 5.35 & 0.737 \\
5 & ... & ... & ... & 19.45 & 24.93 & 0.674 & 21.96 & 23.37 & 0.508 & 47.41 & 5.44 & 0.737 \\
\hline
\end{tabular}
\end{table*}

\clearpage
\newpage

\begin{table*}
\caption{Multi-Gaussian Expansion (MGE) models of the 238 (E1-Sdm) galaxies.}
\begin{tabular}{|c|ccc|ccc|ccc|ccc|}
\hline
&\multicolumn{3}{|c|}{NGC0257}&\multicolumn{3}{|c|}{NGC0429}&\multicolumn{3}{|c|}{NGC0444}&\multicolumn{3}{|c|}{NGC0499}\\ 
\hline
$j$ & $I_{0,j}$ & $\xi_{j}'$ & $q'$ & $I_{0,j}$ & $\xi_{j}'$ & $q'$ & $I_{0,j}$ & $\xi_{j}'$ & $q'$ & $I_{0,j}$ & $\xi_{j}'$ & $q'$ \\
\hline
&(L$_{\odot}$\,pc$^{-2}$)&(arcsec)&&(L$_{\odot}$\,pc$^{-2}$)&(arcsec)&&(L$_{\odot}$\,pc$^{-2}$)&(arcsec)&&(L$_{\odot}$\,pc$^{-2}$)&(arcsec) & \\ 
\hline
1 & 1457.57 & 1.53 & 0.671 & 3246.45 & 1.47 & 0.221 & 169.66 & 1.08 & 0.371 & 4082.86 & 1.06 & 0.719 \\
2 & 490.95 & 4.22 & 0.671 & 2219.39 & 1.92 & 0.510 & 129.74 & 3.59 & 0.371 & 1836.50 & 1.98 & 0.719 \\
3 & 51.29 & 17.87 & 0.671 & 424.35 & 5.44 & 0.510 & 17.44 & 8.62 & 0.371 & 886.93 & 3.62 & 0.595 \\
4 & 102.00 & 19.28 & 0.613 & 228.43 & 9.58 & 0.221 & 55.51 & 22.52 & 0.215 & 668.59 & 4.95 & 0.719 \\
5 & 29.95 & 30.60 & 0.613 & 2.59 & 26.32 & 0.510 & 9.51 & 23.26 & 0.371 & 265.45 & 10.33 & 0.595 \\
\hline
\end{tabular}
\end{table*}

\clearpage
\newpage

\begin{table*}
\caption{Multi-Gaussian Expansion (MGE) models of the 238 (E1-Sdm) galaxies.}
\begin{tabular}{|c|ccc|ccc|ccc|ccc|}
\hline
&\multicolumn{3}{|c|}{NGC0810}&\multicolumn{3}{|c|}{NGC0932}&\multicolumn{3}{|c|}{NGC1056}&\multicolumn{3}{|c|}{NGC1060}\\ 
\hline
$j$ & $I_{0,j}$ & $\xi_{j}'$ & $q'$ & $I_{0,j}$ & $\xi_{j}'$ & $q'$ & $I_{0,j}$ & $\xi_{j}'$ & $q'$ & $I_{0,j}$ & $\xi_{j}'$ & $q'$ \\
\hline
&(L$_{\odot}$\,pc$^{-2}$)&(arcsec)&&(L$_{\odot}$\,pc$^{-2}$)&(arcsec)&&(L$_{\odot}$\,pc$^{-2}$)&(arcsec)&&(L$_{\odot}$\,pc$^{-2}$)&(arcsec) & \\ 
\hline
1 & 1175.94 & 2.27 & 0.488 & 1703.45 & 1.11 & 0.872 & 3709.09 & 0.66 & 0.351 & 2117.10 & 1.06 & 1.000 \\
2 & 356.12 & 4.72 & 0.488 & 1857.41 & 1.61 & 0.972 & 462.74 & 3.07 & 0.958 & 2572.95 & 1.99 & 0.749 \\
3 & 37.04 & 5.26 & 0.488 & 103.22 & 2.02 & 0.872 & 3033.90 & 3.26 & 0.351 & 1149.40 & 3.86 & 0.749 \\
4 & 228.11 & 7.29 & 0.665 & 660.01 & 3.11 & 0.972 & 547.61 & 6.43 & 0.351 & 154.47 & 6.23 & 1.000 \\
5 & 85.13 & 14.97 & 0.665 & 412.26 & 5.92 & 0.972 & 87.30 & 7.52 & 0.958 & 660.57 & 7.02 & 0.749 \\
\hline
\end{tabular}
\end{table*}

\clearpage
\newpage

\begin{table*}
\caption{Multi-Gaussian Expansion (MGE) models of the 238 (E1-Sdm) galaxies.}
\begin{tabular}{|c|ccc|ccc|ccc|ccc|}
\hline
&\multicolumn{3}{|c|}{NGC2553}&\multicolumn{3}{|c|}{NGC2554}&\multicolumn{3}{|c|}{NGC2592}&\multicolumn{3}{|c|}{NGC2604}\\ 
\hline
$j$ & $I_{0,j}$ & $\xi_{j}'$ & $q'$ & $I_{0,j}$ & $\xi_{j}'$ & $q'$ & $I_{0,j}$ & $\xi_{j}'$ & $q'$ & $I_{0,j}$ & $\xi_{j}'$ & $q'$ \\
\hline
&(L$_{\odot}$\,pc$^{-2}$)&(arcsec)&&(L$_{\odot}$\,pc$^{-2}$)&(arcsec)&&(L$_{\odot}$\,pc$^{-2}$)&(arcsec)&&(L$_{\odot}$\,pc$^{-2}$)&(arcsec) & \\ 
\hline
1 & 316.92 & 0.62 & 0.933 & 5075.34 & 1.34 & 0.996 & 4289.02 & 1.18 & 0.897 & 281.08 & 1.17 & 1.000 \\
2 & 1767.05 & 1.12 & 0.933 & 479.98 & 1.75 & 0.737 & 2625.76 & 2.07 & 0.897 & 145.08 & 2.95 & 1.000 \\
3 & 902.56 & 2.45 & 0.487 & 910.14 & 3.36 & 0.996 & 62.81 & 3.87 & 0.897 & 24.39 & 6.47 & 1.000 \\
4 & 449.85 & 4.01 & 0.933 & 297.51 & 6.34 & 0.996 & 489.21 & 4.66 & 0.625 & 56.11 & 19.23 & 1.000 \\
5 & 61.04 & 9.69 & 0.933 & 87.88 & 11.18 & 0.737 & 132.08 & 5.55 & 0.897 & 24.99 & 33.55 & 0.930 \\
\hline
\end{tabular}
\end{table*}

\clearpage
\newpage

\begin{table*}
\caption{Multi-Gaussian Expansion (MGE) models of the 238 (E1-Sdm) galaxies.}
\begin{tabular}{|c|ccc|ccc|ccc|ccc|}
\hline
&\multicolumn{3}{|c|}{NGC3811}&\multicolumn{3}{|c|}{NGC3815}&\multicolumn{3}{|c|}{NGC3994}&\multicolumn{3}{|c|}{NGC4003}\\ 
\hline
$j$ & $I_{0,j}$ & $\xi_{j}'$ & $q'$ & $I_{0,j}$ & $\xi_{j}'$ & $q'$ & $I_{0,j}$ & $\xi_{j}'$ & $q'$ & $I_{0,j}$ & $\xi_{j}'$ & $q'$ \\
\hline
&(L$_{\odot}$\,pc$^{-2}$)&(arcsec)&&(L$_{\odot}$\,pc$^{-2}$)&(arcsec)&&(L$_{\odot}$\,pc$^{-2}$)&(arcsec)&&(L$_{\odot}$\,pc$^{-2}$)&(arcsec) & \\ 
\hline
1 & 710.09 & 0.93 & 0.932 & 1138.99 & 1.40 & 0.529 & 4748.50 & 1.31 & 0.602 & 3821.92 & 0.42 & 0.947 \\
2 & 4045.09 & 1.13 & 0.675 & 244.46 & 5.35 & 0.529 & 671.72 & 4.04 & 0.602 & 339.47 & 1.65 & 0.947 \\
3 & 513.49 & 3.14 & 0.932 & 115.61 & 6.90 & 0.324 & 571.59 & 7.53 & 0.379 & 650.42 & 2.17 & 0.444 \\
4 & 121.30 & 14.24 & 0.675 & 132.88 & 9.41 & 0.529 & 16.83 & 9.43 & 0.379 & 343.99 & 2.88 & 0.947 \\
5 & 63.01 & 16.21 & 0.932 & 117.02 & 15.95 & 0.324 & 163.56 & 9.76 & 0.602 & 170.22 & 5.75 & 0.444 \\
\hline
\end{tabular}
\end{table*}

\clearpage
\newpage

\begin{table*}
\caption{Multi-Gaussian Expansion (MGE) models of the 238 (E1-Sdm) galaxies.}
\begin{tabular}{|c|ccc|ccc|ccc|ccc|}
\hline
&\multicolumn{3}{|c|}{NGC5056}&\multicolumn{3}{|c|}{NGC5218}&\multicolumn{3}{|c|}{NGC5378}&\multicolumn{3}{|c|}{NGC5480}\\ 
\hline
$j$ & $I_{0,j}$ & $\xi_{j}'$ & $q'$ & $I_{0,j}$ & $\xi_{j}'$ & $q'$ & $I_{0,j}$ & $\xi_{j}'$ & $q'$ & $I_{0,j}$ & $\xi_{j}'$ & $q'$ \\
\hline
&(L$_{\odot}$\,pc$^{-2}$)&(arcsec)&&(L$_{\odot}$\,pc$^{-2}$)&(arcsec)&&(L$_{\odot}$\,pc$^{-2}$)&(arcsec)&&(L$_{\odot}$\,pc$^{-2}$)&(arcsec) & \\ 
\hline
1 & 902.38 & 1.19 & 1.000 & 160.93 & 4.32 & 0.551 & 1304.94 & 1.31 & 0.745 & 3071.30 & 0.41 & 0.694 \\
2 & 198.84 & 5.28 & 1.000 & 528.92 & 9.22 & 0.551 & 474.77 & 2.24 & 0.745 & 1483.65 & 1.17 & 1.000 \\
3 & 101.77 & 11.79 & 0.455 & 82.36 & 21.83 & 0.551 & 375.79 & 5.46 & 0.767 & 434.51 & 1.92 & 1.000 \\
4 & 11.17 & 20.67 & 1.000 & 32.89 & 27.79 & 0.820 & 22.28 & 17.70 & 0.767 & 204.72 & 3.72 & 1.000 \\
5 & 48.10 & 26.37 & 0.455 & ... & ... & ... & 54.67 & 31.78 & 0.745 & 199.21 & 18.24 & 0.694 \\
\hline
\end{tabular}
\end{table*}

\clearpage
\newpage

\begin{table*}
\caption{Multi-Gaussian Expansion (MGE) models of the 238 (E1-Sdm) galaxies.}
\begin{tabular}{|c|ccc|ccc|ccc|ccc|}
\hline
&\multicolumn{3}{|c|}{NGC5908}&\multicolumn{3}{|c|}{NGC5971}&\multicolumn{3}{|c|}{NGC5980}&\multicolumn{3}{|c|}{NGC5987}\\ 
\hline
$j$ & $I_{0,j}$ & $\xi_{j}'$ & $q'$ & $I_{0,j}$ & $\xi_{j}'$ & $q'$ & $I_{0,j}$ & $\xi_{j}'$ & $q'$ & $I_{0,j}$ & $\xi_{j}'$ & $q'$ \\
\hline
&(L$_{\odot}$\,pc$^{-2}$)&(arcsec)&&(L$_{\odot}$\,pc$^{-2}$)&(arcsec)&&(L$_{\odot}$\,pc$^{-2}$)&(arcsec)&&(L$_{\odot}$\,pc$^{-2}$)&(arcsec) & \\ 
\hline
1 & 1691.19 & 1.47 & 0.167 & 827.69 & 1.02 & 0.739 & 457.15 & 1.58 & 0.619 & 4154.75 & 1.23 & 0.695 \\
2 & 943.11 & 1.63 & 0.697 & 171.41 & 1.65 & 0.379 & 662.30 & 3.03 & 0.619 & 1205.81 & 2.88 & 0.695 \\
3 & 275.39 & 4.57 & 0.697 & 483.81 & 1.68 & 0.739 & 271.29 & 8.76 & 0.329 & 454.34 & 7.33 & 0.259 \\
4 & 702.55 & 10.18 & 0.167 & 212.07 & 3.21 & 0.739 & 79.96 & 11.90 & 0.619 & 197.94 & 8.76 & 0.695 \\
5 & 104.89 & 17.54 & 0.697 & 150.65 & 6.24 & 0.379 & 230.72 & 20.09 & 0.329 & 125.08 & 21.23 & 0.259 \\
\hline
\end{tabular}
\end{table*}

\clearpage
\newpage

\begin{table*}
\caption{Multi-Gaussian Expansion (MGE) models of the 238 (E1-Sdm) galaxies.}
\begin{tabular}{|c|ccc|ccc|ccc|ccc|}
\hline
&\multicolumn{3}{|c|}{NGC6186}&\multicolumn{3}{|c|}{NGC6278}&\multicolumn{3}{|c|}{NGC6301}&\multicolumn{3}{|c|}{NGC6310}\\ 
\hline
$j$ & $I_{0,j}$ & $\xi_{j}'$ & $q'$ & $I_{0,j}$ & $\xi_{j}'$ & $q'$ & $I_{0,j}$ & $\xi_{j}'$ & $q'$ & $I_{0,j}$ & $\xi_{j}'$ & $q'$ \\
\hline
&(L$_{\odot}$\,pc$^{-2}$)&(arcsec)&&(L$_{\odot}$\,pc$^{-2}$)&(arcsec)&&(L$_{\odot}$\,pc$^{-2}$)&(arcsec)&&(L$_{\odot}$\,pc$^{-2}$)&(arcsec) & \\ 
\hline
1 & 5530.20 & 0.67 & 0.361 & 3313.01 & 1.30 & 0.884 & 173.80 & 1.52 & 0.551 & 724.83 & 1.48 & 0.773 \\
2 & 1175.02 & 2.79 & 0.888 & 640.40 & 2.26 & 0.550 & 90.21 & 3.50 & 0.685 & 60.18 & 2.63 & 0.248 \\
3 & 117.20 & 7.72 & 0.361 & 1576.68 & 2.50 & 0.884 & 30.85 & 10.42 & 0.685 & 301.58 & 3.67 & 0.773 \\
4 & 234.81 & 17.12 & 0.361 & 677.42 & 6.79 & 0.550 & 73.95 & 22.30 & 0.551 & 243.98 & 19.65 & 0.248 \\
5 & 32.20 & 25.63 & 0.888 & 151.71 & 23.19 & 0.550 & 15.48 & 30.59 & 0.685 & 27.70 & 30.58 & 0.248 \\
\hline
\end{tabular}
\end{table*}

\clearpage
\newpage

\begin{table*}
\caption{Multi-Gaussian Expansion (MGE) models of the 238 (E1-Sdm) galaxies.}
\begin{tabular}{|c|ccc|ccc|ccc|ccc|}
\hline
&\multicolumn{3}{|c|}{NGC7321}&\multicolumn{3}{|c|}{NGC7364}&\multicolumn{3}{|c|}{NGC7466}&\multicolumn{3}{|c|}{NGC7489}\\ 
\hline
$j$ & $I_{0,j}$ & $\xi_{j}'$ & $q'$ & $I_{0,j}$ & $\xi_{j}'$ & $q'$ & $I_{0,j}$ & $\xi_{j}'$ & $q'$ & $I_{0,j}$ & $\xi_{j}'$ & $q'$ \\
\hline
&(L$_{\odot}$\,pc$^{-2}$)&(arcsec)&&(L$_{\odot}$\,pc$^{-2}$)&(arcsec)&&(L$_{\odot}$\,pc$^{-2}$)&(arcsec)&&(L$_{\odot}$\,pc$^{-2}$)&(arcsec) & \\ 
\hline
1 & 2182.33 & 0.39 & 0.558 & 340.73 & 1.17 & 1.000 & 1193.24 & 1.36 & 0.825 & 470.54 & 1.07 & 0.994 \\
2 & 680.69 & 1.80 & 0.558 & 2724.63 & 1.33 & 0.658 & 163.99 & 5.29 & 0.825 & 4.56 & 2.83 & 0.390 \\
3 & 356.47 & 2.32 & 0.963 & 217.75 & 2.87 & 1.000 & 57.69 & 14.98 & 0.392 & 43.89 & 4.25 & 0.390 \\
4 & 225.77 & 7.17 & 0.963 & 276.85 & 3.76 & 0.658 & 48.43 & 23.03 & 0.392 & 95.90 & 8.64 & 0.994 \\
5 & 86.35 & 17.77 & 0.558 & 582.00 & 5.76 & 0.658 & ... & ... & ... & 98.84 & 13.81 & 0.390 \\
\hline
\end{tabular}
\end{table*}

\clearpage
\newpage

\begin{table*}
\caption{Multi-Gaussian Expansion (MGE) models of the 238 (E1-Sdm) galaxies.}
\begin{tabular}{|c|ccc|ccc|ccc|ccc|}
\hline
&\multicolumn{3}{|c|}{NGC7671}&\multicolumn{3}{|c|}{NGC7683}&\multicolumn{3}{|c|}{NGC7711}&\multicolumn{3}{|c|}{NGC7716}\\ 
\hline
$j$ & $I_{0,j}$ & $\xi_{j}'$ & $q'$ & $I_{0,j}$ & $\xi_{j}'$ & $q'$ & $I_{0,j}$ & $\xi_{j}'$ & $q'$ & $I_{0,j}$ & $\xi_{j}'$ & $q'$ \\
\hline
&(L$_{\odot}$\,pc$^{-2}$)&(arcsec)&&(L$_{\odot}$\,pc$^{-2}$)&(arcsec)&&(L$_{\odot}$\,pc$^{-2}$)&(arcsec)&&(L$_{\odot}$\,pc$^{-2}$)&(arcsec) & \\ 
\hline
1 & 3621.75 & 1.19 & 0.559 & 2738.97 & 1.28 & 0.735 & 3167.12 & 1.01 & 0.776 & 2103.31 & 1.13 & 0.863 \\
2 & 2407.55 & 1.77 & 0.955 & 1366.39 & 2.44 & 0.735 & 303.84 & 1.13 & 0.431 & 1301.06 & 1.47 & 0.574 \\
3 & 619.89 & 3.22 & 0.955 & 365.18 & 3.86 & 0.735 & 2464.53 & 1.87 & 0.776 & 890.10 & 2.29 & 0.574 \\
4 & 25.16 & 7.30 & 0.955 & 114.56 & 5.19 & 0.471 & 748.74 & 3.17 & 0.776 & 1167.56 & 2.60 & 0.863 \\
5 & 201.15 & 7.95 & 0.559 & 449.36 & 6.80 & 0.735 & 400.63 & 6.27 & 0.776 & 138.33 & 5.83 & 0.863 \\
\hline
\end{tabular}
\end{table*}

\clearpage
\newpage

\begin{table*}
\caption{Multi-Gaussian Expansion (MGE) models of the 238 (E1-Sdm) galaxies.}
\begin{tabular}{|c|ccc|ccc|ccc|ccc|}
\hline
&\multicolumn{3}{|c|}{UGC01057}&\multicolumn{3}{|c|}{UGC01271}&\multicolumn{3}{|c|}{UGC02222}&\multicolumn{3}{|c|}{UGC02229}\\ 
\hline
$j$ & $I_{0,j}$ & $\xi_{j}'$ & $q'$ & $I_{0,j}$ & $\xi_{j}'$ & $q'$ & $I_{0,j}$ & $\xi_{j}'$ & $q'$ & $I_{0,j}$ & $\xi_{j}'$ & $q'$ \\
\hline
&(L$_{\odot}$\,pc$^{-2}$)&(arcsec)&&(L$_{\odot}$\,pc$^{-2}$)&(arcsec)&&(L$_{\odot}$\,pc$^{-2}$)&(arcsec)&&(L$_{\odot}$\,pc$^{-2}$)&(arcsec) & \\ 
\hline
1 & 636.26 & 1.30 & 0.400 & 3650.03 & 0.53 & 0.507 & 206.86 & 0.39 & 0.716 & 1035.69 & 1.63 & 0.524 \\
2 & 126.65 & 8.12 & 0.400 & 1699.29 & 1.23 & 0.889 & 2988.24 & 1.32 & 0.716 & 392.45 & 3.73 & 0.680 \\
3 & 101.35 & 8.26 & 0.223 & 522.71 & 2.31 & 0.507 & 588.20 & 3.01 & 0.716 & 163.35 & 8.86 & 0.524 \\
4 & 49.05 & 14.48 & 0.400 & 647.88 & 3.37 & 0.889 & 274.27 & 7.07 & 0.404 & 36.76 & 25.35 & 0.524 \\
5 & 39.91 & 22.27 & 0.223 & 123.73 & 8.49 & 0.889 & 118.59 & 7.76 & 0.716 & ... & ... & ... \\
\hline
\end{tabular}
\end{table*}

\clearpage
\newpage

\begin{table*}
\caption{Multi-Gaussian Expansion (MGE) models of the 238 (E1-Sdm) galaxies.}
\begin{tabular}{|c|ccc|ccc|ccc|ccc|}
\hline
&\multicolumn{3}{|c|}{UGC05598}&\multicolumn{3}{|c|}{UGC05771}&\multicolumn{3}{|c|}{UGC05990}&\multicolumn{3}{|c|}{UGC06036}\\ 
\hline
$j$ & $I_{0,j}$ & $\xi_{j}'$ & $q'$ & $I_{0,j}$ & $\xi_{j}'$ & $q'$ & $I_{0,j}$ & $\xi_{j}'$ & $q'$ & $I_{0,j}$ & $\xi_{j}'$ & $q'$ \\
\hline
&(L$_{\odot}$\,pc$^{-2}$)&(arcsec)&&(L$_{\odot}$\,pc$^{-2}$)&(arcsec)&&(L$_{\odot}$\,pc$^{-2}$)&(arcsec)&&(L$_{\odot}$\,pc$^{-2}$)&(arcsec) & \\ 
\hline
1 & 274.73 & 1.93 & 0.539 & 624.99 & 0.72 & 0.873 & 309.85 & 5.37 & 0.207 & 1151.88 & 0.94 & 0.790 \\
2 & 46.27 & 2.03 & 0.242 & 1384.77 & 0.77 & 0.681 & 140.91 & 7.48 & 0.342 & 1145.73 & 1.87 & 0.790 \\
3 & 34.94 & 9.39 & 0.539 & 1846.83 & 1.27 & 0.873 & 52.44 & 16.93 & 0.207 & 328.17 & 3.50 & 0.790 \\
4 & 124.95 & 12.92 & 0.242 & 392.76 & 2.90 & 0.873 & 15.41 & 23.07 & 0.342 & 62.87 & 6.24 & 0.790 \\
5 & 32.77 & 25.08 & 0.242 & 207.18 & 3.68 & 0.681 & 14.17 & 23.07 & 0.207 & 106.35 & 10.64 & 0.171 \\
\hline
\end{tabular}
\end{table*}

\clearpage
\newpage

\begin{table*}
\caption{Multi-Gaussian Expansion (MGE) models of the 238 (E1-Sdm) galaxies.}
\begin{tabular}{|c|ccc|ccc|ccc|ccc|}
\hline
&\multicolumn{3}{|c|}{UGC09665}&\multicolumn{3}{|c|}{UGC09873}&\multicolumn{3}{|c|}{UGC09892}&\multicolumn{3}{|c|}{UGC10097}\\ 
\hline
$j$ & $I_{0,j}$ & $\xi_{j}'$ & $q'$ & $I_{0,j}$ & $\xi_{j}'$ & $q'$ & $I_{0,j}$ & $\xi_{j}'$ & $q'$ & $I_{0,j}$ & $\xi_{j}'$ & $q'$ \\
\hline
&(L$_{\odot}$\,pc$^{-2}$)&(arcsec)&&(L$_{\odot}$\,pc$^{-2}$)&(arcsec)&&(L$_{\odot}$\,pc$^{-2}$)&(arcsec)&&(L$_{\odot}$\,pc$^{-2}$)&(arcsec) & \\ 
\hline
1 & 378.74 & 2.24 & 0.481 & 496.13 & 0.39 & 0.917 & 739.25 & 0.39 & 0.231 & 4872.94 & 0.94 & 0.927 \\
2 & 56.10 & 10.74 & 0.481 & 187.98 & 0.88 & 0.917 & 174.38 & 1.14 & 0.625 & 1820.24 & 1.93 & 0.927 \\
3 & 93.41 & 10.76 & 0.178 & 345.34 & 1.50 & 0.206 & 485.77 & 1.76 & 0.231 & 412.54 & 2.97 & 0.927 \\
4 & 190.68 & 20.95 & 0.178 & 44.55 & 2.29 & 0.917 & 107.45 & 4.61 & 0.625 & 81.94 & 3.99 & 0.624 \\
5 & 2.12 & 24.97 & 0.481 & 43.02 & 3.46 & 0.917 & 6.81 & 13.02 & 0.625 & 114.81 & 5.89 & 0.927 \\
\hline
\end{tabular}
\end{table*}

\clearpage
\newpage

\begin{table*}
\caption{Multi-Gaussian Expansion (MGE) models of the 238 (E1-Sdm) galaxies.}
\begin{tabular}{|c|ccc|ccc|ccc|ccc|}
\hline
&\multicolumn{3}{|c|}{UGC10811}&\multicolumn{3}{|c|}{UGC10905}&\multicolumn{3}{|c|}{UGC10972}&\multicolumn{3}{|c|}{UGC11228}\\ 
\hline
$j$ & $I_{0,j}$ & $\xi_{j}'$ & $q'$ & $I_{0,j}$ & $\xi_{j}'$ & $q'$ & $I_{0,j}$ & $\xi_{j}'$ & $q'$ & $I_{0,j}$ & $\xi_{j}'$ & $q'$ \\
\hline
&(L$_{\odot}$\,pc$^{-2}$)&(arcsec)&&(L$_{\odot}$\,pc$^{-2}$)&(arcsec)&&(L$_{\odot}$\,pc$^{-2}$)&(arcsec)&&(L$_{\odot}$\,pc$^{-2}$)&(arcsec) & \\ 
\hline
1 & 1263.04 & 0.91 & 0.336 & 2354.99 & 1.96 & 0.617 & 737.84 & 0.71 & 0.882 & 2682.61 & 0.93 & 0.703 \\
2 & 621.93 & 1.99 & 1.000 & 49.16 & 4.49 & 0.357 & 466.79 & 1.15 & 0.233 & 1229.42 & 1.83 & 0.703 \\
3 & 29.79 & 5.46 & 1.000 & 472.88 & 4.59 & 0.617 & 237.91 & 2.06 & 0.882 & 841.74 & 2.67 & 0.703 \\
4 & 95.83 & 16.29 & 0.336 & 99.84 & 12.81 & 0.617 & 181.68 & 16.10 & 0.233 & 279.28 & 5.97 & 0.512 \\
5 & 19.42 & 23.00 & 0.336 & 32.41 & 15.36 & 0.357 & 47.78 & 33.61 & 0.233 & 77.57 & 10.84 & 0.703 \\
\hline
\end{tabular}
\end{table*}

\begin{table*}
\begin{tabular}{|c|ccc|ccc|}
\hline
\multicolumn{1}{|c|}{}&\multicolumn{3}{|c|}{UGC12723}&\multicolumn{3}{|c|}{UGC12857}\\ 
\hline
j & $I_{0,j}$ & $\xi_{j}'$ & $q'$ & $I_{0,j}$ & $\xi_{j}'$ & $q'$ \\
\hline
& (L$_{\odot}$\,pc$^{-2}$)&(arcsec)&& (L$_{\odot}$\,pc$^{-2}$)&(arcsec) &\\ 
\hline
1 & 152.03 & 13.34 & 0.147 & 50.74 & 1.53 & 0.386 \\
2 & 23.23 & 19.60 & 0.234 & 39.82 & 4.18 & 0.386 \\
3 & ... & ... & ... & 105.85 & 6.15 & 0.386 \\
4 & ... & ... & ... & 102.44 & 11.90 & 0.146 \\
5 & ... & ... & ... & 36.15 & 13.13 & 0.386 \\
6 & ... & ... & ... & 33.82 & 21.48 & 0.386 \\
7 & ... & ... & ... & 125.91 & 24.07 & 0.146 \\
\hline
\end{tabular}
\end{table*}

\clearpage
\newpage

\section{Circular velocity curves of the 238 CALIFA galaxies (Online Material) }
\label{A:CVCs}

In Table \ref{tab:CVC1}$-$Table \ref{tab:CVC60}, we present the circular velocity curves as a function of the sampling radius $R$, normalized on the effective radius $R_{\mathrm{e}}$ of the 238 CALIFA (E1--Sdm) galaxies from JAM-MCMC method together with their uncertaities, calculated from the 75th and 25th percentile of the distribution of the dynamical mass-to-light ratio $(M/L)_{\mathrm{dyn}}$ (see Section \ref{SS:dyn}).  

\begin{table*}
\caption{Circular velocity curves of the 238 (E1--Sdm) CALIFA galaxies with uncertainties corresponding to 75th and 25th percentiles of the data .}
\begin{tabular}{|c|ccc|ccc|ccc|ccc|}
\hline
&\multicolumn{3}{|c|}{IC0480}&\multicolumn{3}{|c|}{IC0540}&\multicolumn{3}{|c|}{IC0674}&\multicolumn{3}{|c|}{IC0944}\\ 
\hline
$R/R_\mathrm{e}$ & $V_c$ & $\Delta V_c^{75th}$ & $\Delta V_c^{25th}$ & $V_c$ & $\Delta V_c^{75th}$ & $\Delta V_c^{25th}$ & $V_c$ & $\Delta V_c^{75th}$ & $\Delta V_c^{25th}$ & $V_c$ & $\Delta V_c^{75th}$ & $\Delta V_c^{25th}$ \\
\hline
0.05 & 12.51 & 0.03 & 0.04 & 29.17 & 0.09 & 0.08 & 94.59 & 0.17 & 0.16 & 116.85 & 0.18 & 0.15 \\
0.054 & 13.31 & 0.04 & 0.04 & 31.04 & 0.09 & 0.08 & 100.18 & 0.18 & 0.17 & 124.58 & 0.19 & 0.16 \\
0.057 & 14.18 & 0.04 & 0.04 & 33.04 & 0.10 & 0.09 & 105.97 & 0.19 & 0.18 & 132.73 & 0.21 & 0.17 \\
0.062 & 15.12 & 0.04 & 0.05 & 35.19 & 0.11 & 0.09 & 111.97 & 0.20 & 0.19 & 141.30 & 0.22 & 0.18 \\
0.066 & 16.14 & 0.05 & 0.05 & 37.49 & 0.12 & 0.10 & 118.16 & 0.22 & 0.20 & 150.26 & 0.23 & 0.19 \\
0.071 & 17.24 & 0.05 & 0.05 & 39.95 & 0.12 & 0.11 & 124.56 & 0.23 & 0.21 & 159.62 & 0.25 & 0.20 \\
\hline
\end{tabular}
\label{tab:CVC1}
\end{table*}

\clearpage
\newpage

\begin{table*}
\caption{Circular velocity curves of the 238 (E1--Sdm) CALIFA galaxies with 75th and 25th percentile uncertainties.}
\begin{tabular}{|c|ccc|ccc|ccc|ccc|}
\hline
&\multicolumn{3}{|c|}{IC1079}&\multicolumn{3}{|c|}{IC1151}&\multicolumn{3}{|c|}{IC1256}&\multicolumn{3}{|c|}{IC1528}\\ 
\hline
$R/R_\mathrm{e}$ & $V_c$ & $\Delta V_c^{75th}$ & $\Delta V_c^{25th}$ & $V_c$ & $\Delta V_c^{75th}$ & $\Delta V_c^{25th}$ & $V_c$ & $\Delta V_c^{75th}$ & $\Delta V_c^{25th}$ & $V_c$ & $\Delta V_c^{75th}$ & $\Delta V_c^{25th}$ \\
\hline
0.05 & 230.73 & 0.45 & 0.41 & 42.20 & 0.13 & 0.12 & 38.08 & 0.13 & 0.12 & 27.17 & 0.08 & 0.05 \\
0.054 & 240.10 & 0.47 & 0.43 & 44.49 & 0.14 & 0.13 & 40.54 & 0.14 & 0.13 & 28.99 & 0.08 & 0.06 \\
0.057 & 248.96 & 0.49 & 0.45 & 46.81 & 0.14 & 0.14 & 43.13 & 0.15 & 0.14 & 30.94 & 0.09 & 0.06 \\
0.062 & 257.16 & 0.50 & 0.46 & 49.16 & 0.15 & 0.15 & 45.85 & 0.15 & 0.15 & 33.01 & 0.10 & 0.07 \\
0.066 & 264.57 & 0.52 & 0.47 & 51.51 & 0.16 & 0.15 & 48.70 & 0.16 & 0.16 & 35.21 & 0.10 & 0.07 \\
0.071 & 271.06 & 0.53 & 0.49 & 53.84 & 0.17 & 0.16 & 51.67 & 0.17 & 0.17 & 37.55 & 0.11 & 0.07 \\
\hline
\end{tabular}
\end{table*}

\clearpage
\newpage

\begin{table*}
\caption{Circular velocity curves of the 238 (E1--Sdm) CALIFA galaxies with 75th and 25th percentile uncertainties.}
\begin{tabular}{|c|ccc|ccc|ccc|ccc|}
\hline
&\multicolumn{3}{|c|}{IC1652}&\multicolumn{3}{|c|}{IC1755}&\multicolumn{3}{|c|}{IC2101}&\multicolumn{3}{|c|}{IC2247}\\ 
\hline
$R/R_\mathrm{e}$ & $V_c$ & $\Delta V_c^{75th}$ & $\Delta V_c^{25th}$ & $V_c$ & $\Delta V_c^{75th}$ & $\Delta V_c^{25th}$ & $V_c$ & $\Delta V_c^{75th}$ & $\Delta V_c^{25th}$ & $V_c$ & $\Delta V_c^{75th}$ & $\Delta V_c^{25th}$ \\
\hline
0.05 & 125.06 & 0.27 & 0.25 & 93.51 & 0.14 & 0.14 & 18.42 & 0.07 & 0.06 & 125.15 & 0.28 & 0.28 \\
0.054 & 129.23 & 0.28 & 0.26 & 99.72 & 0.15 & 0.14 & 19.66 & 0.07 & 0.06 & 126.20 & 0.29 & 0.29 \\
0.057 & 132.97 & 0.29 & 0.26 & 106.27 & 0.16 & 0.15 & 21.01 & 0.08 & 0.07 & 127.68 & 0.29 & 0.29 \\
0.062 & 136.22 & 0.30 & 0.27 & 113.16 & 0.17 & 0.16 & 22.45 & 0.08 & 0.07 & 129.68 & 0.29 & 0.30 \\
0.066 & 138.91 & 0.30 & 0.28 & 120.39 & 0.18 & 0.17 & 23.99 & 0.09 & 0.08 & 132.23 & 0.30 & 0.30 \\
0.071 & 140.99 & 0.31 & 0.28 & 127.95 & 0.19 & 0.19 & 25.65 & 0.10 & 0.09 & 135.29 & 0.31 & 0.31 \\
\hline
\end{tabular}
\end{table*}

\clearpage
\newpage

\begin{table*}
\caption{Circular velocity curves of the 238 (E1--Sdm) CALIFA galaxies with 75th and 25th percentile uncertainties.}
\begin{tabular}{|c|ccc|ccc|ccc|ccc|}
\hline
&\multicolumn{3}{|c|}{IC2487}&\multicolumn{3}{|c|}{IC4566}&\multicolumn{3}{|c|}{IC5309}&\multicolumn{3}{|c|}{IC5376}\\ 
\hline
$R/R_\mathrm{e}$ & $V_c$ & $\Delta V_c^{75th}$ & $\Delta V_c^{25th}$ & $V_c$ & $\Delta V_c^{75th}$ & $\Delta V_c^{25th}$ & $V_c$ & $\Delta V_c^{75th}$ & $\Delta V_c^{25th}$ & $V_c$ & $\Delta V_c^{75th}$ & $\Delta V_c^{25th}$ \\
\hline
0.05 & 72.62 & 0.13 & 0.20 & 97.03 & 0.26 & 0.14 & 46.98 & 0.26 & 0.21 & 79.91 & 0.21 & 0.17 \\
0.054 & 75.80 & 0.13 & 0.21 & 103.26 & 0.28 & 0.15 & 49.89 & 0.27 & 0.22 & 85.20 & 0.22 & 0.19 \\
0.057 & 78.85 & 0.14 & 0.22 & 109.79 & 0.29 & 0.16 & 52.93 & 0.29 & 0.23 & 90.79 & 0.24 & 0.20 \\
0.062 & 81.71 & 0.14 & 0.23 & 116.60 & 0.31 & 0.17 & 56.08 & 0.31 & 0.25 & 96.67 & 0.25 & 0.21 \\
0.066 & 84.33 & 0.15 & 0.23 & 123.67 & 0.33 & 0.18 & 59.33 & 0.33 & 0.26 & 102.83 & 0.27 & 0.22 \\
0.071 & 86.64 & 0.15 & 0.24 & 130.98 & 0.35 & 0.19 & 62.67 & 0.34 & 0.28 & 109.28 & 0.28 & 0.24 \\
\hline
\end{tabular}
\end{table*}

\clearpage
\newpage

\begin{table*}
\caption{Circular velocity curves of the 238 (E1--Sdm) CALIFA galaxies with 75th and 25th percentile uncertainties.}
\begin{tabular}{|c|ccc|ccc|ccc|ccc|}
\hline
&\multicolumn{3}{|c|}{MCG-01-54-016}&\multicolumn{3}{|c|}{MCG-02-02-030}&\multicolumn{3}{|c|}{MCG-02-02-040}&\multicolumn{3}{|c|}{MCG-02-03-015}\\ 
\hline
$R/R_\mathrm{e}$ & $V_c$ & $\Delta V_c^{75th}$ & $\Delta V_c^{25th}$ & $V_c$ & $\Delta V_c^{75th}$ & $\Delta V_c^{25th}$ & $V_c$ & $\Delta V_c^{75th}$ & $\Delta V_c^{25th}$ & $V_c$ & $\Delta V_c^{75th}$ & $\Delta V_c^{25th}$ \\
\hline
0.05 & 41.83 & 0.33 & 0.36 & 64.64 & 0.13 & 0.11 & 30.89 & 0.14 & 0.11 & 120.34 & 0.26 & 0.35 \\
0.054 & 43.21 & 0.34 & 0.37 & 67.89 & 0.13 & 0.12 & 32.65 & 0.15 & 0.11 & 128.02 & 0.27 & 0.38 \\
0.057 & 44.43 & 0.35 & 0.38 & 71.14 & 0.14 & 0.13 & 34.47 & 0.15 & 0.12 & 136.04 & 0.29 & 0.40 \\
0.062 & 45.48 & 0.36 & 0.39 & 74.36 & 0.15 & 0.13 & 36.33 & 0.16 & 0.13 & 144.40 & 0.31 & 0.42 \\
0.066 & 46.32 & 0.36 & 0.40 & 77.51 & 0.15 & 0.14 & 38.22 & 0.17 & 0.13 & 153.06 & 0.33 & 0.45 \\
0.071 & 46.94 & 0.37 & 0.40 & 80.57 & 0.16 & 0.14 & 40.12 & 0.18 & 0.14 & 161.99 & 0.34 & 0.47 \\
\hline
\end{tabular}
\end{table*}

\clearpage
\newpage

\begin{table*}
\caption{Circular velocity curves of the 238 (E1--Sdm) CALIFA galaxies with 75th and 25th percentile uncertainties.}
\begin{tabular}{|c|ccc|ccc|ccc|ccc|}
\hline
&\multicolumn{3}{|c|}{MCG-02-51-004}&\multicolumn{3}{|c|}{NGC0001}&\multicolumn{3}{|c|}{NGC0023}&\multicolumn{3}{|c|}{NGC0155}\\ 
\hline
$R/R_\mathrm{e}$ & $V_c$ & $\Delta V_c^{75th}$ & $\Delta V_c^{25th}$ & $V_c$ & $\Delta V_c^{75th}$ & $\Delta V_c^{25th}$ & $V_c$ & $\Delta V_c^{75th}$ & $\Delta V_c^{25th}$ & $V_c$ & $\Delta V_c^{75th}$ & $\Delta V_c^{25th}$ \\
\hline
0.05 & 78.16 & 0.11 & 0.12 & 88.85 & 0.28 & 0.20 & 112.32 & 1.06 & 0.55 & 134.64 & 0.27 & 0.23 \\
0.054 & 83.28 & 0.12 & 0.13 & 94.69 & 0.30 & 0.21 & 119.82 & 1.13 & 0.59 & 143.13 & 0.29 & 0.25 \\
0.057 & 88.66 & 0.12 & 0.14 & 100.84 & 0.32 & 0.22 & 127.75 & 1.21 & 0.63 & 151.99 & 0.30 & 0.26 \\
0.062 & 94.30 & 0.13 & 0.14 & 107.29 & 0.34 & 0.24 & 136.09 & 1.28 & 0.67 & 161.18 & 0.32 & 0.28 \\
0.066 & 100.19 & 0.14 & 0.15 & 114.04 & 0.36 & 0.25 & 144.85 & 1.37 & 0.71 & 170.68 & 0.34 & 0.29 \\
0.071 & 106.32 & 0.15 & 0.16 & 121.08 & 0.38 & 0.27 & 154.03 & 1.45 & 0.76 & 180.45 & 0.36 & 0.31 \\
\hline
\end{tabular}
\end{table*}

\clearpage
\newpage

\begin{table*}
\caption{Circular velocity curves of the 238 (E1--Sdm) CALIFA galaxies with 75th and 25th percentile uncertainties.}
\begin{tabular}{|c|ccc|ccc|ccc|ccc|}
\hline
&\multicolumn{3}{|c|}{NGC0160}&\multicolumn{3}{|c|}{NGC0171}&\multicolumn{3}{|c|}{NGC0177}&\multicolumn{3}{|c|}{NGC0192}\\ 
\hline
$R/R_\mathrm{e}$ & $V_c$ & $\Delta V_c^{75th}$ & $\Delta V_c^{25th}$ & $V_c$ & $\Delta V_c^{75th}$ & $\Delta V_c^{25th}$ & $V_c$ & $\Delta V_c^{75th}$ & $\Delta V_c^{25th}$ & $V_c$ & $\Delta V_c^{75th}$ & $\Delta V_c^{25th}$ \\
\hline
0.05 & 186.33 & 0.54 & 0.26 & 102.28 & 1.03 & 0.35 & 57.93 & 0.12 & 0.14 & 145.93 & 0.21 & 0.17 \\
0.054 & 196.53 & 0.57 & 0.27 & 107.17 & 1.08 & 0.37 & 61.90 & 0.13 & 0.15 & 154.17 & 0.22 & 0.18 \\
0.057 & 206.87 & 0.60 & 0.28 & 111.96 & 1.13 & 0.38 & 66.12 & 0.14 & 0.16 & 162.54 & 0.24 & 0.19 \\
0.062 & 217.26 & 0.63 & 0.30 & 116.59 & 1.18 & 0.40 & 70.61 & 0.15 & 0.17 & 171.00 & 0.25 & 0.20 \\
0.066 & 227.62 & 0.66 & 0.31 & 120.97 & 1.22 & 0.41 & 75.38 & 0.16 & 0.18 & 179.44 & 0.26 & 0.21 \\
0.071 & 237.84 & 0.69 & 0.33 & 125.04 & 1.26 & 0.43 & 80.43 & 0.17 & 0.19 & 187.78 & 0.27 & 0.22 \\
\hline
\end{tabular}
\end{table*}

\clearpage
\newpage

\begin{table*}
\caption{Circular velocity curves of the 238 (E1--Sdm) CALIFA galaxies with 75th and 25th percentile uncertainties.}
\begin{tabular}{|c|ccc|ccc|ccc|ccc|}
\hline
&\multicolumn{3}{|c|}{NGC0214}&\multicolumn{3}{|c|}{NGC0216}&\multicolumn{3}{|c|}{NGC0217}&\multicolumn{3}{|c|}{NGC0237}\\ 
\hline
$R/R_\mathrm{e}$ & $V_c$ & $\Delta V_c^{75th}$ & $\Delta V_c^{25th}$ & $V_c$ & $\Delta V_c^{75th}$ & $\Delta V_c^{25th}$ & $V_c$ & $\Delta V_c^{75th}$ & $\Delta V_c^{25th}$ & $V_c$ & $\Delta V_c^{75th}$ & $\Delta V_c^{25th}$ \\
\hline
0.05 & 97.44 & 0.28 & 0.25 & 18.76 & 0.10 & 0.10 & 103.70 & 0.15 & 0.13 & 40.95 & 0.12 & 0.11 \\
0.054 & 101.92 & 0.30 & 0.26 & 19.81 & 0.10 & 0.11 & 110.26 & 0.16 & 0.14 & 43.50 & 0.13 & 0.12 \\
0.057 & 106.34 & 0.31 & 0.27 & 20.95 & 0.11 & 0.11 & 117.11 & 0.17 & 0.15 & 46.16 & 0.14 & 0.12 \\
0.062 & 110.68 & 0.32 & 0.29 & 22.18 & 0.12 & 0.12 & 124.23 & 0.18 & 0.16 & 48.94 & 0.15 & 0.13 \\
0.066 & 114.92 & 0.34 & 0.30 & 23.49 & 0.13 & 0.13 & 131.60 & 0.19 & 0.17 & 51.82 & 0.16 & 0.14 \\
0.071 & 119.04 & 0.35 & 0.31 & 24.89 & 0.14 & 0.14 & 139.19 & 0.20 & 0.18 & 54.80 & 0.17 & 0.15 \\
\hline
\end{tabular}
\end{table*}

\clearpage
\newpage

\begin{table*}
\caption{Circular velocity curves of the 238 (E1--Sdm) CALIFA galaxies with 75th and 25th percentile uncertainties.}
\begin{tabular}{|c|ccc|ccc|ccc|ccc|}
\hline
&\multicolumn{3}{|c|}{NGC0257}&\multicolumn{3}{|c|}{NGC0429}&\multicolumn{3}{|c|}{NGC0444}&\multicolumn{3}{|c|}{NGC0499}\\ 
\hline
$R/R_\mathrm{e}$ & $V_c$ & $\Delta V_c^{75th}$ & $\Delta V_c^{25th}$ & $V_c$ & $\Delta V_c^{75th}$ & $\Delta V_c^{25th}$ & $V_c$ & $\Delta V_c^{75th}$ & $\Delta V_c^{25th}$ & $V_c$ & $\Delta V_c^{75th}$ & $\Delta V_c^{25th}$ \\
\hline
0.05 & 99.80 & 0.18 & 0.15 & 44.18 & 0.16 & 0.10 & 42.88 & 0.22 & 0.17 & 259.73 & 1.94 & 0.56 \\
0.054 & 105.89 & 0.19 & 0.16 & 47.23 & 0.17 & 0.11 & 45.03 & 0.23 & 0.18 & 273.82 & 2.05 & 0.59 \\
0.057 & 112.20 & 0.20 & 0.17 & 50.49 & 0.18 & 0.11 & 47.17 & 0.24 & 0.19 & 288.06 & 2.15 & 0.62 \\
0.062 & 118.69 & 0.21 & 0.18 & 54.00 & 0.19 & 0.12 & 49.29 & 0.25 & 0.20 & 302.31 & 2.26 & 0.65 \\
0.066 & 125.33 & 0.22 & 0.19 & 57.75 & 0.21 & 0.13 & 51.35 & 0.26 & 0.21 & 316.46 & 2.37 & 0.68 \\
0.071 & 132.08 & 0.24 & 0.20 & 61.75 & 0.22 & 0.14 & 53.35 & 0.27 & 0.22 & 330.34 & 2.47 & 0.71 \\
\hline
\end{tabular}
\end{table*}

\clearpage
\newpage

\begin{table*}
\caption{Circular velocity curves of the 238 (E1--Sdm) CALIFA galaxies with 75th and 25th percentile uncertainties.}
\begin{tabular}{|c|ccc|ccc|ccc|ccc|}
\hline
&\multicolumn{3}{|c|}{NGC0504}&\multicolumn{3}{|c|}{NGC0517}&\multicolumn{3}{|c|}{NGC0528}&\multicolumn{3}{|c|}{NGC0529}\\ 
\hline
$R/R_\mathrm{e}$ & $V_c$ & $\Delta V_c^{75th}$ & $\Delta V_c^{25th}$ & $V_c$ & $\Delta V_c^{75th}$ & $\Delta V_c^{25th}$ & $V_c$ & $\Delta V_c^{75th}$ & $\Delta V_c^{25th}$ & $V_c$ & $\Delta V_c^{75th}$ & $\Delta V_c^{25th}$ \\
\hline
0.05 & 66.61 & 0.09 & 0.09 & 100.08 & 0.41 & 0.22 & 76.92 & 0.12 & 0.10 & 159.40 & 0.31 & 0.20 \\
0.054 & 71.21 & 0.10 & 0.10 & 106.86 & 0.44 & 0.23 & 82.18 & 0.13 & 0.11 & 169.71 & 0.33 & 0.21 \\
0.057 & 76.12 & 0.11 & 0.11 & 114.05 & 0.47 & 0.25 & 87.77 & 0.13 & 0.12 & 180.53 & 0.35 & 0.22 \\
0.062 & 81.35 & 0.11 & 0.11 & 121.65 & 0.50 & 0.27 & 93.70 & 0.14 & 0.13 & 191.82 & 0.37 & 0.24 \\
0.066 & 86.91 & 0.12 & 0.12 & 129.68 & 0.53 & 0.28 & 99.97 & 0.15 & 0.14 & 203.59 & 0.39 & 0.25 \\
0.071 & 92.83 & 0.13 & 0.13 & 138.13 & 0.56 & 0.30 & 106.61 & 0.16 & 0.14 & 215.78 & 0.41 & 0.27 \\
\hline
\end{tabular}
\end{table*}

\clearpage
\newpage

\begin{table*}
\caption{Circular velocity curves of the 238 (E1--Sdm) CALIFA galaxies with 75th and 25th percentile uncertainties.}
\begin{tabular}{|c|ccc|ccc|ccc|ccc|}
\hline
&\multicolumn{3}{|c|}{NGC0551}&\multicolumn{3}{|c|}{NGC0681}&\multicolumn{3}{|c|}{NGC0741}&\multicolumn{3}{|c|}{NGC0755}\\ 
\hline
$R/R_\mathrm{e}$ & $V_c$ & $\Delta V_c^{75th}$ & $\Delta V_c^{25th}$ & $V_c$ & $\Delta V_c^{75th}$ & $\Delta V_c^{25th}$ & $V_c$ & $\Delta V_c^{75th}$ & $\Delta V_c^{25th}$ & $V_c$ & $\Delta V_c^{75th}$ & $\Delta V_c^{25th}$ \\
\hline
0.05 & 61.41 & 0.21 & 0.14 & 133.42 & 0.50 & 0.41 & 422.04 & 0.56 & 0.49 & 15.33 & 0.06 & 0.06 \\
0.054 & 65.22 & 0.22 & 0.15 & 138.13 & 0.51 & 0.42 & 435.61 & 0.58 & 0.51 & 16.22 & 0.07 & 0.06 \\
0.057 & 69.19 & 0.23 & 0.15 & 142.44 & 0.53 & 0.44 & 448.41 & 0.60 & 0.52 & 17.20 & 0.07 & 0.07 \\
0.062 & 73.30 & 0.25 & 0.16 & 146.29 & 0.54 & 0.45 & 460.42 & 0.61 & 0.54 & 18.26 & 0.08 & 0.07 \\
0.066 & 77.53 & 0.26 & 0.17 & 149.62 & 0.56 & 0.46 & 471.64 & 0.63 & 0.55 & 19.40 & 0.08 & 0.08 \\
0.071 & 81.85 & 0.28 & 0.18 & 152.37 & 0.57 & 0.47 & 482.06 & 0.64 & 0.56 & 20.63 & 0.09 & 0.08 \\
\hline
\end{tabular}
\end{table*}

\clearpage
\newpage

\begin{table*}
\caption{Circular velocity curves of the 238 (E1--Sdm) CALIFA galaxies with 75th and 25th percentile uncertainties.}
\begin{tabular}{|c|ccc|ccc|ccc|ccc|}
\hline
&\multicolumn{3}{|c|}{NGC0768}&\multicolumn{3}{|c|}{NGC0774}&\multicolumn{3}{|c|}{NGC0776}&\multicolumn{3}{|c|}{NGC0781}\\ 
\hline
$R/R_\mathrm{e}$ & $V_c$ & $\Delta V_c^{75th}$ & $\Delta V_c^{25th}$ & $V_c$ & $\Delta V_c^{75th}$ & $\Delta V_c^{25th}$ & $V_c$ & $\Delta V_c^{75th}$ & $\Delta V_c^{25th}$ & $V_c$ & $\Delta V_c^{75th}$ & $\Delta V_c^{25th}$ \\
\hline
0.05 & 46.98 & 0.14 & 0.12 & 53.38 & 0.15 & 0.07 & 106.02 & 0.55 & 0.40 & 44.69 & 0.09 & 0.08 \\
0.054 & 50.24 & 0.15 & 0.13 & 57.08 & 0.16 & 0.08 & 112.02 & 0.58 & 0.43 & 47.66 & 0.10 & 0.09 \\
0.057 & 53.72 & 0.16 & 0.13 & 61.04 & 0.17 & 0.08 & 118.13 & 0.61 & 0.45 & 50.82 & 0.10 & 0.09 \\
0.062 & 57.43 & 0.17 & 0.14 & 65.25 & 0.18 & 0.09 & 124.30 & 0.64 & 0.47 & 54.18 & 0.11 & 0.10 \\
0.066 & 61.37 & 0.18 & 0.15 & 69.74 & 0.20 & 0.10 & 130.45 & 0.67 & 0.50 & 57.74 & 0.12 & 0.11 \\
0.071 & 65.56 & 0.19 & 0.16 & 74.52 & 0.21 & 0.10 & 136.53 & 0.70 & 0.52 & 61.50 & 0.12 & 0.11 \\
\hline
\end{tabular}
\end{table*}

\clearpage
\newpage

\begin{table*}
\caption{Circular velocity curves of the 238 (E1--Sdm) CALIFA galaxies with 75th and 25th percentile uncertainties.}
\begin{tabular}{|c|ccc|ccc|ccc|ccc|}
\hline
&\multicolumn{3}{|c|}{NGC0810}&\multicolumn{3}{|c|}{NGC0932}&\multicolumn{3}{|c|}{NGC1056}&\multicolumn{3}{|c|}{NGC1060}\\ 
\hline
$R/R_\mathrm{e}$ & $V_c$ & $\Delta V_c^{75th}$ & $\Delta V_c^{25th}$ & $V_c$ & $\Delta V_c^{75th}$ & $\Delta V_c^{25th}$ & $V_c$ & $\Delta V_c^{75th}$ & $\Delta V_c^{25th}$ & $V_c$ & $\Delta V_c^{75th}$ & $\Delta V_c^{25th}$ \\
\hline
0.05 & 145.82 & 0.28 & 0.35 & 142.98 & 0.30 & 0.21 & 68.39 & 0.31 & 0.20 & 315.97 & 1.17 & 0.64 \\
0.054 & 155.79 & 0.30 & 0.37 & 151.29 & 0.32 & 0.22 & 71.62 & 0.33 & 0.21 & 332.90 & 1.24 & 0.68 \\
0.057 & 166.36 & 0.32 & 0.40 & 159.81 & 0.34 & 0.23 & 74.82 & 0.34 & 0.22 & 350.03 & 1.30 & 0.71 \\
0.062 & 177.56 & 0.34 & 0.42 & 168.47 & 0.36 & 0.24 & 77.94 & 0.36 & 0.23 & 367.25 & 1.36 & 0.75 \\
0.066 & 189.40 & 0.37 & 0.45 & 177.21 & 0.38 & 0.26 & 80.94 & 0.37 & 0.24 & 384.41 & 1.43 & 0.78 \\
0.071 & 201.88 & 0.39 & 0.48 & 185.96 & 0.40 & 0.27 & 83.81 & 0.38 & 0.25 & 401.39 & 1.49 & 0.82 \\
\hline
\end{tabular}
\end{table*}

\clearpage
\newpage

\begin{table*}
\caption{Circular velocity curves of the 238 (E1--Sdm) CALIFA galaxies with 75th and 25th percentile uncertainties.}
\begin{tabular}{|c|ccc|ccc|ccc|ccc|}
\hline
&\multicolumn{3}{|c|}{NGC1167}&\multicolumn{3}{|c|}{NGC1349}&\multicolumn{3}{|c|}{NGC1542}&\multicolumn{3}{|c|}{NGC1645}\\ 
\hline
$R/R_\mathrm{e}$ & $V_c$ & $\Delta V_c^{75th}$ & $\Delta V_c^{25th}$ & $V_c$ & $\Delta V_c^{75th}$ & $\Delta V_c^{25th}$ & $V_c$ & $\Delta V_c^{75th}$ & $\Delta V_c^{25th}$ & $V_c$ & $\Delta V_c^{75th}$ & $\Delta V_c^{25th}$ \\
\hline
0.05 & 201.54 & 0.30 & 0.30 & 173.32 & 0.49 & 0.54 & 39.34 & 0.17 & 0.12 & 112.47 & 0.14 & 0.13 \\
0.054 & 212.57 & 0.32 & 0.31 & 183.47 & 0.52 & 0.58 & 42.02 & 0.18 & 0.13 & 120.08 & 0.15 & 0.14 \\
0.057 & 223.75 & 0.34 & 0.33 & 193.89 & 0.55 & 0.61 & 44.88 & 0.19 & 0.14 & 128.15 & 0.16 & 0.15 \\
0.062 & 234.99 & 0.36 & 0.34 & 204.51 & 0.58 & 0.64 & 47.93 & 0.21 & 0.15 & 136.68 & 0.18 & 0.16 \\
0.066 & 246.19 & 0.37 & 0.36 & 215.25 & 0.61 & 0.68 & 51.18 & 0.22 & 0.16 & 145.67 & 0.19 & 0.17 \\
0.071 & 257.25 & 0.39 & 0.38 & 226.02 & 0.64 & 0.71 & 54.63 & 0.24 & 0.17 & 155.14 & 0.20 & 0.18 \\
\hline
\end{tabular}
\end{table*}

\clearpage
\newpage

\begin{table*}
\caption{Circular velocity curves of the 238 (E1--Sdm) CALIFA galaxies with 75th and 25th percentile uncertainties.}
\begin{tabular}{|c|ccc|ccc|ccc|ccc|}
\hline
&\multicolumn{3}{|c|}{NGC1677}&\multicolumn{3}{|c|}{NGC2253}&\multicolumn{3}{|c|}{NGC2347}&\multicolumn{3}{|c|}{NGC2410}\\ 
\hline
$R/R_\mathrm{e}$ & $V_c$ & $\Delta V_c^{75th}$ & $\Delta V_c^{25th}$ & $V_c$ & $\Delta V_c^{75th}$ & $\Delta V_c^{25th}$ & $V_c$ & $\Delta V_c^{75th}$ & $\Delta V_c^{25th}$ & $V_c$ & $\Delta V_c^{75th}$ & $\Delta V_c^{25th}$ \\
\hline
0.05 & 26.61 & 0.17 & 0.16 & 99.15 & 0.34 & 0.22 & 129.81 & 0.31 & 0.48 & 83.12 & 0.13 & 0.12 \\
0.054 & 28.25 & 0.19 & 0.17 & 105.83 & 0.37 & 0.24 & 134.76 & 0.32 & 0.50 & 88.63 & 0.14 & 0.13 \\
0.057 & 29.99 & 0.20 & 0.18 & 112.90 & 0.39 & 0.25 & 139.75 & 0.33 & 0.52 & 94.45 & 0.15 & 0.14 \\
0.062 & 31.84 & 0.21 & 0.19 & 120.36 & 0.42 & 0.27 & 144.79 & 0.35 & 0.54 & 100.57 & 0.16 & 0.14 \\
0.066 & 33.79 & 0.23 & 0.20 & 128.22 & 0.44 & 0.29 & 149.86 & 0.36 & 0.55 & 106.98 & 0.17 & 0.15 \\
0.071 & 35.85 & 0.24 & 0.22 & 136.48 & 0.47 & 0.31 & 154.96 & 0.37 & 0.57 & 113.68 & 0.18 & 0.16 \\
\hline
\end{tabular}
\end{table*}

\clearpage
\newpage

\begin{table*}
\caption{Circular velocity curves of the 238 (E1--Sdm) CALIFA galaxies with 75th and 25th percentile uncertainties.}
\begin{tabular}{|c|ccc|ccc|ccc|ccc|}
\hline
&\multicolumn{3}{|c|}{NGC2449}&\multicolumn{3}{|c|}{NGC2476}&\multicolumn{3}{|c|}{NGC2481}&\multicolumn{3}{|c|}{NGC2486}\\ 
\hline
$R/R_\mathrm{e}$ & $V_c$ & $\Delta V_c^{75th}$ & $\Delta V_c^{25th}$ & $V_c$ & $\Delta V_c^{75th}$ & $\Delta V_c^{25th}$ & $V_c$ & $\Delta V_c^{75th}$ & $\Delta V_c^{25th}$ & $V_c$ & $\Delta V_c^{75th}$ & $\Delta V_c^{25th}$ \\
\hline
0.05 & 91.43 & 0.16 & 0.13 & 92.68 & 0.30 & 0.26 & 66.05 & 0.13 & 0.10 & 296.98 & 0.87 & 0.81 \\
0.054 & 96.97 & 0.17 & 0.13 & 98.94 & 0.32 & 0.28 & 70.47 & 0.13 & 0.11 & 297.08 & 0.87 & 0.81 \\
0.057 & 102.71 & 0.18 & 0.14 & 105.57 & 0.34 & 0.30 & 75.17 & 0.14 & 0.12 & 295.68 & 0.86 & 0.81 \\
0.062 & 108.62 & 0.19 & 0.15 & 112.58 & 0.36 & 0.32 & 80.16 & 0.15 & 0.13 & 293.00 & 0.86 & 0.80 \\
0.066 & 114.67 & 0.20 & 0.16 & 119.99 & 0.38 & 0.34 & 85.44 & 0.16 & 0.13 & 289.41 & 0.85 & 0.79 \\
0.071 & 120.84 & 0.21 & 0.17 & 127.78 & 0.41 & 0.36 & 91.02 & 0.18 & 0.14 & 285.33 & 0.83 & 0.78 \\
\hline
\end{tabular}
\end{table*}

\clearpage
\newpage

\begin{table*}
\caption{Circular velocity curves of the 238 (E1--Sdm) CALIFA galaxies with 75th and 25th percentile uncertainties.}
\begin{tabular}{|c|ccc|ccc|ccc|ccc|}
\hline
&\multicolumn{3}{|c|}{NGC2553}&\multicolumn{3}{|c|}{NGC2554}&\multicolumn{3}{|c|}{NGC2592}&\multicolumn{3}{|c|}{NGC2604}\\ 
\hline
$R/R_\mathrm{e}$ & $V_c$ & $\Delta V_c^{75th}$ & $\Delta V_c^{25th}$ & $V_c$ & $\Delta V_c^{75th}$ & $\Delta V_c^{25th}$ & $V_c$ & $\Delta V_c^{75th}$ & $\Delta V_c^{25th}$ & $V_c$ & $\Delta V_c^{75th}$ & $\Delta V_c^{25th}$ \\
\hline
0.05 & 89.83 & 0.21 & 0.21 & 186.52 & 0.25 & 0.19 & 79.94 & 0.07 & 0.07 & 41.29 & 0.21 & 0.20 \\
0.054 & 95.83 & 0.23 & 0.23 & 197.96 & 0.26 & 0.20 & 85.37 & 0.08 & 0.08 & 43.34 & 0.22 & 0.21 \\
0.057 & 102.17 & 0.24 & 0.24 & 209.80 & 0.28 & 0.21 & 91.15 & 0.09 & 0.08 & 45.38 & 0.23 & 0.22 \\
0.062 & 108.87 & 0.26 & 0.26 & 222.00 & 0.29 & 0.23 & 97.30 & 0.09 & 0.09 & 47.40 & 0.24 & 0.23 \\
0.066 & 115.92 & 0.28 & 0.27 & 234.50 & 0.31 & 0.24 & 103.83 & 0.10 & 0.10 & 49.37 & 0.25 & 0.24 \\
0.071 & 123.32 & 0.29 & 0.29 & 247.19 & 0.32 & 0.25 & 110.74 & 0.10 & 0.10 & 51.27 & 0.26 & 0.25 \\
\hline
\end{tabular}
\end{table*}

\clearpage
\newpage

\begin{table*}
\caption{Circular velocity curves of the 238 (E1--Sdm) CALIFA galaxies with 75th and 25th percentile uncertainties.}
\begin{tabular}{|c|ccc|ccc|ccc|ccc|}
\hline
&\multicolumn{3}{|c|}{NGC2639}&\multicolumn{3}{|c|}{NGC2730}&\multicolumn{3}{|c|}{NGC2880}&\multicolumn{3}{|c|}{NGC2906}\\ 
\hline
$R/R_\mathrm{e}$ & $V_c$ & $\Delta V_c^{75th}$ & $\Delta V_c^{25th}$ & $V_c$ & $\Delta V_c^{75th}$ & $\Delta V_c^{25th}$ & $V_c$ & $\Delta V_c^{75th}$ & $\Delta V_c^{25th}$ & $V_c$ & $\Delta V_c^{75th}$ & $\Delta V_c^{25th}$ \\
\hline
0.05 & 106.10 & 0.14 & 0.28 & 29.29 & 0.13 & 0.09 & 106.32 & 0.12 & 0.10 & 116.87 & 0.23 & 0.20 \\
0.054 & 112.92 & 0.14 & 0.30 & 31.22 & 0.14 & 0.10 & 112.91 & 0.12 & 0.11 & 122.40 & 0.24 & 0.21 \\
0.057 & 120.05 & 0.15 & 0.32 & 33.27 & 0.15 & 0.10 & 119.75 & 0.13 & 0.11 & 127.83 & 0.25 & 0.22 \\
0.062 & 127.50 & 0.16 & 0.34 & 35.44 & 0.16 & 0.11 & 126.85 & 0.14 & 0.12 & 133.10 & 0.26 & 0.22 \\
0.066 & 135.25 & 0.17 & 0.36 & 37.72 & 0.17 & 0.12 & 134.15 & 0.15 & 0.13 & 138.14 & 0.27 & 0.23 \\
0.071 & 143.27 & 0.18 & 0.38 & 40.13 & 0.18 & 0.12 & 141.62 & 0.16 & 0.13 & 142.87 & 0.28 & 0.24 \\
\hline
\end{tabular}
\end{table*}

\clearpage
\newpage

\begin{table*}
\caption{Circular velocity curves of the 238 (E1--Sdm) CALIFA galaxies with 75th and 25th percentile uncertainties.}
\begin{tabular}{|c|ccc|ccc|ccc|ccc|}
\hline
&\multicolumn{3}{|c|}{NGC2916}&\multicolumn{3}{|c|}{NGC2918}&\multicolumn{3}{|c|}{NGC3057}&\multicolumn{3}{|c|}{NGC3106}\\ 
\hline
$R/R_\mathrm{e}$ & $V_c$ & $\Delta V_c^{75th}$ & $\Delta V_c^{25th}$ & $V_c$ & $\Delta V_c^{75th}$ & $\Delta V_c^{25th}$ & $V_c$ & $\Delta V_c^{75th}$ & $\Delta V_c^{25th}$ & $V_c$ & $\Delta V_c^{75th}$ & $\Delta V_c^{25th}$ \\
\hline
0.05 & 110.07 & 0.12 & 0.12 & 95.75 & 0.15 & 0.12 & 23.97 & 0.18 & 0.18 & 205.63 & 0.36 & 0.36 \\
0.054 & 116.01 & 0.13 & 0.12 & 102.36 & 0.16 & 0.12 & 25.19 & 0.19 & 0.19 & 216.26 & 0.38 & 0.38 \\
0.057 & 122.02 & 0.14 & 0.13 & 109.40 & 0.17 & 0.13 & 26.43 & 0.21 & 0.20 & 226.86 & 0.40 & 0.40 \\
0.062 & 128.04 & 0.14 & 0.13 & 116.88 & 0.18 & 0.14 & 27.67 & 0.22 & 0.21 & 237.30 & 0.42 & 0.42 \\
0.066 & 134.02 & 0.15 & 0.14 & 124.81 & 0.19 & 0.15 & 28.92 & 0.23 & 0.22 & 247.46 & 0.44 & 0.44 \\
0.071 & 139.91 & 0.16 & 0.15 & 133.21 & 0.20 & 0.16 & 30.16 & 0.24 & 0.23 & 257.16 & 0.45 & 0.46 \\
\hline
\end{tabular}
\end{table*}

\clearpage
\newpage

\begin{table*}
\caption{Circular velocity curves of the 238 (E1--Sdm) CALIFA galaxies with 75th and 25th percentile uncertainties.}
\begin{tabular}{|c|ccc|ccc|ccc|ccc|}
\hline
&\multicolumn{3}{|c|}{NGC3160}&\multicolumn{3}{|c|}{NGC3300}&\multicolumn{3}{|c|}{NGC3381}&\multicolumn{3}{|c|}{NGC3615}\\ 
\hline
$R/R_\mathrm{e}$ & $V_c$ & $\Delta V_c^{75th}$ & $\Delta V_c^{25th}$ & $V_c$ & $\Delta V_c^{75th}$ & $\Delta V_c^{25th}$ & $V_c$ & $\Delta V_c^{75th}$ & $\Delta V_c^{25th}$ & $V_c$ & $\Delta V_c^{75th}$ & $\Delta V_c^{25th}$ \\
\hline
0.05 & 92.37 & 0.21 & 0.25 & 72.53 & 0.16 & 0.10 & 54.22 & 0.40 & 0.32 & 240.93 & 0.83 & 0.59 \\
0.054 & 98.02 & 0.23 & 0.27 & 77.37 & 0.17 & 0.10 & 53.56 & 0.39 & 0.32 & 255.52 & 0.88 & 0.62 \\
0.057 & 103.88 & 0.24 & 0.28 & 82.50 & 0.18 & 0.11 & 52.93 & 0.39 & 0.31 & 270.60 & 0.93 & 0.66 \\
0.062 & 109.92 & 0.25 & 0.30 & 87.92 & 0.19 & 0.12 & 52.35 & 0.38 & 0.31 & 286.08 & 0.99 & 0.70 \\
0.066 & 116.11 & 0.27 & 0.32 & 93.62 & 0.20 & 0.13 & 51.84 & 0.38 & 0.31 & 301.88 & 1.04 & 0.74 \\
0.071 & 122.42 & 0.28 & 0.33 & 99.61 & 0.22 & 0.13 & 51.42 & 0.38 & 0.30 & 317.87 & 1.09 & 0.78 \\
\hline
\end{tabular}
\end{table*}

\clearpage
\newpage

\begin{table*}
\caption{Circular velocity curves of the 238 (E1--Sdm) CALIFA galaxies with 75th and 25th percentile uncertainties.}
\begin{tabular}{|c|ccc|ccc|ccc|ccc|}
\hline
&\multicolumn{3}{|c|}{NGC3811}&\multicolumn{3}{|c|}{NGC3815}&\multicolumn{3}{|c|}{NGC3994}&\multicolumn{3}{|c|}{NGC4003}\\ 
\hline
$R/R_\mathrm{e}$ & $V_c$ & $\Delta V_c^{75th}$ & $\Delta V_c^{25th}$ & $V_c$ & $\Delta V_c^{75th}$ & $\Delta V_c^{25th}$ & $V_c$ & $\Delta V_c^{75th}$ & $\Delta V_c^{25th}$ & $V_c$ & $\Delta V_c^{75th}$ & $\Delta V_c^{25th}$ \\
\hline
0.05 & 171.24 & 0.32 & 0.27 & 54.22 & 0.11 & 0.10 & 71.71 & 0.22 & 0.18 & 170.63 & 0.53 & 0.45 \\
0.054 & 179.48 & 0.34 & 0.28 & 57.75 & 0.12 & 0.10 & 76.59 & 0.24 & 0.19 & 174.74 & 0.55 & 0.47 \\
0.057 & 187.54 & 0.35 & 0.29 & 61.48 & 0.12 & 0.11 & 81.80 & 0.26 & 0.20 & 178.17 & 0.56 & 0.47 \\
0.062 & 195.28 & 0.37 & 0.30 & 65.40 & 0.13 & 0.12 & 87.33 & 0.27 & 0.22 & 180.88 & 0.57 & 0.48 \\
0.066 & 202.55 & 0.38 & 0.31 & 69.51 & 0.14 & 0.12 & 93.19 & 0.29 & 0.23 & 182.90 & 0.57 & 0.49 \\
0.071 & 209.19 & 0.39 & 0.32 & 73.79 & 0.15 & 0.13 & 99.40 & 0.31 & 0.25 & 184.27 & 0.58 & 0.49 \\
\hline
\end{tabular}
\end{table*}

\clearpage
\newpage

\begin{table*}
\caption{Circular velocity curves of the 238 (E1--Sdm) CALIFA galaxies with 75th and 25th percentile uncertainties.}
\begin{tabular}{|c|ccc|ccc|ccc|ccc|}
\hline
&\multicolumn{3}{|c|}{NGC4047}&\multicolumn{3}{|c|}{NGC4149}&\multicolumn{3}{|c|}{NGC4185}&\multicolumn{3}{|c|}{NGC4210}\\ 
\hline
$R/R_\mathrm{e}$ & $V_c$ & $\Delta V_c^{75th}$ & $\Delta V_c^{25th}$ & $V_c$ & $\Delta V_c^{75th}$ & $\Delta V_c^{25th}$ & $V_c$ & $\Delta V_c^{75th}$ & $\Delta V_c^{25th}$ & $V_c$ & $\Delta V_c^{75th}$ & $\Delta V_c^{25th}$ \\
\hline
0.05 & 76.98 & 0.25 & 0.17 & 57.28 & 0.15 & 0.13 & 84.39 & 0.26 & 0.16 & 74.34 & 0.13 & 0.11 \\
0.054 & 81.38 & 0.27 & 0.18 & 61.03 & 0.16 & 0.14 & 88.31 & 0.27 & 0.17 & 78.67 & 0.14 & 0.12 \\
0.057 & 85.87 & 0.28 & 0.19 & 64.97 & 0.17 & 0.15 & 92.38 & 0.29 & 0.18 & 83.12 & 0.14 & 0.13 \\
0.062 & 90.42 & 0.30 & 0.20 & 69.12 & 0.18 & 0.16 & 96.58 & 0.30 & 0.18 & 87.65 & 0.15 & 0.13 \\
0.066 & 95.01 & 0.31 & 0.21 & 73.46 & 0.19 & 0.17 & 100.89 & 0.31 & 0.19 & 92.22 & 0.16 & 0.14 \\
0.071 & 99.58 & 0.33 & 0.22 & 77.99 & 0.20 & 0.18 & 105.27 & 0.33 & 0.20 & 96.80 & 0.17 & 0.15 \\
\hline
\end{tabular}
\end{table*}

\clearpage
\newpage

\begin{table*}
\caption{Circular velocity curves of the 238 (E1--Sdm) CALIFA galaxies with 75th and 25th percentile uncertainties.}
\begin{tabular}{|c|ccc|ccc|ccc|ccc|}
\hline
&\multicolumn{3}{|c|}{NGC4470}&\multicolumn{3}{|c|}{NGC4644}&\multicolumn{3}{|c|}{NGC4711}&\multicolumn{3}{|c|}{NGC4816}\\ 
\hline
$R/R_\mathrm{e}$ & $V_c$ & $\Delta V_c^{75th}$ & $\Delta V_c^{25th}$ & $V_c$ & $\Delta V_c^{75th}$ & $\Delta V_c^{25th}$ & $V_c$ & $\Delta V_c^{75th}$ & $\Delta V_c^{25th}$ & $V_c$ & $\Delta V_c^{75th}$ & $\Delta V_c^{25th}$ \\
\hline
0.05 & 23.42 & 0.16 & 0.10 & 44.59 & 0.07 & 0.08 & 38.69 & 0.09 & 0.08 & 364.02 & 0.47 & 0.47 \\
0.054 & 23.56 & 0.16 & 0.10 & 47.62 & 0.08 & 0.08 & 41.10 & 0.09 & 0.09 & 371.09 & 0.48 & 0.48 \\
0.057 & 23.60 & 0.16 & 0.10 & 50.84 & 0.08 & 0.09 & 43.63 & 0.10 & 0.09 & 377.20 & 0.49 & 0.49 \\
0.062 & 23.54 & 0.16 & 0.10 & 54.27 & 0.09 & 0.09 & 46.26 & 0.11 & 0.10 & 382.41 & 0.50 & 0.49 \\
0.066 & 23.41 & 0.16 & 0.10 & 57.91 & 0.10 & 0.10 & 48.99 & 0.11 & 0.10 & 386.77 & 0.50 & 0.50 \\
0.071 & 23.23 & 0.16 & 0.10 & 61.76 & 0.10 & 0.11 & 51.80 & 0.12 & 0.11 & 390.36 & 0.51 & 0.50 \\
\hline
\end{tabular}
\end{table*}

\clearpage
\newpage

\begin{table*}
\caption{Circular velocity curves of the 238 (E1--Sdm) CALIFA galaxies with 75th and 25th percentile uncertainties.}
\begin{tabular}{|c|ccc|ccc|ccc|ccc|}
\hline
&\multicolumn{3}{|c|}{NGC4956}&\multicolumn{3}{|c|}{NGC4961}&\multicolumn{3}{|c|}{NGC5000}&\multicolumn{3}{|c|}{NGC5029}\\ 
\hline
$R/R_\mathrm{e}$ & $V_c$ & $\Delta V_c^{75th}$ & $\Delta V_c^{25th}$ & $V_c$ & $\Delta V_c^{75th}$ & $\Delta V_c^{25th}$ & $V_c$ & $\Delta V_c^{75th}$ & $\Delta V_c^{25th}$ & $V_c$ & $\Delta V_c^{75th}$ & $\Delta V_c^{25th}$ \\
\hline
0.05 & 67.04 & 0.13 & 0.11 & 34.61 & 0.12 & 0.11 & 38.96 & 0.12 & 0.11 & 239.35 & 0.32 & 0.31 \\
0.054 & 71.64 & 0.14 & 0.12 & 36.83 & 0.13 & 0.12 & 41.35 & 0.12 & 0.12 & 253.00 & 0.34 & 0.33 \\
0.057 & 76.53 & 0.15 & 0.13 & 39.17 & 0.13 & 0.13 & 43.84 & 0.13 & 0.13 & 266.94 & 0.36 & 0.35 \\
0.062 & 81.73 & 0.16 & 0.13 & 41.64 & 0.14 & 0.14 & 46.43 & 0.14 & 0.13 & 281.05 & 0.38 & 0.37 \\
0.066 & 87.25 & 0.17 & 0.14 & 44.24 & 0.15 & 0.15 & 49.10 & 0.15 & 0.14 & 295.22 & 0.40 & 0.39 \\
0.071 & 93.10 & 0.19 & 0.15 & 46.97 & 0.16 & 0.16 & 51.84 & 0.16 & 0.15 & 309.31 & 0.42 & 0.40 \\
\hline
\end{tabular}
\end{table*}

\clearpage
\newpage

\begin{table*}
\caption{Circular velocity curves of the 238 (E1--Sdm) CALIFA galaxies with 75th and 25th percentile uncertainties.}
\begin{tabular}{|c|ccc|ccc|ccc|ccc|}
\hline
&\multicolumn{3}{|c|}{NGC5056}&\multicolumn{3}{|c|}{NGC5218}&\multicolumn{3}{|c|}{NGC5378}&\multicolumn{3}{|c|}{NGC5480}\\ 
\hline
$R/R_\mathrm{e}$ & $V_c$ & $\Delta V_c^{75th}$ & $\Delta V_c^{25th}$ & $V_c$ & $\Delta V_c^{75th}$ & $\Delta V_c^{25th}$ & $V_c$ & $\Delta V_c^{75th}$ & $\Delta V_c^{25th}$ & $V_c$ & $\Delta V_c^{75th}$ & $\Delta V_c^{25th}$ \\
\hline
0.05 & 69.17 & 0.12 & 0.10 & 25.86 & 0.11 & 0.05 & 139.64 & 0.34 & 0.31 & 92.24 & 0.41 & 0.29 \\
0.054 & 73.52 & 0.13 & 0.10 & 27.61 & 0.12 & 0.05 & 146.98 & 0.36 & 0.33 & 93.07 & 0.41 & 0.29 \\
0.057 & 78.06 & 0.14 & 0.11 & 29.49 & 0.13 & 0.06 & 154.33 & 0.38 & 0.35 & 94.17 & 0.42 & 0.30 \\
0.062 & 82.78 & 0.15 & 0.12 & 31.52 & 0.14 & 0.06 & 161.61 & 0.40 & 0.36 & 95.53 & 0.43 & 0.30 \\
0.066 & 87.65 & 0.16 & 0.12 & 33.70 & 0.15 & 0.07 & 168.72 & 0.41 & 0.38 & 97.09 & 0.43 & 0.31 \\
0.071 & 92.66 & 0.17 & 0.13 & 36.03 & 0.16 & 0.07 & 175.57 & 0.43 & 0.39 & 98.77 & 0.44 & 0.31 \\
\hline
\end{tabular}
\end{table*}

\clearpage
\newpage

\begin{table*}
\caption{Circular velocity curves of the 238 (E1--Sdm) CALIFA galaxies with 75th and 25th percentile uncertainties.}
\begin{tabular}{|c|ccc|ccc|ccc|ccc|}
\hline
&\multicolumn{3}{|c|}{NGC5485}&\multicolumn{3}{|c|}{NGC5520}&\multicolumn{3}{|c|}{NGC5614}&\multicolumn{3}{|c|}{NGC5630}\\ 
\hline
$R/R_\mathrm{e}$ & $V_c$ & $\Delta V_c^{75th}$ & $\Delta V_c^{25th}$ & $V_c$ & $\Delta V_c^{75th}$ & $\Delta V_c^{25th}$ & $V_c$ & $\Delta V_c^{75th}$ & $\Delta V_c^{25th}$ & $V_c$ & $\Delta V_c^{75th}$ & $\Delta V_c^{25th}$ \\
\hline
0.05 & 217.52 & 0.27 & 0.31 & 84.94 & 0.20 & 0.16 & 111.82 & 0.22 & 0.19 & 33.22 & 0.12 & 0.11 \\
0.054 & 222.44 & 0.28 & 0.32 & 88.68 & 0.21 & 0.17 & 119.18 & 0.24 & 0.20 & 34.98 & 0.13 & 0.11 \\
0.057 & 227.12 & 0.28 & 0.32 & 92.32 & 0.22 & 0.17 & 126.92 & 0.25 & 0.21 & 36.76 & 0.13 & 0.12 \\
0.062 & 231.63 & 0.29 & 0.33 & 95.80 & 0.23 & 0.18 & 135.05 & 0.27 & 0.22 & 38.56 & 0.14 & 0.13 \\
0.066 & 236.05 & 0.29 & 0.34 & 99.09 & 0.23 & 0.19 & 143.56 & 0.28 & 0.24 & 40.34 & 0.15 & 0.13 \\
0.071 & 240.44 & 0.30 & 0.34 & 102.15 & 0.24 & 0.19 & 152.42 & 0.30 & 0.25 & 42.10 & 0.15 & 0.14 \\
\hline
\end{tabular}
\end{table*}

\clearpage
\newpage

\begin{table*}
\caption{Circular velocity curves of the 238 (E1--Sdm) CALIFA galaxies with 75th and 25th percentile uncertainties.}
\begin{tabular}{|c|ccc|ccc|ccc|ccc|}
\hline
&\multicolumn{3}{|c|}{NGC5633}&\multicolumn{3}{|c|}{NGC5657}&\multicolumn{3}{|c|}{NGC5682}&\multicolumn{3}{|c|}{NGC5720}\\ 
\hline
$R/R_\mathrm{e}$ & $V_c$ & $\Delta V_c^{75th}$ & $\Delta V_c^{25th}$ & $V_c$ & $\Delta V_c^{75th}$ & $\Delta V_c^{25th}$ & $V_c$ & $\Delta V_c^{75th}$ & $\Delta V_c^{25th}$ & $V_c$ & $\Delta V_c^{75th}$ & $\Delta V_c^{25th}$ \\
\hline
0.05 & 27.69 & 0.09 & 0.11 & 43.14 & 0.07 & 0.07 & 28.90 & 0.16 & 0.15 & 269.03 & 0.43 & 0.50 \\
0.054 & 29.39 & 0.09 & 0.12 & 46.08 & 0.08 & 0.07 & 30.84 & 0.18 & 0.16 & 272.87 & 0.43 & 0.51 \\
0.057 & 31.20 & 0.10 & 0.13 & 49.22 & 0.08 & 0.08 & 32.92 & 0.19 & 0.17 & 275.27 & 0.44 & 0.51 \\
0.062 & 33.12 & 0.11 & 0.14 & 52.56 & 0.09 & 0.08 & 35.13 & 0.20 & 0.18 & 276.20 & 0.44 & 0.51 \\
0.066 & 35.15 & 0.11 & 0.15 & 56.13 & 0.10 & 0.09 & 37.49 & 0.22 & 0.19 & 275.75 & 0.44 & 0.51 \\
0.071 & 37.29 & 0.12 & 0.16 & 59.93 & 0.10 & 0.10 & 40.00 & 0.23 & 0.21 & 274.12 & 0.44 & 0.51 \\
\hline
\end{tabular}
\end{table*}

\clearpage
\newpage

\begin{table*}
\caption{Circular velocity curves of the 238 (E1--Sdm) CALIFA galaxies with 75th and 25th percentile uncertainties.}
\begin{tabular}{|c|ccc|ccc|ccc|ccc|}
\hline
&\multicolumn{3}{|c|}{NGC5732}&\multicolumn{3}{|c|}{NGC5784}&\multicolumn{3}{|c|}{NGC5876}&\multicolumn{3}{|c|}{NGC5888}\\ 
\hline
$R/R_\mathrm{e}$ & $V_c$ & $\Delta V_c^{75th}$ & $\Delta V_c^{25th}$ & $V_c$ & $\Delta V_c^{75th}$ & $\Delta V_c^{25th}$ & $V_c$ & $\Delta V_c^{75th}$ & $\Delta V_c^{25th}$ & $V_c$ & $\Delta V_c^{75th}$ & $\Delta V_c^{25th}$ \\
\hline
0.05 & 35.54 & 0.19 & 0.16 & 159.10 & 0.17 & 0.15 & 89.31 & 0.11 & 0.12 & 108.22 & 0.12 & 0.10 \\
0.054 & 37.89 & 0.20 & 0.17 & 169.54 & 0.18 & 0.16 & 95.47 & 0.12 & 0.13 & 115.02 & 0.13 & 0.11 \\
0.057 & 40.40 & 0.21 & 0.18 & 180.53 & 0.19 & 0.17 & 102.03 & 0.13 & 0.14 & 122.11 & 0.14 & 0.11 \\
0.062 & 43.05 & 0.23 & 0.20 & 192.04 & 0.20 & 0.18 & 109.01 & 0.14 & 0.15 & 129.46 & 0.15 & 0.12 \\
0.066 & 45.86 & 0.24 & 0.21 & 204.06 & 0.22 & 0.19 & 116.42 & 0.14 & 0.16 & 137.05 & 0.16 & 0.13 \\
0.071 & 48.83 & 0.26 & 0.22 & 216.58 & 0.23 & 0.20 & 124.27 & 0.15 & 0.17 & 144.84 & 0.17 & 0.14 \\
\hline
\end{tabular}%
\end{table*}

\clearpage
\newpage

\begin{table*}
\caption{Circular velocity curves of the 238 (E1--Sdm) CALIFA galaxies with 75th and 25th percentile uncertainties.}
\begin{center}
%
\label{tab:CVC60}
\end{center}
\end{table*}

\clearpage
\newpage

\bsp 

\label{lastpage}

\end{document}